\documentclass[aps,prl,reprint,superscriptaddress,showkeys,amsmath,amssymb,floatfix,nofootinbib]{revtex4-2}

\usepackage{caption}
\usepackage{multirow, graphicx,amssymb,url,mathrsfs,amsmath}
\usepackage{wrapfig,boxedminipage,setspace,epsfig}
\usepackage{amsxtra,amstext,latexsym,dsfont,amsfonts}
\usepackage{color,eucal}
\usepackage[dvipsnames]{xcolor}
\usepackage{float}
\usepackage{slashed,comment}
\usepackage{cases}
\usepackage{caption}
\usepackage{dcolumn}
\usepackage{bm}

\usepackage[dvipsnames]{xcolor}
\usepackage{amsfonts}
\usepackage{siunitx}[=v2]
\usepackage{setspace}
\usepackage{cancel}
\usepackage{hyperref}

\usepackage{tabularx, array}
\newcolumntype{L}[1]{>{\raggedright\arraybackslash}p{#1}}
\newcolumntype{C}[1]{>{\centering\arraybackslash}p{#1}}
\newcolumntype{R}[1]{>{\raggedleft\arraybackslash}p{#1}}

       \def\e  {\epsilon}

\renewcommand{\a}{\alpha}      
\renewcommand{\d}{\delta}      
\renewcommand{\l}{\lambda}

\newcommand{\ra}{\rightarrow}

\newcommand{\bra}[1]{\mbox{$\langle #1 |$}}
\newcommand{\ket}[1]{\mbox{$| #1 \rangle$}}

\newcommand{\braket}[2]{\mbox{$\langle #1  | #2 \rangle$}}

\newcommand{\Tr}{{\rm Tr}\,}

\newcommand{\sech}{\text{sech}}

\newcommand{\HC}[1]{{\color{red}{#1}}}
\newcommand{\blue}[1]{{\color{blue}{#1}}}
\newcommand{\red}[1]{{\color{red}{#1}}}

\begin{document}


\title{Toward a Refinement of Krylov Complexity:\\ Scrambling, Classical Operator Growth and Replicas}

\author{Hugo A. Camargo}
\email[]{hugo.camargo@phys.ncts.ntu.edu.tw}
\affiliation{Physics Division, National Center for Theoretical Sciences, National Taiwan University, Taipei 106319, Taiwan}
\affiliation{Department of Physics, National Sun Yat-sen University, Kaohsiung 80424, Taiwan}
\affiliation{Department of Physics and Photon Science, Gwangju Institute of Science and Technology, 
123 Cheomdan-gwagiro, Gwangju 61005, Korea}

\author{Yichao Fu}
\email[]{yichao.fu@gm.gist.ac.kr}
\affiliation{Department of Physics and Photon Science, Gwangju Institute of Science and Technology, 
123 Cheomdan-gwagiro, Gwangju 61005, Korea}

\author{Keun-Young Kim}
\email[]{fortoe@gist.ac.kr}
\affiliation{Department of Physics and Photon Science, Gwangju Institute of Science and Technology,
123 Cheomdan-gwagiro, Gwangju 61005, Korea}
\affiliation{Research Center for Photon Science Technology, Gwangju Institute of Science and Technology, 123 Cheomdan-gwagiro, Gwangju 61005, Korea}

\author{Yeong Han Park}
\email[]{yeonghanp@gm.gist.ac.kr}
\affiliation{Department of Physics and Photon Science, Gwangju Institute of Science and Technology,
123 Cheomdan-gwagiro, Gwangju 61005, Korea}

\date{\today}

\begin{abstract}
We propose and test logarithmic Krylov (logK) complexity, an operator growth measure akin to Krylov complexity defined through a replica approach, as a viable probe of early-time operator scrambling without false positives. In finite-dimensional quantum systems, such as the Lipkin--Meshkov--Glick (LMG) model and the mixed-field Ising model at the chaotic point, we provide numerical evidence that logK-complexity discriminates between genuine and saddle-dominated scrambling at early times, correctly avoiding the exponential contribution coming from the unstable saddle in the former case, and closely tracking the conventional Krylov complexity in the latter. In integrable quantum systems admitting infinite-dimensional Krylov subspaces, such as the SYK$_{2}$ model and the quantum inverted harmonic oscillator, we show that by modifying the Krylov spreading operator, obtained through generalizing the analytic continuation procedure in the replica trick, the logK complexity can be refined to capture the integrable properties of the theories. We supplement these analyses by extending the Krylov formalism in classical dynamical systems and defining classical versions of these operator growth measures, showing that the false positives arising from unstable saddles in classical phase space are non-existent.
\end{abstract}


\maketitle

\phantomsection
\label{sec:intro}
I. \textit{Introduction}-- 
Over the past several decades, chaos has been a central topic in a wide range of fields, including physics, mathematics, biology, and even neural networks~\cite{gutzwiller1991chaos, Banks_Dragan_Jones_2003, degn2013chaos,NatureChaos1,NatureChaos2}. In classical dynamical systems, rigorously described within the framework of ergodic theory~\cite{Berry1985SemiclassicalTO, walters2000introduction}, chaos (mixing) reflects the unpredictability of non-integrable dynamics and sensitivity to initial conditions. In classical K-mixing systems, this behavior is quantitatively characterized by the presence of a positive Lyapunov exponent. In contrast, chaos in quantum systems still lacks a rigorous formulation, although more recent efforts have also made progress toward a description of this phenomenon in terms of a quantum ergodic theory (see, e.g.~\cite{Gesteau:2023rrx,Ouseph:2023juq,Camargo:2025zxr}). In practical terms, quantum chaos in closed quantum systems is generally associated with thermalization~\cite{Deutsch:1991msp, Srednicki:1994mfb, Rigol:2007juv} and information scrambling~\cite{Xu:2022vko}, although it is often operationally defined through the presence of Hamiltonian spectral statistics and eigenvector distributions consistent with those of random matrices, following semiclassical arguments~\cite{Bohigas:1983er}. In terms of operator growth, out-of-time order correlators (OTOCs)~\cite{1969JETP...28.1200L} are a well-studied probe of scrambling. These have also provided powerful insights into the physics of black holes in holography~\cite{Sekino:2008he, Shenker:2013pqa, Roberts:2014isa, Shenker:2014cwa,Maldacena:2015waa} and in condensed matter systems~\cite{Xu:2022vko, Fan_2017}.\\
$\phantom{word}$Scrambling generally refers to a phenomenon in interacting quantum systems in which initially localized information spreads over a large number of degrees of freedom by time evolution, essentially rendering it inaccessible by any local measurement~\cite{Xu:2022vko}. In thermalizing quantum systems with a large number of degrees of freedom per site, it is typically identified by an early-time exponential growth of the double commutator 
\begin{eqnarray}
    -\langle[\hat{V}(t),\hat{W}]^2\rangle&=&\langle \hat{W}^\dagger \hat{V}^\dagger(t)\hat{V}(t)\hat{W}\rangle+\langle \hat{V}^\dagger(t)\hat{W}^\dagger \hat{W} \hat{V}(t) \rangle\nonumber\\
    &&-2\mathrm{Re}(\langle \hat{V}^\dagger(t)\hat{W}^\dagger \hat{V}(t)\hat{W} \rangle )~, 
\end{eqnarray}
where $\hat{V},\,\hat{W}$ are local operators and where $\langle \hat{V}^\dagger(t)\hat{W}^\dagger \hat{V}(t)\hat{W} \rangle\equiv \mathrm{OTOC}(t)$. In this context, the exponent is considered to be the quantum counterpart of the classical Lyapunov exponent.~\footnote{We should remark, nonetheless, that some authors refer to scrambling as the late-time vanishing of out-of-time order correlators (OTOCs), which is independent of an early-time exponential decay.} Recently, a different notion of operator growth known as Krylov complexity~\cite{Parker:2018yvk} has also been shown to have a regime of exponential growth in non-integrable many-body quantum systems. A generalization of this notion to quantum states, Krylov state (or spread) complexity~\cite{Balasubramanian:2022tpr} has also been shown to acquire characteristic features in such systems ~\cite{Espanol:2022cqr, Hashimoto:2023swv, Camargo:2023eev, Bhattacharjee:2024yxj, Baggioli:2024wbz, Camargo:2024deu, Alishahiha:2024vbf, Grabarits:2025xys}. However, scrambling is not unique to quantum systems with random matrix spectral statistics, since it can also emerge in integrable quantum systems that possess isolated unstable saddle points \cite{Xu:2019lhc, Bhattacharjee:2022vlt, Huh:2023jxt, Aguilar-Gutierrez:2025hbf}, a finding usually summarized in the statement that ``scrambling is necessary but not sufficient for chaos''~\cite{Dowling:2023hqc}. Given the close connection between the early-time dynamics of OTOCs and the semiclassical butterfly effect, as well as the intimate relation between the former and Krylov complexity, it is an important endeavor to understand how to avoid these false positives when probing the early-time dynamics of non-integrable quantum systems using different notions of operator growth.\\
$\phantom{word}$Motivated by recent work that tackles this problem by generalizing OTOCs using a replica approach~\cite{Trunin:2023rwm,Trunin:2023xmw}, the main goal of this manuscript is to propose and test an operator growth measure akin to Krylov complexity, \emph{logarithmic Krylov complexity} (\emph{logK-complexity}), as a viable probe of early-time scrambling without false positives. The key idea being that by performing a
correct average over phase space, it should be possible to avoid the exponentially growing contributions from the classically unstable saddle points in otherwise integrable systems. This quantity is defined through a replica approach to a higher-order generalization of Krylov complexity. \\
$\phantom{word}$ To test this measure, we consider quantum systems with finite- and infinite-dimensional Krylov spaces exhibiting early-time scrambling, as captured by the exponential growth of Krylov complexity. This allows us to benchmark and test logarithmic Krylov complexity and to refine its definition in infinite-dimensional systems. To complement this analysis, we also consider the Lanczos algorithm in classical dynamical systems, which in the past was restricted to the study of autocorrelation functions and Lanczos coefficients using a classical version of the recursion method~\cite{RecursionBook}. We extend the existing framework and define classical versions of Krylov and logK complexity using a classical analogue of the spreading operator in the Krylov basis. Our analytical and numerical results suggest that logarithmic Krylov complexity can be refined to serve as a viable probe of early-time scrambling in finite and infinite-dimensional Krylov subspaces. \\
$\phantom{word}$Our manuscript is organized as follows: In Section.~\hyperref[sec:ReviewKrylov]{II.}, we briefly review the basic properties of Krylov complexity and provide general details of the construction of the Krylov complexity. In Section.~\hyperref[sec:LogKandElogK]{III.}, we define the logarithmic Krylov complexity and its exponentiated form, \emph{elogK-complexity}, by applying the replica trick to higher-order versions of the usual Krylov complexity. In Section.~\hyperref[sec:ExamplelogKCFT]{IV.}, we analytically compute logK complexity in the low energy (conformal) limit of the $q$-body Sachdev--Ye--Kitaev (SYK) model. Then, in Section.~\hyperref[sec:NumericalAnalysis]{V.}, we numerically examine the logK complexity in the Lipkin--Meshkov--Glick (LMG) model and the mixed-field Ising model at the chaotic point. In Section.~\hyperref[sec:QuantKrylovIHO]{VI.}, we extend our analytical analysis to the inverted harmonic oscillator, finding similarity to the SYK case. In Sections~\hyperref[sec:ClassicalPSKrylov]{VII.} and~\hyperref[sec:ClassKrylovExampleSaddle]{VIII.}, we formulate the Krylov approach in classical phase space and apply it to the study of saddle-dominated scrambling in classical systems, finding that both classical Krylov and elogK complexity exhibit sub-exponential growth in these cases. In Section.~\hyperref[sec:ReplicaGeneralization]{IX.} we define a refined Krylov spreading operator which includes information about the system and operator while retaining the universal information from Krylov complexity. We show how this notion resolves the tension found in the integrable $q=2$ SYK and in the inverted harmonic oscillator and propose a way in which it could arise from a generalization of the standard analytic continuation used in the replica trick. Finally, in Sections~\hyperref[sec:Discussion]{X.} and ~\hyperref[sec:Conclusion]{XI.} we discuss our results and offer some conclusions.~\footnote{We work in units where $\hbar=1=k_{B}$.}\\
\textbf{Notation}: We denote the usual operators in the algebra $\mathcal{A}(\mathcal{\mathcal{H}}):=\lbrace \hat{A}:\mathcal{H}\rightarrow \mathcal{H}\rbrace $ with a hat $\hat{\phantom{c}}$, and superoperators in the GNS algebra $\mathcal{A}(\mathcal{H}_{\mathrm{GNS}})=\lbrace \breve{B}:\mathcal{H}_{\mathrm{GNS}}\rightarrow \mathcal{H}_{\mathrm{GNS}}\rbrace$ using a \emph{breve} accent $\breve{\phantom{c}}$.

\phantomsection
II. \textit{Overview of Krylov Complexity}
\label{sec:ReviewKrylov}--
Krylov complexity~\cite{Parker:2018yvk} is a growth measure for operators undergoing Heisenberg time evolution $\hat{\mathcal{O}}(t)=e^{it\hat{H}}\hat{\mathcal{O}}\,e^{-it\hat{H}}$ generated by a time-independent Hamiltonian $\hat{H}$. Through the Gelfand--Naimark--Segal (GNS) construction, operators $\hat{\mathcal{O}}$ are promoted to GNS states $\vert \mathcal{O})$ whose time evolution is naturally constrained to the Krylov subspace $\mathcal{K}$, a subspace of the full GNS Hilbert space $\mathcal{K}\leq \mathcal{H}_{\textrm{GNS}}$ generated by successive commutation of the initial operator with the Hamiltonian. A requirement for the GNS construction involves choosing an inner product. In systems at finite temperature, a common choice is the Wightman inner product, defined by
\begin{equation}
    \left(A|B\right)^{W}:=\frac{\Tr(e^{-\beta \hat{H}/2} \hat{A}^\dagger e^{-\beta \hat{H}/2} \hat{B})  }{\Tr(e^{-\beta \hat{H}})}~,
    \label{eq:WightmanInProd}
\end{equation}
where $\beta=T^{-1}$ is the inverse temperature, $\hat{H}$ is the Hamiltonian of the system, and $\vert A),\vert B )$ are GNS states associated with the operators $\hat{A},\hat{B}$ respectively~\footnote{Different choices of inner products lead to different GNS Hilbert spaces and the Wightman inner product is not the only possibility at finite temperature. For details on the GNS construction and details on different inner products at finite temperature, the reader may refer to Appendix A of~\cite{Magan:2020iac}.}. The dimension of the Krylov subspace $D_\mathcal{K}=\textrm{dim}(\mathcal{K})$ will generally depend on the dynamics, as well as on the choice of the initial operator $\hat{\mathcal{O}}$. As argued in~\cite{Rabinovici:2020ryf}, in thermalizing quantum systems, the dimension of Krylov space is generally bounded by 
\begin{equation}
   1\leq D_\mathcal{K}=\mathrm{dim(\mathcal{K})}\leq D^2-D+1~,
   \label{eq:DimKrylovQuantum}
\end{equation}
where $D=\mathrm{dim}(\mathcal{H})$ is the dimension of the physical Hilbert space $\mathcal{H}$, which we assume to be finite at this time. However, the dimension of the Krylov subspace $D_{\mathcal{K}}$ could be much smaller than the maximum allowed value if the quantum system is integrable or if the operator is close to being a conserved quantity.\\
$\phantom{word}$Using the Gram--Schmidt orthogonalization procedure, in this context called the Lanczos algorithm, one can recursively construct a complete orthonormal basis, called the Krylov basis $\lbrace \vert \mathcal{K}_{n})\rbrace$, in the Krylov subspace $\mathcal{K}$ with respect to a choice of inner product. This essentially translates the problem of solving the Heisenberg equation into solving a Schr\"{o}dinger-like equation in a one-dimensional chain, called the Krylov chain (see, e.g.~\cite{Nandy:2024evd,Rabinovici:2025otw} for details). After obtaining the Krylov basis, one can expand the time-evolved GNS state as follows:
\begin{equation}
    |\mathcal{O}(t))=e^{it\breve{\mathcal{L}}}|\mathcal{O}_{0})=\sum^{D_\mathcal{K}-1}_{n=0}i^n \varphi_n(t) |\mathcal{K}_n)~,
    \label{eq:TimeEvolOpKrylovBasis}
\end{equation}
where $\breve{\mathcal{L}}$ is the Liouvillian (GNS Hamiltonian) $\breve{\mathcal{L}}\equiv [\hat{H},\cdot]$ and $\vert \mathcal{O})=\vert \mathcal{O}_{0})$ is taken as the initial GNS state in the Lanczos algorithm. Since the Krylov basis is complete and orthonormal, the functions $\varphi_n(t):=i^{-n}(\mathcal{K}_{n}\vert e^{it\breve{\mathcal{L}}}\vert\mathcal{O}_{0})$ can be interpreted as wavefunctions, whose squared sum is normalized to $1$ and which describe the hopping of a particle in the Krylov chain. In particular, the initial wavefunction $\varphi_{0}(t):=(\mathcal{O}_{0}\vert \mathcal{O}(t))$, called the \emph{autocorrelation function}, plays a key role in this approach, as all subsequent $\varphi_{n}(t)$ can be obtained from it through a recursion relation. The Krylov complexity $K(t)$ associated with the growth of the operator $\vert \mathcal{O})\equiv \vert\mathcal{O}_{0})$ is defined as 
\begin{equation}
\label{eq:KrylovComp}
    K(t):=(\mathcal{O}_0|e^{-i\breve{\mathcal{L}}t}\breve{n}\, e^{i\breve{\mathcal{L}}t} |\mathcal{O}_0)=\sum^{D_\mathcal{K}-1}_{n=0} n |\varphi_n(t)|^2~,
\end{equation}
where $\breve{n}:=\sum^{D_\mathcal{K}-1}_{n=0} n|\mathcal{K}_n)(\mathcal{K}_n|$ is the Krylov ``spreading'' superoperator, which measures the average position of the GNS state $\vert \mathcal{O}(t))$ in the Krylov chain at time $t$. \\
$\phantom{word}$From its inception~\cite{Parker:2018yvk}, Krylov complexity was proposed to probe features of non-integrable quantum dynamics. In quantum many-body systems, such as the large-$N_{f}$ limit of the $q=4$ Sachdev--Ye--Kitaev (SYK) model~\cite{Sachdev:1992fk,KitaevTalks} with $N_{f}$ Majorana fermions, this is characterized by a time window of exponential growth $K(t)\sim e^{\lambda_{K}t}$, similar to how OTOCs probe scrambling through a similar exponential decay~\footnote{In finite dimensional quantum systems, this exponential behavior is found around early-times $0\lesssim t\lesssim \log(S)$, where $S$ is the number of degrees of freedom in the system, which is usually related to the dimension of the Hilbert space $D$ via $D\sim e^{S}$. For the SYK$_{q}$, this result was originally found in the thermodynamic limit $N_{f}\rightarrow \infty$, where strictly speaking $\log(S)\rightarrow \infty$.}. This exponential growth of the Krylov complexity is intimately tied to the growth rate of the elements of the Liouvillian in the Krylov basis: $(\mathcal{K}_{n-1}\vert\breve{\mathcal{L}}\vert \mathcal{K}_{n})=b_{n}$. The operator growth hypothesis~\cite{Parker:2018yvk} states that in non-integrable many-body quantum systems, the $b_{n}$ should grow as fast as possible, which due to locality constraints should be linear in $n$, namely
\begin{align}
    \label{eq:LinGrowthbn}
    b_{n}\sim \alpha \,n +\gamma+ O(1)\,\,\,\textrm{for}\,\,\, 1\ll n\lesssim S~,
\end{align}
where $\gamma$ is a constant that depends on the operator. The connection being that, under general conditions, $\lambda_{K}=2\alpha$, and therefore, a regime of linear growth of the Lanczos coefficients implies the existence of a regime of exponential growth of the Krylov complexity~\footnote{There are many subtleties with this statement that we will not discuss in this work. For example, the presence of IR scales may induce ``staggering'' of the Lanczos sequences~\cite{Avdoshkin:2022xuw,Camargo:2022rnt}, leading to a different relation between the Krylov exponent $\lambda_{K}$ and the growth rate(s) of the $b_{n}$. The reader is encouraged to see~\cite{Nandy:2024evd,Rabinovici:2025otw} for details.}. However, subsequent works showed that this behavior can also occur in integrable systems with instabilities, especially those dominated by unstable saddle points~\cite{Xu:2019lhc, Bhattacharjee:2022vlt, Huh:2023jxt, Aguilar-Gutierrez:2025hbf}. Although the authors in~\cite{Aguilar-Gutierrez:2025hbf} provided evidence that saddle-dominated scrambling can be distinguished from genuine scrambling by examining the long-time behavior of Krylov complexity, it remains necessary to understand how Krylov complexity can be refined to discriminate between saddle-dominated and genuine scrambling at early times ($t\lesssim O(\log(S))$) in finite-dimensional systems.


\phantomsection
III. \textit{Logarithmic Krylov Complexity}\label{sec:LogKandElogK} -- To address this issue and motivated by the definition of logarithmic OTOC~\cite{Trunin:2023xmw, Trunin:2023rwm}, we introduce a closely related quantity, which we call \textit{logarithmic Krylov (logK) Complexity} $\mathbf{L}_K(t)$, which we heuristically define through the following expression involving the expectation value of the (matrix) logarithm of the spreading superoperator $\breve{n}$
\begin{equation}
    \mathbf{L}_K(t)\doteq(\mathcal{O}_0|\log(\breve{n}(t))  |\mathcal{O}_0)~,
    \label{eq:LogKDef}
\end{equation}
where $\breve{n}(t):=e^{-i\breve{\mathcal{L}}t}\,\breve{n}\,e^{i\breve{\mathcal{L}}t}$ is the time-evolved spreading superoperator~\footnote{Note that $\breve{n}$ and $e^{\pm i\breve{\mathcal{L}}t}$ do not commute, and therefore $\log(\breve{n}(t))= e^{-i\breve{\mathcal{L}t}}\log(\breve{n})e^{i\breve{\mathcal{L}}t} \neq \log(\breve{n})$.}. Since it is in general difficult to evaluate $\log(\breve{n})$ explicitly, we instead consider a different working definition for logK-complexity. We think of logK-complexity as arising from the application of the replica trick to the expectation value of the higher-order spreading superoperator, namely 
\begin{equation}
    \mathbf{L}_K(t):=\left.\frac{\partial}{\partial m}(\mathcal{O}_0|\breve{n}^{m}(t)|\mathcal{O}_0)\right\vert_{m\rightarrow 0}=\left.\frac{\partial}{\partial m} K^{(m)}(t)\right\vert_{m\rightarrow 0}~,
    \label{eq-Log-K replica}
\end{equation}
where $K^{(m)}(t)$ is the operator analogue of the higher-order spread complexity for integer $m$ proposed in~\cite{Fu:2024fdm, Camargo:2024rrj, Fan:2023ohh}~\footnote{See \cite{Fu:2025kkh} for a discussion of the holographic duals of higher-order and logK state complexities.}, defined by
\begin{align}
    \label{eq:HigherOrdKcomp}
    K^{(m)}(t):=(\mathcal{O}_0|\breve{n}^{m}(t)|\mathcal{O}_0)=\sum_{n=0}^{D_{\mathcal{K}}-1}n^{m}\vert \varphi_{n}(t)\vert^{2}~,
\end{align}
and where $\breve{n}^{m}(t)=e^{-i\breve{\mathcal{L}t}}(\breve{n}^{m})e^{i\breve{\mathcal{L}}t}$. Thus, in our approach, we consider the higher-order Krylov complexities with integer $m$ to be replica copies of the standard Krylov complexity. Importantly, to compute the derivative of the higher-order Krylov complexity $K^{(m)}(t)$ and subsequently the limit $m\rightarrow 0$, we need to \emph{analytically continue} $K^{(m)}(t)$ with integer $m$ to real $m$, a step that will be crucial in our discussion for systems with infinite-dimensional Hilbert spaces. \\
$\phantom{word}$To make a closer comparison with Krylov complexity~\eqref{eq:KrylovComp}, we also define the exponentiated logK-complexity, which we call \emph{elogK-complexity}, by
\begin{equation}
    \mathbf{E}_{K}(t):=e^{\mathbf{L}_K(t)}-1~,
    \label{eq:ElogK}
\end{equation}
where we chose the constant factor so that $\mathbf{E}_K(0)=0$. Our expectation, following the works by D. Trunin~\cite{Trunin:2023rwm,Trunin:2023xmw}, being that whenever we have a truly chaotic system, we should find a time window where $K(t)\propto \mathbf{E}_{K}(t)\sim e^{\lambda_{K}t}+\ldots$ where the ellipsis denotes subleading terms, and therefore $\log(K(t))\propto \mathbf{L}_{K}(t)\sim \lambda_{K}t+\ldots$, whereas in non-chaotic quantum systems we instead expect in general $\log(K(t))\neq \mathbf{L}_{K}(t)$, or equivalently $\log((\mathcal{O}_{0}\vert\breve{n}(t)\vert \mathcal{O}_{0}))\neq (\mathcal{O}_{0}\vert\log(\breve{n}(t))\vert \mathcal{O}_{0})$.\\
$\phantom{word}$We can ask whether~\eqref{eq:ElogK} defines a cost function whose minimization with respect to different bases in the Krylov space yields a notion of quantum complexity of the GNS state $\vert \mathcal{O}_{0} )$. Since $\mathbf{E}_{K}(t)$ is a monotonic function of $\mathbf{L}_{K}(t)$, we can restate the question in terms of $\mathbf{L}_{K}(t)$ and ask whether it is minimized by the Krylov basis, at least for small times. To achieve this, we first perform the replica trick (\ref{eq-Log-K replica}) with the simplifying assumption that the GNS Hilbert space is finite dimensional~\footnote{Otherwise, we have to assume that the series of derivatives converges uniformly so that we can exchange the derivative over the replica index and the sum.}. Applying the replica trick to the spreading superoperator, we find
$\partial_m \breve{n}^m\vert_{m\rightarrow 0}=\sum_{D_{\mathcal{K}}-1\geq n \geq 0}\log (n) |\mathcal{K}_n)( \mathcal{K}_n |$, where we performed the standard analytic continuation $\breve{n}^{m}=e^{m\log(\breve{n})}$ to real $m$. This leads to a logarithmic divergence at $n=0$, which we need to remove to obtain a finite quantity. Assuming the following holds,
\begin{align}
    \label{eq:OrderDerivativeHighOrdSpreadOp}
    \frac{\partial}{\partial m}K^{(m)}(t)\Big\vert_{m\rightarrow 0}\equiv \left( \mathcal{O}_{0}(t)\left\vert  \left(\frac{\partial}{\partial m}\breve{n}^{m} \right)\Big\vert_{m\rightarrow 0} \right\vert \mathcal{O}_{0}(t)\right)~,
\end{align}
one way to perform the regularization of $\mathbf{L}_{K}(t)$ is by subtracting the divergent contribution at $n=0$ from $\partial_m \breve{n}^m\vert_{m\rightarrow 0}$: 
\begin{align}
    \label{eq:ShiftedSpreadingOp}
    \begin{split}
    \partial_m \breve{n}^m\vert_{m\rightarrow 0}\rightarrow \partial_m \breve{n}^m\vert_{m\rightarrow 0}-(\log (n) |\mathcal{K}_0)( \mathcal{K}_0 \vert)\vert_{n\rightarrow 0}~.
    \end{split}
\end{align}
This results in the following regularized expression of logK-complexity:
\begin{equation}
\begin{split}
&\mathbf{L}_K(t)\equiv\sum^{D_\mathcal{K}-1}_{n=1} \log(n)(\mathcal{O}_0|e^{-i\breve{\mathcal{L}}t}|\mathcal{K}_n)( \mathcal{K}_n |e^{i\breve{\mathcal{L}}t}|\mathcal{O}_0)~.
\end{split}
    \label{eq-NaiveLogK}
\end{equation}
This expression is equivalent to (\ref{eq-Log-K replica}), after subtracting the logarithmic divergence at $n\rightarrow 0$, in finite-dimensional systems or in infinite-dimensional systems whenever the derivative of~\eqref{eq:HigherOrdKcomp} with respect to the index $m$ is absolutely convergent. 

Although we focus on operator complexity, several arguments derived for the optimality of the Krylov basis at small times for spread complexity~\cite{Balasubramanian:2022tpr} also hold for K-complexity, after fixing the inner product. In particular, Corollary $1 $ in~\cite{Balasubramanian:2022tpr} shows the optimality of the Krylov basis around $t=0$ (minimization of~\eqref{eq:KrylovComp}), where $n$ is replaced by any sequence of monotonically increasing coefficients $\kappa_n$. In our case, namely~\eqref{eq-NaiveLogK}, $\kappa_n=\log(n)$ is a monotonically increasing sequence in $n$, and thus we also expect the Krylov basis to yield the minimum of logK-complexity around $t=0$ over all possible choices of bases and for a fixed inner product~\footnote{Some authors view the ambiguity of the inner product as an additional parameter over which one should further minimize the complexity functionals. We do not take that approach in this section and we instead restrict ourselves to the choice of Wightman inner product, which yields the slowest growth for Krylov complexity over all other finite-temperature inner products~\cite{Magan:2020iac}.}. In the Supplemental Material~\hyperref[app:InitiallogK]{A}, we show that the universal initial time growth of logK-complexity is given by 
\begin{equation}
\label{eq:LogKInitialt}
    \mathbf{L}_K(t) \approx\frac{\log(2)}{4}b_1^2b_2^2t^4+O(t^{6})~,\quad (\textrm{small  \,}t)~,
\end{equation}
where $b_1,b_2$ are Lanczos coefficients. We verify this in our numerical calculations for various physical systems, and it is the first difference from the conventional Krylov complexity, which has a universal early-time growth $\propto b^{2}_{1}t^{2}$. It is also worthwhile analyzing the long-time regime in thermalizing systems, following~\cite{Rabinovici:2021qqt,Rabinovici:2022beu}. After the exponential growth regime $0\lesssim t \lesssim O(\log(S))$, Krylov complexity transitions to a linear growth for $O(\log(S))\lesssim t \lesssim O(e^{2S})$ before saturating to a value $K(t_{s})\approx K_{s}\propto O(D_{\mathcal{K}}/2)$ around $t\approx t_s \sim O(e^{2S})$. At this time, the initial GNS state $\vert \mathcal{O}_{0})$ would have become fully delocalized in the Krylov space $\mathcal{K}$. A way to study what happens here is to consider the long-time behavior of logK-complexity 
\begin{align}
    \label{eq:LongTimeLogK}
    \overline{\mathbf{L}}_{K}:=\lim_{T\rightarrow \infty}\frac{1}{T}\int_{0}^{T}\textrm{d}t \,\mathbf{L}_{K}(t)~.
\end{align}
In the Supplemental Material~\hyperref[app:LongTimelogK]{B}, we show that for a fully delocalizing initial operator $\hat{\mathcal{O}}_{0}$, this average is given by $\frac{1}{D_\mathcal{K}} \log((2)_{D_\mathcal{K}-2})$, where $(x)_{n}:=\Gamma(x+n)/\Gamma(x)$ is the Pochhammer symbol. In this same limit, we find the long-time average of the elogK-complexity~\eqref{eq:ElogK} to be $\overline{\mathbf{E}}_{K}\sim\Gamma(D_{\mathcal{K}})^{(1/D_{\mathcal{K}})}$. Comparing this with the long-time average of the Krylov complexity $\overline{K}\sim D_{\mathcal{K}}/2$, we see that generically, Krylov complexity will saturate to a higher value than the elogK-complexity, although their ratio is constant and given by $\mathfrak{e}/2\approx 1.35914$.


\phantomsection
IV. \textit{Logarithmic Krylov Complexity in the Conformal Limit of the SYK}~\label{sec:ExamplelogKCFT} -- To have a better idea of the behavior of logK-complexity, it is very illustrative to compute it explicitly in the paradigmatic example of the Sachdev--Ye--Kitaev (SYK) model~\cite{Sachdev:1992fk,KitaevTalks} at low energies, where the theory exhibits conformal invariance. Here we follow~\cite{Parker:2018yvk} and~\cite{Caputa:2021sib,Dymarsky:2021bjq}. The SYK$_{q}$ model is a model of $N_{f}$ Majorana fermions with $q$-body interactions described by the Hamiltonian 
\begin{align}
    \label{eq:SYKqHam}
\hat{H}^{(q)}_{\mathrm{SYK}}=i^{q/2}\sum_{1\leq i_{1}\leq \ldots \leq i_{q}\leq N_{f}}J_{i_{1}\ldots i_{q}}\hat{\psi}_{i_{1}}\cdots \hat{\psi}_{i_{q}}~,
\end{align}
where $[\hat{\psi}_{i},\hat{\psi}_{j}]_{+}:=\hat{\psi}_{i}\hat{\psi}_{j}+\hat{\psi}_{j}\hat{\psi}_{i}=\delta_{ij}$ are $N_{f}$ Majorana fermions and where $J_{i_{1}\ldots i_{q}}$ are antisymmetric random couplings sampled from a Gaussian distribution with zero mean and variance $\sigma_{q}^{2}:=\overline{J_{i_{1}\ldots i_{q}}^{2}}=N_{f}^{-(q-1)}(q-1)!\mathcal{J}^{2}$. As is well-known (see e.g.~\cite{Maldacena:2016hyu}), this model is solvable in the large-$N_{f}$ limit. Moreover, at finite and low temperatures $\beta \mathcal{J}\gg 1$, this model exhibits an emergent conformal invariance accurately described by a chiral conformal field theory (CFT) with $\mathrm{SL}(2,\mathbb{R})$ symmetry. For an initial operator of the form $\hat{\mathcal{O}}_{0}=\sqrt{2}\hat{\psi}_{1}$ and using the Wightman inner-product~\eqref{eq:WightmanInProd}, the autocorrelation function $\varphi^{W}_{0}(t):=2(\hat{\psi}_{1}\vert \hat{\psi}_{1}(t))^{W}$ becomes
\begin{align}
    \label{eq:AutocSYKTHermalLowT}
    \varphi^{W}_{0}(t)=(\cosh(\pi t/\beta))^{-2/q}~,
\end{align}
with associated Lanczos coefficients $b_{n}$ $=(\pi/\beta)$ $\times$ $\sqrt{n(n-1+2/q)}$. This is a particular example of a more general class of autocorrelation functions of CFT$_{2}$ primary fields that transform under specific representations of the $\mathrm{SL}(2,\mathbb{R})$ algebra. For example, the thermal (Wightman) autocorrelation function of a chiral primary field $\hat{\mathcal{O}}_{0}$ with scaling dimension $\Delta=h$ is given precisely by~\eqref{eq:AutocSYKTHermalLowT}, where $h=1/q$. More generally, for an autocorrelation function of the form $\varphi_{0}(t)=(\cosh(\alpha t))^{-\eta}$, it is possible to find a closed form expression for the wavefunctions $\varphi_{n}(t)$, which satisfy the recursion relation
\begin{align}
    \label{eq:RecursionPhin}
    \frac{\mathrm{d}\varphi_{n}(t)}{\mathrm{d}t}=b_{n}\varphi_{n-1}(t)-b_{n+1}\varphi_{n+1}(t)~,
\end{align}
where the Lanczos coefficients are $b_{n}$$=\alpha\sqrt{n(n-1+\eta)}$. Here, $\alpha$ and $\eta$ are related to the representation of primary $\mathrm{SL}(\mathbb{R},2)$ states using $\mathrm{SU}(1,1)$ Perelomov coherent states~\cite{Caputa:2021sib}, with $\alpha$ related to the energy scale of the Liouvillian and $\eta$ to the weight $h$ of the primary. In this case, the $\varphi_{n}(t)$ are given by
\begin{align}
    \label{eq:PhinCFT}
    \varphi_{n}(t)=\sqrt{\frac{\Gamma(n+\eta)}{n!\Gamma(\eta)}}\frac{\tanh^{n}(\alpha t)}{\cosh^{\eta}(\alpha t)}~.
\end{align}
Setting $\eta=2h=2/q$, $\alpha=\pi/\beta$, we recover the wavefunctions for the conformal limit of the SYK at low temperatures~\eqref{eq:AutocSYKTHermalLowT}. With them, we can directly compute the Krylov complexity~\eqref{eq:KrylovComp}, which is given by
\begin{align}
    \label{eq:KrylovCFT}
    K(t)=\sum_{n=0}^{\infty}n\vert \varphi_{n}(t)\vert^{2}=\eta \sinh^{2}(\alpha t)~,
\end{align}
and which shows the late-time exponential growth $K(t)\propto e^{2\alpha(t-t_{\ast})}$ discussed in the \hyperlink{sec:intro}{Introduction} in the context of the universal operator growth hypothesis, where $t_{\ast}=\log(4\eta)/2\alpha$. Using~\eqref{eq:PhinCFT} and the definition~\eqref{eq:HigherOrdKcomp}, we find that the $m$-th order Krylov complexity for integer $1\leq\eta\in \mathbb{Z}^{+}$ has the following formal expression for $1\leq m \in \mathbb{Z}^{+}$
\begin{align}
    \label{eq:HigherOrdKrylovCFT}
    \begin{split}
    &K^{(m)}(t)=\sum_{n=0}^{\infty}n^{m}\vert \varphi_{n}(t)\vert^{2}\\
    &=\sech(\alpha t)^{2\eta}\sum_{i=0}^{\eta-1}\kappa(\eta,i)\mathrm{Li}_{-(m+i)}(\tanh^{2}(\alpha t))~,
    \end{split}
\end{align}
where $\mathrm{Li}_{n}(z)$ is the polylogarithm (Jonqui\`{e}re's function), which is defined for any complex $n$ and $z$ as the analytic continuation of the Dirichlet series $\sum_{k=1}^{\infty}z^{k}/k^{n}$. The polylogarithm $\mathrm{Li}_{n}(z)$ is absolutely convergent for all $n \in \mathbb{C}$ and for all $z \in \mathbb{C}$ inside the unit disk $\vert z\vert <1$~\cite{NIST:DLMF}. In~\eqref{eq:HigherOrdKrylovCFT}, the $\kappa(\eta,i)$ are numerical coefficients given by
\begin{align}
    \label{eq:CoeffHigherOrdKrylovCFT}
    \kappa(\eta,i)=\frac{\vert S_{1}(\eta,i+1)\vert}{(\eta-1)!}\,\,\,,\,\,\,\eta\geq 1 \,\,\,\&\,\,\,0\leq i \leq \eta-1~,
\end{align}
where $ S_{1}(n,k)$ are Stirling numbers of the first kind, which have the following generating function
\begin{align}
    \label{eq:StirlingNumbers}
    \frac{1}{k!}\left(\log(1+x)\right)^{k}=\sum_{n=k}^{\infty}S_{1}(n,k)\frac{x^{n}}{n!}~.
\end{align}
Before computing the logK-complexity, it is worth examining the behavior of $K^{(m)}(t)$ given by~\eqref{eq:HigherOrdKrylovCFT} for fixed index $m \geq 1$. Let us momentarily focus on $\eta=1$ for simplicity, corresponding to a chiral primary with scaling dimension $\Delta =h=1/2$. In this case, the sum in the $m$-th order Krylov complexity contains only one term: $K^{(m)}(t)\vert_{\eta=1}=\mathrm{Li}_{-m}(\tanh^{2}(\alpha t))\sech^{2}(\alpha t)$. At early times, it behaves as $K^{(m)}(t)\vert_{\eta=1}\approx \alpha^{2}t^{2}+(2^{m}-5/3)\alpha^{4}t^{4}-O(\alpha^{6}t^{6})$, whereas at late times it has a \emph{leading} exponential growth given by $K^{(m)}(t)\vert_{\eta=1}\propto e^{2\alpha m t}$, where we directly note that the index $m$ enhances the late-time growth rate of the usual Krylov complexity~\eqref{eq:KrylovCFT}. To be more precise, the $m$-th order Krylov complexity for $\eta=1$ and integer $1 \leq m\in \mathbb{Z}^{+}$ can be written as
\begin{align}
    \label{eq:HigherOrdKrylovCFTeta1}
    \begin{split}
    &K^{(m)}(t)\vert_{\eta=1}=\sum_{j=-m}^{m}\tilde{\kappa}^{(1)}_{m,j}e^{2\alpha j t}\\
    &=\tilde{\kappa}^{(1)}_{m,-m}e^{-2m\alpha t}+\cdots + \left(-\frac{1}{2}\right)^{m}+\cdots+\tilde{\kappa}^{(1)}_{m,m}e^{2m\alpha t}~,
    \end{split}
\end{align}
with $\tilde{\kappa}^{(1)}_{m,m}=\tilde{\kappa}^{(1)}_{m,-m}=2^{-2m}\Gamma(m+1)/\Gamma(1)$ and where $\ldots$ denote terms containing terms $e^{2\alpha jt}$ with powers $1\leq \vert j \vert \leq (m-1)$. From~\eqref{eq:HigherOrdKrylovCFTeta1} we find that the leading behavior at large $t$ is dominated by the term with the largest exponent, given by $j=m$. Note that the representation~\eqref{eq:HigherOrdKrylovCFTeta1} holds only when $1\leq m \in \mathbb{Z}^{+}$. In fact, this leading late-time behavior $\propto e^{2m\alpha t}$ is also present for larger integer $\eta \geq 1$, namely for heavier $\mathrm{SL}(2,\mathbb{R})$ primaries. This can be seen from the fact that, in general, the sum of polylogarithms in~\eqref{eq:HigherOrdKcomp} contributes at late times with a leading factor proportional to $e^{2(m+\eta)(\alpha t)}$. Thus, the \emph{leading} large-time behavior of the higher-order Krylov complexity for general integer $\eta \geq 1$ is given by
\begin{align}
    \label{eq:HigherOrdKrylovCFTLarget}
    \begin{split}
    &K^{(m)}(t) \sim 2^{2\eta}e^{-2\alpha \eta t}\times e^{2(m+\eta)(\alpha t)} \propto e^{2\alpha m t}~,
    \end{split}
\end{align}
where the first factor comes from the late-time behavior of $\sech^{2\eta}(\alpha t)$. Similarly, the initial-time behavior of $K^{(m)}(t)$ for general integer $\eta\geq 1$ is of the form
\begin{align}
    \label{eq:HigherOrdKrylovCFTSmallt}
    \begin{split}
    &K^{(m)}(t)\approx \eta\,\alpha^{2}t^{2}\\
    &+\frac{\eta}{3}\left(\frac{3}{2}\left(1+\eta\right)2^{m}-(2+3\eta)\right)\alpha^{4}t^{4}-O(\alpha^{6}t^{6})~.
    \end{split}
\end{align}
$\phantom{word}$To find the logK complexity, we take the higher-order Krylov complexities with index $m$~\eqref{eq:HigherOrdKrylovCFT} as replica copies of the Krylov complexity~\eqref{eq:KrylovCFT}. Note that the polylogarithm $\mathrm{Li}_{-(m+i)}(\tanh^{2}(\alpha t))$ is well defined and absolutely convergent for any $(m+i)$ given that $\vert \tanh^{2}(\alpha t)\vert \leq 1$ and in particular, for finite $t$ and for any real and finite $\alpha$, we have $\vert \tanh^{2}(\alpha t)\vert < 1$. This is because the polylogarithm $\mathrm{Li}_{n}(z)$ is an analytic function of the power $n$ for fixed argument $z$. As a consequence, $K^{(m)}(t)$ can be \emph{analytically continued} to non-integer $m$ by the standard analytic continuation $n^{m}\mapsto e^{m\log(n)}$. Thus, we are able to differentiate $K^{(m)}(t)$~\eqref{eq:HigherOrdKrylovCFT} with respect to the replica index $m$ and take the limit $m\rightarrow 0$ to find the logK-complexity, which has the following formal expression
\begin{align}
\label{eq:LogKCFT}
\begin{split}
\mathbf{L}_K(t) &=\lim_{m\to0}\frac{\partial}{\partial m}\left(K^{(m)}(t)\right)\\
&=-\sech(\a t)^{2\eta} \sum^{\eta-1}_{\ell=0} c(\eta,\ell)\mathrm{Li}'_{-\ell}(\tanh^{2}(\a t))~,\\
\end{split}
\end{align}
where $\mathrm{Li}'_{n}(z):=\partial \mathrm{Li}_{n}(z) /\partial n$ is the derivative of the polylogarithm $\mathrm{Li}_{n}(z)$ with respect to the power $n$, and where $c(\eta,\ell)$ are numerical coefficients given by
\begin{equation}
\label{eq:CoeffLogKCFT}
    c(\eta,\ell):= \sum_{1\leq j_1\leq j_2 \dots, j_\ell\leq \eta-1}\frac{1}{j_1j_2\dots j_\ell}~,
\end{equation}
where $\{j_1,j_2,\dots,j_\ell\}$ are sequences of integers of length $\ell$, which represents all possible subsets from ${1,2,\dots,\eta-1}$\footnote{By convention, we set $c(\eta,0)=1$ for all $\eta$. For example, when $\eta=1$, and $\ell=0$, $c(1,0)=1$. For $\eta=2$, although $\ell$ can be $0,1$ but the only subset is $\{1\}$. Therefore, $c(2,0)=c(3,0)=1$. The first non-trivial example is $\eta=3$. When $\ell=0$, it is assumed to be $1$. When $\ell=1$, all subsets are $\{\{1\}, \{2\}\}$, which leads to $c(3,1)=1+\frac{1}{2}=\frac{3}{2}$. When $\ell=2$, there is only one subset: $\{\{1,2\}\}$, which gives $c(3,2)=\frac{1}{1\cdot 2}=\frac{1}{2}$. Similarly, we can find: $c(4,0)=1$, $c(4,1)=1+\frac{1}{2}+\frac{1}{3}=\frac{11}{6}$, $c(4,2)=\frac{1}{1\cdot 2}+\frac{1}{2\cdot 3}+\frac{1}{1\cdot 3}=1$, $c(4,3)=\frac{1}{1\cdot 2\cdot 3}=\frac{1}{6}$}. Note that we would not have obtained this result if we had used an expression such as~\eqref{eq:HigherOrdKrylovCFTeta1}, which is only valid for integer $m$. We also remark that, since the polylogarithm is absolutely convergent, we could have found the same expression if we used the regularized expression~\eqref{eq-NaiveLogK} instead. A careful analysis of the logK complexity~\eqref{eq:LogKCFT} shows that its large-time behavior is given in general by
\begin{align}
\label{eq:LogKCFTLarget}
\mathbf{L}_K(t) \sim 2\alpha t-2^{2-\eta} \,\,\,,\,\,\,(\mathrm{large} \,\,t)~,
\end{align}
whereas its early-time behavior is given by
\begin{align}
\label{eq:LogKCFTInitialt}
\mathbf{L}_K(t) \approx \frac{\eta(1+\eta)}{2}\log(2)\alpha^{4}t^{4}- O(\alpha^{6}t^{6})~,
\end{align}
which is consistent with~\eqref{eq:LogKInitialt}.

\begin{figure}[htbp]
\begin{center}
\includegraphics[width=0.6\linewidth]{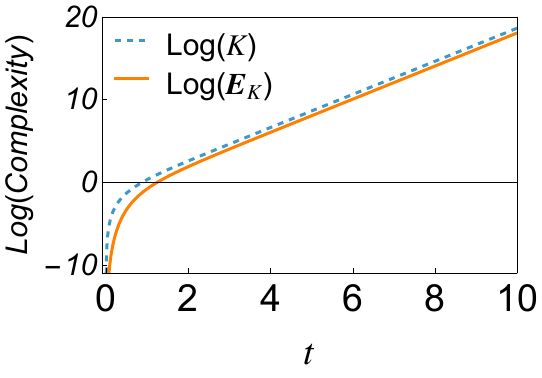}
\end{center}
\caption{Comparison of $\log(K(t))$~\eqref{eq:KrylovCFT} (blue, dashed) and $\log(\mathbf{E}_{K}(t))$~\eqref{eq:eLogKCFT} (orange) for $\eta=\alpha=1$.}
\label{fig:LogKvsKrylovCFT}
\end{figure}

It is clear that in this case we have the relation $\log(K(t))\sim \mathbf{L}_{K}(t)$ at late times. In fact, they give the same late-time growth rate, or \emph{Krylov exponent}
\begin{align}
    \label{eq:KrylovExponentConsistenctCFT}
    \begin{split}
   \lambda_{K}:&= \lim_{t\rightarrow \infty}\frac{\mathrm{d}}{\mathrm{d}t}\log(K(t))\equiv2\alpha\\
   &\equiv \lim_{t\rightarrow \infty}\frac{\mathrm{d}}{\mathrm{d}t}\mathbf{L}_{K}(t):= \lambda_{\mathbf{L}_{K}}~.
   \end{split}
\end{align}
Thus, this is an example where both $\log(K(t))$ and $\mathbf{L}_{K}(t)$ agree in the same leading, linear late-time behavior. Note that the Krylov exponent nor their late-time linear-growth rely on the details of the CFT, and this behavior holds also for integrable CFTs, and by natural extension, to the low-energy/low temperature limit of SYK$_{2}$ ($\eta = 1$), which is integrable. We discuss this result in more detail in Sec.~\hyperref[sec:ReplicaGeneralization]{IX.} and in the~\hyperref[sec:Discussion]{Discussion}. We can also compare the Krylov complexity and the exponentiated logK complexity, the elogK-complexity~\eqref{eq:ElogK}
\begin{align}
    \label{eq:eLogKCFT}
    \mathbf{E}_{K}(t)=e^{\mathbf{L}_{K}(t)}-1\sim e^{2\alpha( t-t_{\ast}^{\mathbf{E}})}\,\,,\,\,(\mathrm{large}\,\,t)~,
\end{align}
where $t_{\ast}^{\mathbf{E}}=2^{2-\eta}/2\alpha$. In Figure~\ref{fig:LogKvsKrylovCFT}, we compare the elogK~\eqref{eq:eLogKCFT} and usual Krylov complexity~\eqref{eq:KrylovCFT} for specific values of $\eta$ and $\alpha$.

We end this discussion with a remark. The authors in~\cite{Dymarsky:2021bjq} relied on the Toda hierarchy method to compute the Lanczos coefficients for the $d$-dimensional free massless boson and free fermion at finite temperature. In this approach, an analytical expression for the Lanczos coefficients is obtained through the Toda functions, which can be used to find the late-time $t\gg t_{\ast}$ growth rate of the Krylov complexity, assuming a specific pole-structure of the autocorrelation function. However, it does not automatically yield the wavefunctions. Since we do not know the precise relation between the pole structure of the autocorrelation function and the linear growth of logK-complexity, we believe that possessing only the Lanczos coefficients, as given by the Toda method, is insufficient for determining logK-complexity. We leave this analysis for future work.


\phantomsection
V. \textit{Logarithmic Krylov Complexity in Many-Body Models}\label{sec:NumericalAnalysis} -- 
In this section, we present numerical comparisons of Krylov and elogK complexities in finite-dimensional quantum systems that contain classically unstable saddles, but which are otherwise both classically Liouville integrable and quantum mechanically integrable (although not necessarily Bethe--Ansatz-integrable). In such systems, it is known that both the OTOC and Krylov complexity grow exponentially up to the scrambling time $t_{\ast}\sim\log(S)$~\cite{Trunin:2023xmw, Trunin:2023rwm,Bhattacharjee:2022vlt, Bhattacharjee:2022vlt, Aguilar-Gutierrez:2025hbf, Bhattacharya:2023xjx}.
 
Following the introduction, we expect logK complexity $\mathbf{L}_K(t)$ to display integrable behavior in saddle-dominated systems while retaining its sensitivity to chaos in truly chaotic systems.  In order to show this numerically, we consider the Lipkin--Meshkov--Glick (LMG) model~\cite{Lipkin:1964yk} and the mixed-field Ising model at the chaotic point. To better compare it with the Krylov complexity $K(t)$, we will compute the elogK complexity $\mathbf{E}_{K}(t)$ defined in (\ref{eq:ElogK}). As we will see later in this section, the elogK-complexity $\mathbf{E}_{K}(t)$ displays a dampened growth in this saddle-dominated model, in contrast to the conventional Krylov complexity $K(t)$, whereas at the chaotic point of the mixed-field Ising model, $\mathbf{E}_{K}(t)$ exhibits exponential growth, closely matching $K(t)$ at early times. 
%

\phantomsection
A. \textit{The LMG Model}\label{subsec:NumericalAnalysisLMG} -- 
We first compute the $\mathbf{E}_{K}(t)$ in the LMG model, which is described by the following quantum Hamiltonian:
\begin{align}
\hat{H}_{\mathrm{LMG}}=\frac{\hat{S}_{x}}{s}+J\left(\frac{\hat{S}_{z}}{s}\right)^{2}=\hat{x}+J\hat{z}^2~,
\end{align}
where $\hat{S}_{x}$ and $\hat{S}_{z}$ are spin operators in the spin-$s$ representation of $\mathrm{SU}(2)$, which can be given as spin operators of a collection of $2J$ self interacting particles using Pauli matrices: $\hat{S}_\alpha=\sum_i \hat{\sigma}^i_{\alpha}/2$, and \(J\) is a coupling constant, and $\{\hat{x},\hat{y},\hat{z}\}=\{\hat{S}_{x}/s,\hat{S}_{y}/s,\hat{S}_{z}/s\}$. In the classical limit, it is described by the Hamiltonian $H_{\mathrm{LMG}}=x+2z^2$, where $x,y,z$ satisfy the Lie-Poisson algebra $\{x,y\}=z$ and the constraint $x^2+y^2+z^2=1$. From the classical phase space analysis, one can find an unstable classical saddle point at $(x,y,z)=(1,0,0)$ with exponent \(\lambda_{\mathrm{LMG}}=\sqrt{2J-1}\). For later discussions, we follow the convention in \cite{Xu:2019lhc} by setting $J=2$. In this context, the parameter $1/s$ can be used as an effective $\hbar_{\mathrm{eff}}$ whose limit $\hbar_{\mathrm{eff}}\rightarrow 0$ ($s\rightarrow \infty$) leads to a semiclassical limit.

\begin{figure}[htbp]
    \centering
        {\includegraphics[width=4.0cm]{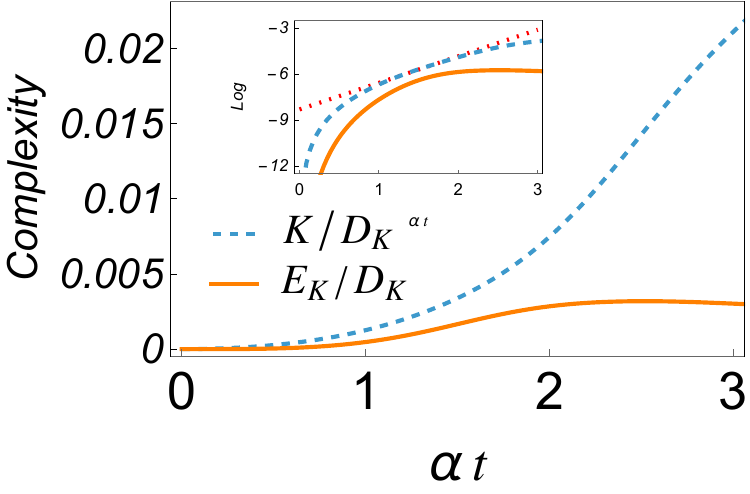}}\hspace{3mm}
        {\includegraphics[width=4.0cm]{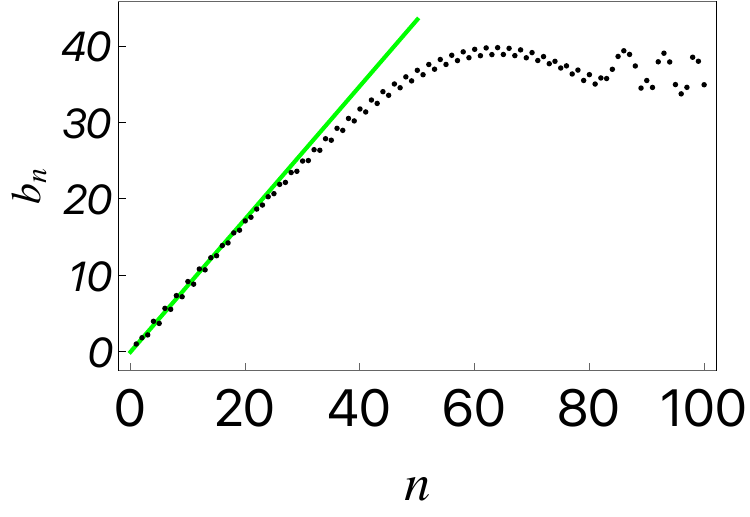}}\vspace{3mm}
        {\includegraphics[width=4.0cm]{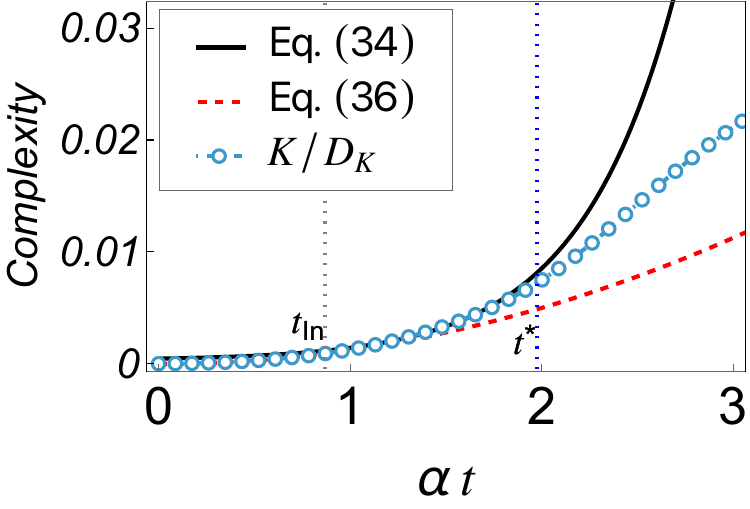}}\hspace{3mm}
        {\includegraphics[width=4.0cm]{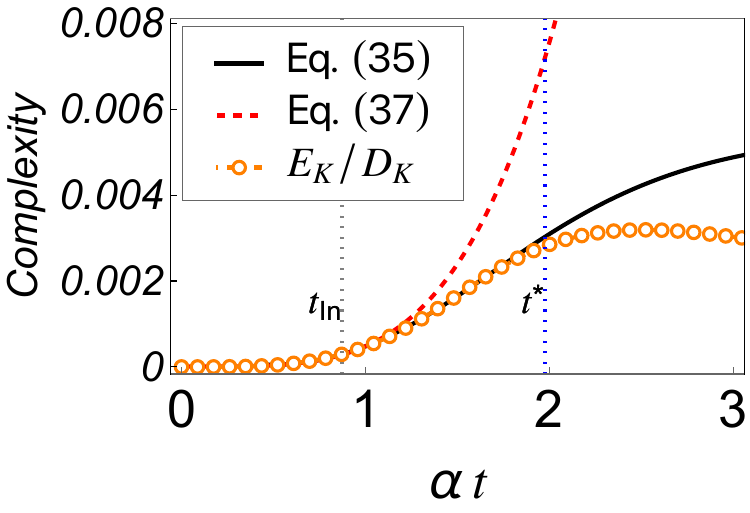}}\vspace{3mm}
        {\includegraphics[width=4.0cm]{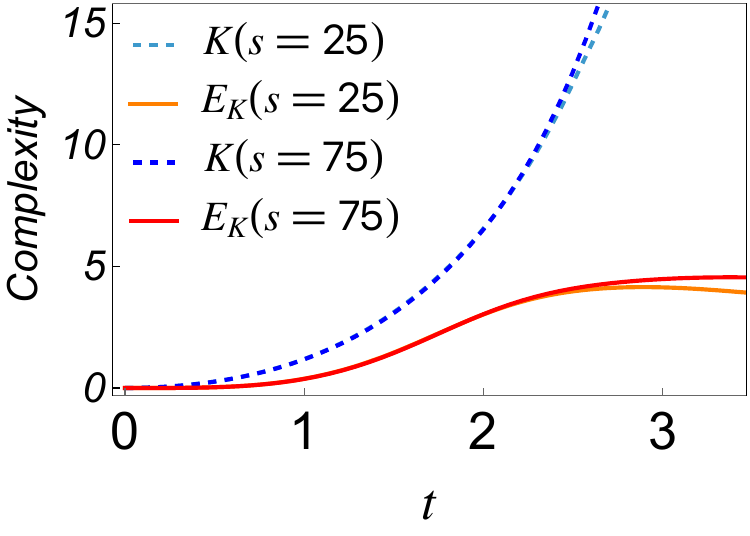}}\hspace{3mm}
        {\includegraphics[width=4.0cm]{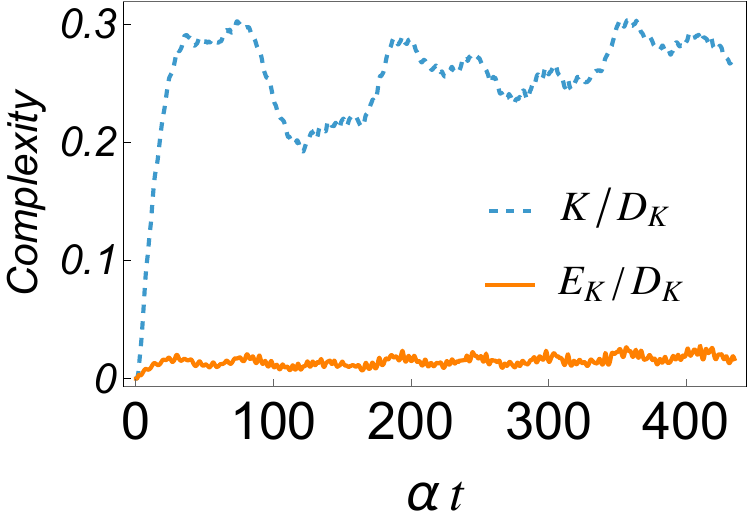}}
    \caption{Krylov complexity $K(t)$ and elogK complexity $\mathbf{E}_{K}(t)$ in the LMG model for initial operator $\hat{z}$ with classical exponent $\lambda_{\mathrm{saddle}}=\sqrt{3}$. The horizontal axis is scaled with the slope of $b_{n}$ by $\alpha$, whereas the vertical axis is scaled by the Krylov dimension $D_{\mathcal{K}}=1300$. \textbf{Upper left panel:} $K(t)$ (blue, dashed) and $\mathbf{E}_{K}(t)$ (orange) for $s=25$. The inset is the log plot where the red dashed line is of slope $\lambda_{\mathrm{saddle}}=\sqrt{3}$. \textbf{Upper right panel:} The corresponding Lanczos coefficients with linear fitting by $\alpha=0.8701$. \textbf{Middle left panel:} $K(t)$ for $s=25$. The blue dot-dashed data are from numerical calculation. The red dashed curve is the initial time fitting of the quadratic polynomial function at $0<t_{\mathrm{In}}\lesssim1$. The black curve is the exponential function fitting up to the scrambling time $t^*$. \textbf{Middle right panel:} $\mathbf{E}_{K}(t)$ for $s=25$. The orange dot-dashed data are from numerical calculation. The red dashed curve is the initial time fitting of the quartic polynomial function at $0<t_{\mathrm{In}}\lesssim1$. The black curve is the fitting up to the scrambling time $t^*$. \textbf{Lower left panel:} Comparison of $K(t)$ and $\mathbf{E}_{K}(t)$ for $s=25$ and $s=75$. \textbf{Lower right panel:} Late-time saturation of $K(t)$ and $\mathbf{E}_{K}(t)$ for $s=25$. For $s=25$, the least numerical precision of the Krylov basis is on the order of $10^{-600}$, whereas for $s=75$, it is on the order of $10^{-100}$.}
    \label{fig-LMG} 
\end{figure}

For numerical calculations, we choose the operator $\hat{z}$, which grows exponentially around the unstable saddle. As shown in Figure \ref{fig-LMG}, $\mathbf{E}_{K}(t)$ deviates significantly from $K(t)$ at early times. While $K(t)$ exhibits exponential growth at early times, $\mathbf{E}_{K}(t)$ follows a sub-exponential behavior, with both fitting functions given by:
\begin{eqnarray}
    \frac{K(t)}{D_{\mathcal{K}}}\vert_{\alpha\lesssim\alpha t\lesssim\alpha t^*}&\approx &0.00015e^{1.99(\alpha t)}~,\\
    \frac{\mathbf{E}_{K}(t)}{D_{\mathcal{K}}}\vert_{\alpha\lesssim\alpha t\lesssim\alpha t^*}&\approx &0.0059e^{-13.91e^{-1.99(\alpha t)}(\alpha t)}~,
\end{eqnarray}
where $\alpha$ is the slope of the linear growth of $b_{n}$ and $\a t^*\propto\a\log(2s+1)/\l_{\mathrm{saddle}}\approx1.9752$ is the scrambling time. The functional dependence of $\mathbf{E}_{K}(t)/D_{\mathcal{K}}$ in terms of $\alpha t$ was found numerically and currently we do not have an understanding whether it could arise as a consequence of first principles or whether it also holds for other saddle-dominated systems. We leave this study for future work.

From Figure \ref{fig-LMG}, we confirm our analytical initial-time predictions for $K(t)$ and $\mathbf{E}_{K}(t)$, that is, for $0\lesssim t$, $K(t)$ grows quadratically and $\mathbf{E}_{K}(t)$ grows quartically. This behavior is given by:
\begin{eqnarray}
    \frac{K(t)}{D_{\mathcal{K}}}\vert_{0\lesssim\alpha t\lesssim\alpha}&\approx &0.0013(\alpha t)^2~,\\
    \frac{\mathbf{E}_{K}(t)}{D_{\mathcal{K}}}\vert_{0\lesssim\alpha t\lesssim\alpha}&\approx & 0.00047(\alpha t)^4~.
\end{eqnarray}


\begin{figure}[htbp]
    \centering
        {\includegraphics[width=4.0cm]{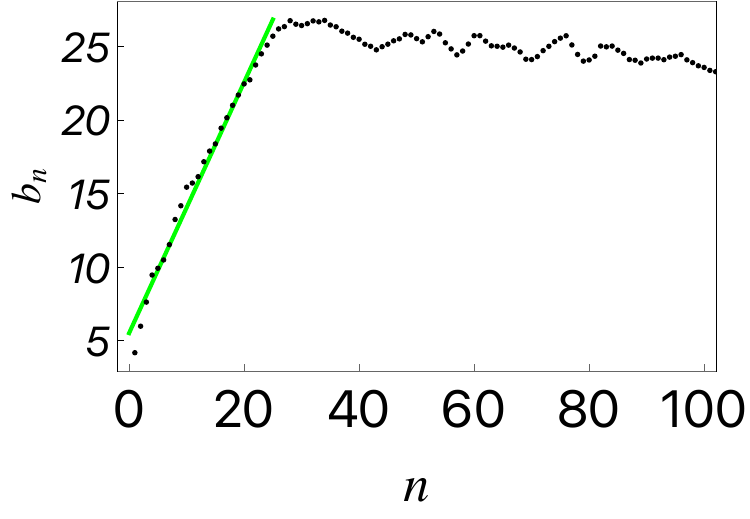}}\hspace{3mm}
        {\includegraphics[width=4.0cm]{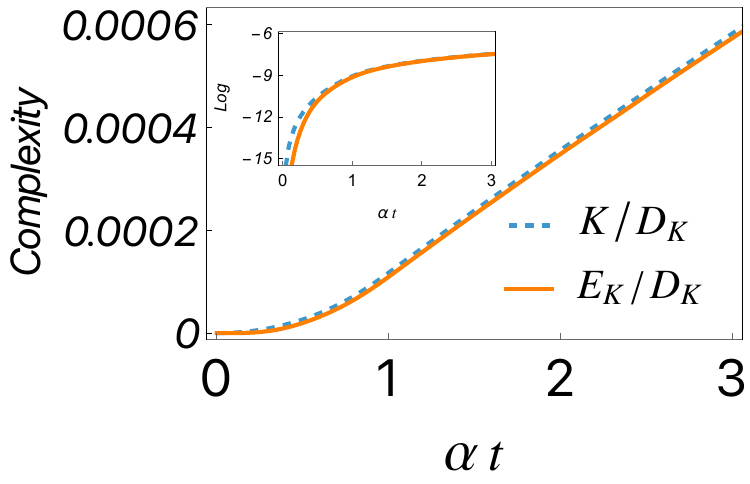}}\vspace{3mm}
        {\includegraphics[width=4.5cm]{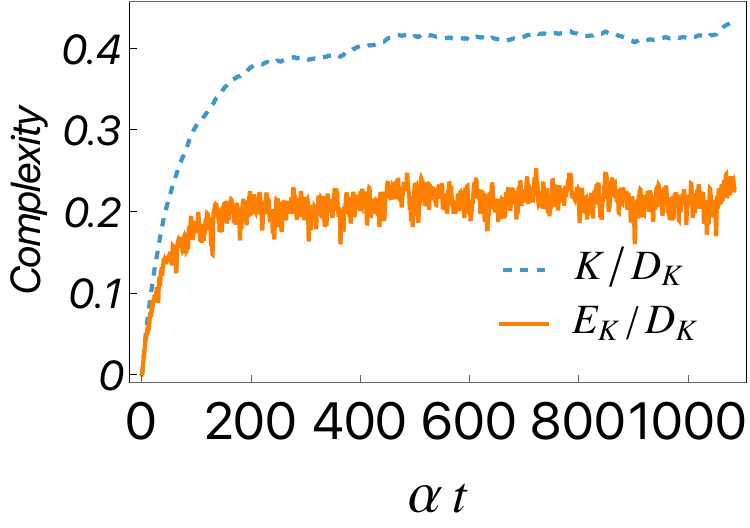}}
    \caption{\textbf{Upper left panel:} Lanczos coefficients for the operator $\hat{S}_{5}^{z}+\hat{S}_{6}^{z}$ in the mixed-field Ising model with $h_{x}=-1.05$, $h_{z}=0.5$ with sites $L=10$ in the negative parity sector. The linear growth with slope $\alpha=0.8516$ indicates the exponential growth of Krylov complexity. \textbf{Upper right panel:} $K(t)$ (blue, dashed) and $\mathbf{E}_{K}(t)$ (orange) for the same parameters and operator in the negative parity sector. The horizontal axis is scaled with the slope of $b_{n}$ by $\alpha$, whereas the vertical axis is scaled by the Krylov dimension $D_{\mathcal{K}}=245521$. The inset is the log plot. \textbf{Lower panel:} Late-time saturation of $K(t)$ and $\mathbf{E}_{K}(t)$ in the negative parity sector with Krylov dimension $D_{\mathcal{K}}=3081$ for sites $L=7$ and operator $\hat{S}_{3}^{z}+\hat{S}_{5}^{z}$ with slope $\alpha=1.0848$. For $L=10$, the least numerical precision of the Krylov basis is on the order of $10^{-360}$, whereas for $L=7$, it is on the order of $10^{-820}$.}
    \label{fig: mixedFising}
\end{figure}

\phantomsection
B. \textit{The Mixed-Field Ising Model}\label{subsec:NumericalAnalysisMFI} -- 
In this subsection, we numerically compute the logK complexity in a genuinely chaotic system: the mixed-field Ising model at the chaotic point. In this case, we expect that the elogK complexity should behave similarly to the conventional Krylov complexity, at least in terms of early-time exponential growth. As we will see below, this expectation is supported by our numerical calculations. We consider a one-dimensional mixed-field Ising model with open boundary conditions, described by the following Hamiltonian:
\begin{align}
\hat{H}_{\mathrm{MFI}}=-\sum_{i=1}^{L-1} \hat{S}_{i}^{z}\hat{S}_{i+1}^{z}-\sum_{i=1}^{L} \left(h_{x}\hat{S}_{i}^{x}+h_{z}\hat{S}_{i}^{z}\right)~,
\end{align}
where $L$ is the total number of sites, \(h_{x}\) and \(h_{z}\) are parameters controlling the strength of transverse and longitudinal magnetic fields, respectively. This model is integrable when either parameter vanishes; otherwise, it enters a non-integrable phase for finite values of the parameters. In particular, its energy spectrum exhibits Wigner--Dyson statistics when the parameters are chosen as \(h_{x}=-1.05\) and \(h_{z}=0.5\) \cite{Craps:2019rbj, Ba_uls_2011}. In our numerical evaluation, since the model preserves a parity symmetry, we restrict our analysis to an individual parity sector.

In our numerical implementation, we set $L=10$ and study the operator $\hat{S}_{5}^{z}+\hat{S}_{6}^{z}$ at the chaotic point at early times. As shown in the upper right panel in Figure \ref{fig: mixedFising}, the elogK complexity and the conventional Krylov complexity exhibit similar behavior at early times. The linear growth of $b_n$ from the upper left panel in Figure \ref{fig: mixedFising} indicates the exponential growth of the Krylov complexity in the form $\sim e^{2\alpha t}$. The parallel growth of the elogK complexity with the Krylov complexity at early times suggests the same form of exponential growth for $\mathbf{E}_K(t)$, as well as the linear growth of $\mathbf{L}_K(t)$, summarized in the Table. \ref{tab:KrylovTab}. This numerical analysis supports our expectation that the elogK complexity has a window of exponential growth in chaotic finite-dimensional many-body systems. We also show the late-time saturation of both with sites $L=7$ in the lower panel in Figure \ref{fig: mixedFising}. They all saturate lower than the upper bounds (\ref{eq-K-longtime}) and (\ref{eq-EK-longtime}) as expected, since this is not a maximally chaotic/thermalizing model.\\
$\phantom{word}$To summarize, our numerical analysis in this Section showed that elogK complexity is highly suppressed in integrable systems dominated by saddles, while exhibiting exponential growth in genuine chaotic systems. The behaviors of logK and elogK complexities are listed in Table \ref{tab:KrylovTab} and compared with the conventional Krylov complexity.

\begin{widetext}
\begin{center}
\begingroup
\setlength{\tabcolsep}{7pt} 
\renewcommand{\arraystretch}{1.5} 
\begin{table}[htbp]
    \centering
    \begin{tabular}{|c|c|c|c|c|c}
        \hline
       \multirow{2}{*}{Time Scale} &  \multicolumn{2}{c|}{Complexity Measure}& \multicolumn{2}{c|}{Logarithmic form}\\ \cline{2-5}
        &\hspace{10pt} $K(t)$ \hspace{10pt} & $\mathbf{E}_K(t)$ & $\log(K(t))$     &  $\mathbf{L}_K(t)$ \\ \cline{1-5}
        Initial growth $( 0\lesssim t )$ & $ b_1^2 t^2 $ & $ \frac{\log(2)}{4}b_1^2 b_2^2 t^4 $ & $ b_1^2 t^2$ & $\frac{\log(2)}{4}b_1^2 b_2^2 t^4 $   \\ \cline{1-5}
        Early time growth $(t<t_*)$ in the LMG model& \multirow{2}{*}{$ e^{\lambda_{\mathrm{cl}} t}$} & \multirow{2}{*}{$e^{e^{-\lambda_{\mathrm{cl}} t}t}$} & \multirow{2}{*}{\red{$\lambda_{\mathrm{cl}} t$}} & \multirow{2}{*}{\red{$e^{-\lambda_{\mathrm{cl}} t}t$ }} \\
        (from numerical fittings in the Figure. \ref{fig-LMG})&&&&\\ \cline{1-5}
        Early time growth $(t<t_*)$ in the chaotic mixed-field& \multirow{2}{*}{$ e^{2\lambda t}$} & \multirow{2}{*}{$e^{2\lambda t}$ $^{[\textcolor{green}{*}]}$}& \multirow{2}{*}{\blue{$2\lambda t$}} & \multirow{2}{*}{\blue{$2\lambda t$} $^{[\textcolor{green}{*}]}$} \\ 
        Ising model (from numerical data in the Figure. \ref{fig: mixedFising})&&&&\\
        \cline{1-5}
        Late-time saturation value in chaotic systems& $D_{\mathcal{K}}/2$&$\Gamma(D_{\mathcal{K}})^{(1/D_{\mathcal{K}})}$ &$\log(D_{\mathcal{K}}/2)$ &$\frac{\log((2)_{D_\mathcal{K}-2})}{D_\mathcal{K}} $\\
        \hline
    \end{tabular}
    \caption{Summary of initial and early-time growth, and late-time saturation of complexity measures $K(t)$ and $\mathbf{E}_K(t)$ and their logarithmic forms $\log(K(t))$ and $\mathbf{L}_K(t)$ from analytical analyzes based on perturbation solutions for the Schrodinger equation (\ref{App-eq-SchrodingerEq}), numerical analysis in Section. \hyperref[sec:NumericalAnalysis]{V.}, and late-time average (\ref{eq:LongTimeLogK}), respectively. Detailed explanations of [\textcolor{green}{*}] are given in Section \hyperref[subsec:NumericalAnalysisMFI]{V.B.}}
    \label{tab:KrylovTab}
\end{table}
\endgroup
\end{center}
\end{widetext}

%

\phantomsection
VI. \textit{Logarithmic Krylov Complexity in the Quantum Inverted Harmonic Oscillator}\label{sec:QuantKrylovIHO} --
Another interesting setup for studying logK complexity is the inverted harmonic oscillator in quantum mechanics. This setup has been used in the past as another simple model for saddle-dominated scrambling in which probes such as OTOCs and Krylov complexity exhibit exponential growth~\cite{Hashimoto:2020xfr,Hashimoto:2023swv}. Similarly to the conformal/low energy limit of the SYK model, discussed in Section.~\hyperref[sec:ExamplelogKCFT]{IV.}, we want to compute the higher-order Krylov complexities and the logK and elogK complexity in this setup. The starting point is the Hamiltonian
\begin{align}
    \label{eq:QuantumIHOHam}
    \hat{H}_{\mathrm{IHO}}=\frac{\lambda}{2} \left(\hat{x}\hat{p}+\hat{p}\hat{x}\right)~,
\end{align}
where $[\hat{x},\hat{p}]=i$, $\lambda \in \mathbb{R}$ and where $\lbrace \hat{x},\hat{p}\rbrace$ are related to the canonical position and momentum operators in quantum mechanics by $\lbrace\hat{X}=(\hat{x}-\hat{p})/\sqrt{2}, \hat{P}=(\hat{x}+\hat{p})/\sqrt{2}\rbrace$. The energy spectrum of the Hamiltonian~\eqref{eq:QuantumIHOHam}, which can be written as a generator of dilatations $\hat{H}_{\mathrm{IHO}}=-i\lambda(\hat{x}\partial_{x}+1/2)$ with $\hat{p}=-i\partial_{x}$, is continuous and unbounded from above and below. As a consequence, the Gibbs/KMS thermal state $e^{-\beta \hat{H}_{\mathrm{IHO}}}$ does not exist in this case for any $\beta>0$, and the partition function $\mathcal{Z}(\beta)=\mathrm{Tr}(e^{-\beta \hat{H}_{\mathrm{IHO}}})$ diverges. There are a few ways to overcome this difficulty. While some involve restringing the operator algebra or introducing cutoffs in the energy spectrum, an alternative is to consider the infinite temperature limit $ \beta\rightarrow 0$ of the Wightman inner product (the Hilbert--Schmidt (HS) inner product) and consider operators that are normalizable with respect to it. Such a class of operators includes Gaussian operators of the form
\begin{align}
    \label{eq:GaussianOpIHO}
    \hat{\mathcal{O}}_{0}= \left(\frac{2}{\pi\alpha}\right)^{1/4}\,e^{-\frac{\hat{x}^{2}}{\alpha}}~,
\end{align}
where $\alpha>0$ is a constant that labels different Gaussian operators and is related to the variance of the Gaussian probability distribution $\langle x \vert \hat{\mathcal{O}}_{0} \rangle$~\footnote{Of course,~\eqref{eq:GaussianOpIHO} is not properly normalized as a probability distribution but rather as an operator with respect to the HS norm}. The class of operators ~\eqref{eq:GaussianOpIHO} are normalized with respect to the Hilbert--Schmidt (HS) norm where the trace can be expanded in position eigenstates
\begin{align}
    \label{eq:GaussianOpNormaIHO}
    \begin{split}
  &\vert \vert  \hat{\mathcal{O}}_{0} \vert \vert^{2}=(\hat{\mathcal{O}}_{0}\vert \hat{\mathcal{O}}_{0})^{\mathrm{HS}}=\mathrm{Tr}\left(\hat{\mathcal{O}}_{0}^{\dagger}\hat{\mathcal{O}}_{0}\right)\\
  &=\int_{-\infty}^{+\infty}\mathrm{d}x\,\hat{\mathcal{O}}_{0}^{\dagger}(x)\hat{\mathcal{O}}_{0}(x)=1~,
  \end{split}
\end{align}
and where we omitted the divergent factor $1/\mathcal{Z}(0)$ in the inner product. Using the solutions to the classical equations of motion $x(t)=e^{\lambda t}x(0)$, the unitary time evolution of $\hat{O}_{0}$ can be found in a straightforward way to be given by
\begin{align}
    \label{eq:GaussianOpIHOTimeEvol}
     \hat{\mathcal{O}}_{0}(t)= e^{it\hat{H}_{\mathrm{IHO}}}\hat{\mathcal{O}}_{0}e^{-it\hat{H}_{\mathrm{IHO}}}=\left(\frac{2}{\pi\alpha}\right)^{1/4}\,e^{-\frac{e^{2\lambda t}\hat{x}^{2}}{\alpha}}~,
\end{align}
which we can normalize for all $t$ with respect to the HS inner product by redefining it to $\hat{\mathcal{O}}_{0}(t):=(2/\pi\alpha)^{1/4}e^{\lambda t/2}\,e^{-e^{2\lambda t}\hat{x}^{2}/\alpha}$. As noted in~\cite{Hashimoto:2023swv}, the Krylov basis of operators can be written in terms of even Hermite polynomials $H_{2n}(z)$ as
\begin{align}
    \label{eq:KrylovBasisHermiteIHO}
    \hat{\mathcal{K}}_{n}=\left(\frac{2}{\pi\alpha}\right)^{1/4}\,\frac{1}{\sqrt{(2n)!}\,2^{n}}\,H_{2n}\left(\sqrt{\frac{2}{\alpha}}\hat{x}\right)e^{-\frac{\hat{x}^{2}}{\alpha}}~.
\end{align}
In the Supplemental Material~\hyperlink{app:MathDetailsIHO}{C.} we show that the wavefunctions $\varphi_{n}(t):=i^{-n}(\hat{\mathcal{K}}_{n}\vert \hat{\mathcal{O}}_{0}(t))^{\mathrm{HS}}$ take the general form
\begin{align}
    \label{eq:ProbAmpQuantIHO}
    \varphi_{n}(t)=(-1)^{n}\frac{\sqrt{(2n)!}}{2^{n}n!}\frac{\tanh^{n}(\lambda t)}{(\cosh(\lambda t))^{1/2}}~,
\end{align}
and are thus formally a particular case of the wavefunctions discussed in the conformal limit of the SYK~\eqref{eq:PhinCFT}. Setting $\eta = 1/2$ in the latter, these wavefunctions coincide up to the factor $(-1)^{n}$. For example, the Krylov complexity computed from the wavefunctions~\eqref{eq:ProbAmpQuantIHO} is given by
\begin{align}
    \label{eq:KrylovCompQuantumIHO}
    K_{\mathcal{O}}(t)=\frac{1}{2}\sinh^{2}(\lambda t)~,
\end{align}
which matches~\eqref{eq:KrylovCFT} for $\eta=1/2$ and $\alpha=\lambda$\footnote{Note that the parameter $\alpha$ appearing in the Gaussian operator~\eqref{eq:GaussianOpIHO} disappears from the wavefunctions $\varphi_{n}(t)$, the Lanczos coefficients $b_{n}=\lambda\sqrt{n(n-1/2)}$ and from the Krylov complexity~\eqref{eq:KrylovCompQuantumIHO}, because the integrals of products of Hermite polynomials through the HS inner product are independent of the scale of $x$.}.\\
$\phantom{word}$Using the definition~\eqref{eq:HigherOrdKcomp}, we can formally write the higher-order Krylov complexity for integer $1\leq m \in \mathbb{Z}^{+}$ as
\begin{align}
\label{eq:HighOrdKIHO}
\begin{split}
    K^{(m)}_{\mathcal{O}}(t)=&\frac{1}{2} {}_{m}F_{m-1}\left(\frac{3}{2},\underbrace{2,\dots,2}_{m-1};\underbrace{1,\dots,1}_{m-1}; \tanh^2(\lambda t)\right)\\
    &\times \sech(\lambda t)\tanh^2(\lambda t)~,
    \end{split}
\end{align}
where $_{p}F_{q}(a;b;z)$ is the generalized hypergeometric function, where $a=(a_{1},\ldots , a_{p})$, and $b=(b_{1},\ldots,b_{q})$. Note that this is different from~\eqref{eq:HigherOrdKrylovCFT} in general, since that expression was valid only for $\eta = 1$. However, in general, it also has a similar formal expression to~\eqref{eq:HigherOrdKrylovCFTeta1}. In fact, for integer $1 \leq m \in \mathbb{Z}^{+}$, Eq.~\eqref{eq:HighOrdKIHO} can be written as 
\begin{align}
\label{eq:HighOrdKIHOSimple}
\begin{split}
    &K^{(m)}_{\mathcal{O}}(t)=\sech(\lambda t)\cosh^{1+2m}(\lambda t)\\
    & \times \sum_{k=1}^{m} \vert S_{2}(m,k)\vert\,\frac{\Gamma(1/2+k)}{\Gamma(1/2)}\tanh^{2k}(\lambda t)\sech^{2(m-k)} (\lambda t)~,
    \end{split}
\end{align}
where $S_{2}(m,k)$ are Stirling numbers of the second kind, which have the following generating function
\begin{align}
    \label{eq:StirlingSecondKindGenFunc}
    \frac{1}{k!}\left(e^{x}-1\right)^{k}=\sum_{n=k}^{\infty}S_{2}(n,k)\frac{x^{n}}{n!}~.
\end{align}
From~\eqref{eq:HighOrdKIHOSimple} we see that for integer $1 \leq m \in \mathbb{Z}^{+}$, the $m$-th order Krylov complexity has the form
\begin{align}
    \label{eq:HigherOrdKIHOExp}
    \begin{split}
    &K^{(m)}_{\mathcal{O}}(t)=\sum_{j=-m}^{m}\tilde{\nu}_{m,j}e^{2\lambda j t}\\
    &=\frac{\Gamma(\frac{1}{2}+m)}{2^{1+2m}\Gamma(\frac{3}{2})}e^{-2m\lambda t}+\cdots +\frac{\Gamma(\frac{1}{2}+m)}{2^{1+2m}\Gamma(\frac{3}{2})}e^{2m\lambda t}~,
    \end{split}
\end{align}
with $\tilde{\nu}_{m,-m}=\tilde{\nu}_{m,m}=2^{-1-2m}\Gamma(1/2+m)/\Gamma(3/2)$, and where $\ldots$ denotes terms $e^{2j\lambda t}$ with powers $1\leq \vert j \vert \leq (m-1)$. So, similarly to Section.~\hyperref[sec:ExamplelogKCFT]{IV.}, Eq.~\eqref{eq:HigherOrdKrylovCFTeta1}, the $m$-th order Krylov complexity has a leading large-$t$ behavior dominated by $\propto e^{2m\lambda t}$. The initial-time behavior for integer $m$ is instead given by
\begin{align}
    \label{eq:HigherOrdKIHOInitialy}
    \begin{split}
    &K^{(m)}_{\mathcal{O}}(t)\approx\frac{\lambda^{2} t^{2}}{2}+ \frac{\left(9\cdot2^{m}-14\right)}{24}\lambda^{4} t^{4} + O( \lambda^{6} t^{6})~,
    \end{split}
\end{align}
which exactly matches the early-time behavior~\eqref{eq:HigherOrdKrylovCFTSmallt} for $\eta=1$. This strengthens our intuition that this analysis is mathematically very close to the $\eta = 1$ CFT case. \\
In order to analytically continue the higher-order complexities~\eqref{eq:HighOrdKIHOSimple} to real $m\in \mathbb{R}$ and find the logK complexity, it is useful to rewrite the higher-order complexities for integer $m$ in the following way:
\begin{align}
    \label{eq:HighOrdKIHOy}
    \begin{split}
    K^{(m)}_{\mathcal{O}}(t)&=\sech(\lambda t)\sum_{n=0}^{\infty}\frac{n^{m}(2n)!}{2^{2n}(n!)^2}\tanh(\lambda t)^{2n}\\
    &=\sech(\lambda t)\sum_{n=0}^{\infty}n^{m}\begin{pmatrix}{}2n\\
    n
    \end{pmatrix}\frac{y^{n}}{4^n}~,
    \end{split}
\end{align}
where $y=\tanh^{2}(\lambda t)$ and $\vert y\vert \leq 1$. Since this series~\eqref{eq:HighOrdKIHOy} is absolutely convergent for $m\in \mathbb{Z}^{+}$, we can perform the standard analytic continuation $n^{m}\mapsto e^{m\log(n)}$ and perform the replica trick, yielding a formal series expression for logK
\begin{align}
    \label{eq:LogKIHOlogKFormalSeries}
    \begin{split}
    &\mathbf{L}_{\mathcal{O}}(t)=\lim_{m\rightarrow 0}\frac{\partial}{\partial m}K_{\mathcal{O}}^{(m)}\\
    &=\sech(\lambda t)\sum_{n=1}^{\infty}\log(n)\begin{pmatrix}{}2n\\
    n
    \end{pmatrix}\frac{\tanh(\lambda t)^{2n}}{4^{n}}~,
    \end{split}
\end{align}
where we arrived at a regularized expression by subtracting the logarithmic divergence at $n=0$ following ~\eqref{eq:ShiftedSpreadingOp}. 

To see the large $\lambda t$ behavior of~\eqref{eq:LogKIHOlogKFormalSeries}, corresponding to the limit $y\rightarrow 1$, we can perform an expansion of the formal series using
\begin{align}
    \label{eq:LogSeriesLimity1}
    \begin{split}
    &\sum_{n=1}^{\infty}\log(n)\begin{pmatrix}{}2n\\
    n
    \end{pmatrix}y^{n}\\
    &\approx\frac{1}{2}(1-y)^{-1/2}\left(\log\left(\frac{1}{1-y}\right)-\gamma - \log(2)\right)\\
    &+O\left((1-y)^{1/2}\log(1-y)\right)~,
    \end{split}
\end{align}
where $\gamma$ is the Euler--Mascheroni constant. Multiplying this expression by $\sech(\lambda t)$ and using the identities $1-y=\sech(\lambda t)^{2}$, $\log((1-y)^{-1})$ = $\log(\cosh^{2}(\lambda t))$ = $2\log(\cosh(\lambda t))$, and $\sech(\lambda t)(1-y)^{-1/2}$ = $\sech(\lambda t)\cosh(\lambda t)=1$, we find the asymptotic large $\lambda t$ behavior of~\eqref{eq:LogKIHOlogKFormalSeries} to be
\begin{align}
    \label{eq:LogKIHOlogKlarget}
    \mathbf{L}_{\mathcal{O}}(t)\sim \lambda t - \frac{\gamma}{2}-2\log(2)\,\,\,,\,\,\,(\mathrm{large } \,\,\,t)~.
\end{align}
This shows that, similarly to the SYK$_{q}$ at low energies and temperatures, in this case, and for the specific choice of Gaussian initial operator~\eqref{eq:GaussianOpIHO}, the logarithmic complexity itself is incapable of accurately avoiding the instability from the classical unstable saddle. Although this could be due to the fact that the Krylov subspace is also infinite dimensional for this particular choice of operator, an additional exponential growth is already present in $\hat{O}_{0}(t)\propto e^{\lambda t/2}e^{-e^{2\lambda t}\hat{x}^{2}/2}$. Nevertheless, this brings into question the use of logK complexity to resolve the unstable saddle issue in infinite dimensional systems. In Sec.~\hyperref[sec:ReplicaGeneralization]{IX.} we return to this issue with a possible resolution. However, it would be also useful to understand whether classical notions of Krylov and logK complexity are able to bypass this problem. This is the focus of the next sections.


\phantomsection
VII. \textit{Classical Phase Space Analysis}\label{sec:ClassicalPSKrylov} -- To understand the main difference between logK complexity and K-complexity as probes of scrambling in finite-dimensional dynamical systems, it is illustrative to study their behavior in classical systems with unstable saddles. Before doing so, in this section we review the Krylov formalism in classical phase space. We emphasize that, as noted in~\cite{Parker:2018yvk}, the recursion method has been extensively discussed in the context of classical dynamics~\cite{RecursionBook,Lee:1982ort} to study various properties of many-body systems such as transport coefficients and even classical Lyapunov exponents~\cite{FlorencioLee:1985har,Liu:1995xyz}. However, to a large extent this classical analysis has focused on the properties of classical autocorrelation functions and their related moments, Lanczos coefficients and continued fraction expansions. To bridge the gap between the quantum formalism of Krylov complexity and the recursion method in classical dynamics, we give an overview of the basic framework and define classical notions of Krylov and logK complexity.\\
\phantomsection
A. \textit{The Algebra of Functions in Classical Phase Space}\label{subsec:ClassicalBasics} -- In (bosonic) classical mechanical systems, phase space $\mathcal{S}$ is a real $2N$-dimensional symplectic manifold~\cite{BookArnold}, where the dynamics (phase flow) are generated by a Hamiltonian vector field $X_{H}$, related to the system's Hamiltonian $H$, through the Poisson brackets $\lbrace \cdot,\cdot \rbrace_{\mathrm{PB}}$
\begin{equation}
    X_{H}:=\sum_{k=1}^{N}\left(\frac{\partial H}{\partial p_{k}}\frac{\partial \,\cdot}{\partial q^{k}}-\frac{\partial H}{\partial q^{k}}\frac{\partial\,\cdot }{\partial p_{k}}\right)\equiv - \lbrace H,\cdot\,\rbrace_{\mathrm{PB}}~,
    \label{eq:PoissonBracket}
\end{equation}
where $\lbrace \mathbf{q}=(q^{1},\ldots,q^{N}) ,\mathbf{p}=(p_{1},\ldots,p_{N})\rbrace $ are canonical (Darboux) coordinates in phase space $\mathcal{S}$, $H=H(\mathbf{q},\mathbf{p})$, and where the symplectic form is $\omega=\sum_{k}\mathrm{d}p_{k}\wedge \mathrm{d}q^{k}$~\footnote{For simplicity, here we focus on bosonic phase space. Classical fermionic phase spaces are equipped instead with a symmetric positive-definite bilinear form, a metric $g_{ab}$. See e.g.~\cite{Hackl:2020ken} for details and references therein.}. In this sense, the Hamiltonian vector field $X_{H}$ is the classical analogue of the Liouvillian: $\hat{\mathcal{L}} := [H,\cdot]\rightarrow -\{H,\cdot\}_{\mathrm{PB}}\equiv  X_{H}$. Similarly, the classical analogue of quantum observables (Hermitian operators) in the operator algebra $\mathcal{A}(\mathcal{H})$ are real-valued and smooth functions $C^{\infty}(\mathcal{S})$ in phase space $\mathcal{S}$, which together with pointwise multiplication, define an associative and commutative algebra: the algebra of classical observables $\mathcal{A}(\mathcal{S})$. Time evolution in phase space is given by the $1$-parameter family of diffeomorphisms generated by $X_{H}$, $h_{t}:\mathcal{S}\rightarrow \mathcal{S}$, $t\in \mathbb{R}$, such that $h_{t}(\mathbf{q},\mathbf{p})=(\mathbf{q}(t),\mathbf{p}(t))$ , where $(\mathbf{q}(t),\mathbf{p}(t))$ satisfy Hamilton's equations: $\dot{q}^{k}:=\textrm{d}q^{k}/\textrm{d}t=X_{H}(q^{k})=\partial H / \partial p_{k}$ and $\dot{p}_{k}:=\textrm{d}p_{k}/\textrm{d}t=X_{H}(p_{k})=-\partial H / \partial q^{k}$, for all $k=1,\ldots, N$, and where $(\mathbf{q}(0)=\mathbf{q},\mathbf{p}(0)=\mathbf{p})$. In other words:
\begin{align}
    \label{eq:TimeEvolPhase}
    \frac{\textrm{d}}{\textrm{d}t}(h_{t}(\mathbf{q},\mathbf{p}))\Big\vert_{t=0}=X_{H}(\mathbf{q},\mathbf{p})~.
\end{align}
This defines the Hamiltonian phase flow in $\mathcal{S}$. Intuitively, this means that along the integral curves of $X_{H}$ we have $X_{H}=\textrm{d}/\textrm{d}t$, and consequently, the integral curves of $X_{H}$ satisfy Hamilton's equations. In general, given a classical observable $f:\mathcal{S}\rightarrow \mathbb{R}$, $f \in \mathcal{A}(\mathcal{S})$, with $f(h_{t}(\mathbf{q},\mathbf{p}))=f(\mathbf{q}(t),\mathbf{p}(t)):=f_{t}(\mathbf{q},\mathbf{p})$ that does not depend explicitly on time, its time evolution is described by the differential equation
\begin{align}
    \label{eq:TimeEvolClassic}
    \frac{\textrm{d} }{\textrm{d}t}f_t = X_{H}(f_{t})=- \lbrace H,f_t \rbrace_{\mathrm{PB}}~.
\end{align}
One may think of~\eqref{eq:TimeEvolClassic} as a differential equation for the family of functions $f_{t}\in \mathcal{A}(\mathcal{S})$ with initial condition $f_{t}(\mathbf{q},\mathbf{p})\vert_{t=0}=f(\mathbf{q},\mathbf{p})$. Importantly, we can enhance the algebraic structure of \emph{measurable} classical observables $\Sigma_{\mathcal{S}}\subset \mathcal{A}(\mathcal{S})$ by introducing an inner-product. If $\mathbf{x}=\lbrace x^{1},\ldots,x^{2N}\rbrace=\lbrace q^{1},\ldots,q^{N},p_{1},\ldots,p_{N}\rbrace$ are canonical coordinates in $\mathcal{S}$ and $f\in \Sigma_{\mathcal{S}}$ is a measurable function, we can define its phase-space average by
\begin{align}
    \label{eq:PhaseSpaceAve}
    \langle f\rangle_{\mathcal{S}}:=\frac{1}{\mu(\mathcal{S})}\int_{\mathcal{S}} \textrm{d}\mu(\mathbf{x})\,f(\mathbf{x})~,\,
\end{align}
where $\mu:\mathcal{S}\rightarrow \mathbb{R}$ is a measure in $\mathcal{S}$ and $\mu(\mathcal{S}):=\int_{\mathcal{S}}\textrm{d}\mu(\mathbf{x})$, such that $\langle \mathbb{I}\rangle_{\mathcal{S}}=1$, where $\mathbb{I}(\mathbf{x}):=1$ for all $\mathbf{x}\in \mathcal{S}$. We also require that the measure $\mu$ is preserved under Hamiltonian flow: $\mu(h_{t}(F))=\mu(F)$ for all $F\in \Gamma$, where $\Gamma\subseteq \Sigma_{\mathcal{S}}$ is any measurable subset of $\Sigma_{\mathcal{S}}$. In particular, this implies that the phase space average~\eqref{eq:PhaseSpaceAve} is invariant under phase flow, namely
\begin{align}
    \label{eq:TimeEvolPhaseSpaceAve}
    \begin{split}
    \langle f_{t}\rangle_{\mathcal{S}}&=\frac{1}{\mu(\mathcal{S})}\int_{\mathcal{S}}\,\mathrm{d}\mu(\mathbf{x})f(h_{t}(\mathbf{x}))\\
    &=\frac{1}{\mu(\mathcal{S})}\int_{\mathcal{S}}\,\mathrm{d}\mu(\mathbf{y})f(\mathbf{y})=\langle f\rangle_{\mathcal{S}}~,
    \end{split}
\end{align}
where $\mathbf{y}=h_{t}(\mathbf{x})$, since $\mu(h_{t}(F))=\mu(F)=\mu(h_{t}^{-1}(F))$ for all measurable $F\in \Gamma\subset \Sigma_{\mathcal{S}}$ implies $\mu(\mathbf{y})=\mu(\mathbf{x})$. This allows us to define an inner product of classical observables as follows: If $f,g\in \Sigma_{\mathcal{S}}$ then we define their inner product (with respect to $\mu$) as:
\begin{align}
    \label{eq:PhaseSpaceInnProd}
    \langle f,g\rangle:=\langle f\cdot g\rangle_{\mathcal{S}}-\langle f\rangle_{\mathcal{S}}\langle g\rangle_{\mathcal{S}}~.
\end{align}
Note that here we defined $\langle f,g\rangle \equiv \langle g,f \rangle$ as the connected two-point correlation function between $f$ and $g$. Thus, the positive definiteness of~\eqref{eq:PhaseSpaceInnProd} is equivalent to the condition that $\vert \vert f\vert \vert^{2}:=\langle f,f \rangle=\langle f^{2}\rangle_{\mathcal{S}}-\langle f\rangle_{\mathcal{S}}^{2} \geq 0$, which can be interpreted as the requirement that the variance of $f$ in $\mathcal{S}$ with respect to the measure $\mu$ be non-negative~\footnote{Using this definition of inner product, functions of the form $f(\mathbf{x})=k$ where $k\in \mathbb{R}$ will trivially have zero norm. This includes the unit function $\mathbb{I}(\mathbf{x}):=1$ with unit expectation value. To avoid this, we could simply consider $\langle f,g\rangle:=\langle f\cdot g\rangle_{\mathcal{S}}$ . Alternatively, one could focus on non-vanishing ``centered'' observables $f\rightarrow f'=f-\langle f\rangle_{\mathcal{S}}$}. Finally, in order for~\eqref{eq:PhaseSpaceInnProd} to properly define a positive-definite inner product, we take the quotient of $\Sigma_{\mathcal{S}}$ by the null space of observables with vanishing variance: $N_{\mu}:=\lbrace f \in \Sigma_{S}\,\vert\,f\neq0\,\,,\,\,\vert \vert f \vert \vert^{2}\equiv0\rbrace$. Thus, the algebra of measurable observables will be the (completion under the norm topology of the) quotient of $\Sigma_{\mathcal{S}}$ by the null space $N_{\mu}$, $\hat{\Sigma}_{\mathcal{S}}:=\widetilde{(\Sigma_{\mathcal{S}}/N_{\mu})}$, which is essentially equivalent to $\mathbb{L}^{2}(\Sigma_{S},\mu)$. \\
B. \textit{The Recursion Method}\label{subsec:ClassicalRecursion} -- Given a measurable function $f\in \hat{\Sigma}_{\mathcal{S}}$, we now consider the following sequence of nested Poisson brackets for $f$:
\begin{align}
    \label{eq:SequencefPB}
    \begin{split}
  & \lbrace \tilde{f}_{0}=f\,\,,\,\,\tilde{f}_{1}=-\lbrace H,f\rbrace_{\mathrm{PB}}\,\,,\,\ldots\rbrace~.
   \end{split}
\end{align}
This sequence is comprised of observables $\tilde{f}_{n}$ that contain all mixed derivatives of $f$ up to order $n$ in phase space combined with derivatives of the Hamiltonian $H$ of the same order, with
\begin{align}
    \label{eq:SequencefPBn}
    \begin{split}
    \tilde{f}_{n}:=(-1)^n&\lbrace \underbrace{H,\,\ldots,\lbrace H,\lbrace H, f\rbrace_{PB}\rbrace_{PB}\ldots}\rbrace_{PB}~.\\
    &  \quad\quad\quad\quad\,\quad (n \textrm{ times})
    \end{split}
\end{align}
Given this sequence, we construct an orthonormal basis of functions $\lbrace \mathfrak{K}_{n}\rbrace$ in $\hat{\Sigma}_{S}$ using the inner product~\eqref{eq:PhaseSpaceInnProd} via the Gramm--Schmidt procedure:
\begin{align}
    \label{eq:GrammSchmidtFunctions}
    \begin{split}
    &k_{0}=\tilde{f}_{0}\,\,,\,\,\mathfrak{K}_{0}=\frac{k_{0}}{\vert \vert k_{0}\vert \vert}~,\\
    &k_{1}=\tilde{f}_{1}-\tilde{m}_{10}k_{0}\,\,,\,\,\tilde{m}_{10}=\frac{\langle\tilde{f}_{1},k_{0}\rangle}{\langle k_{0},k_{0}\rangle},\,\,\mathfrak{K}_{1}=\frac{k_{1}}{\vert \vert k_{1}\vert \vert}~,\\
    &\,\,\,\,\,\,\,\,\vdots\quad \quad \quad \quad \quad \quad \quad \quad \quad  \vdots \quad \quad\quad \quad \quad \quad \quad \vdots\\
    &k_{\ell}=\tilde{f}_{\ell}-\sum_{i=0}^{\ell-1}\tilde{m}_{\ell i}k_{i}\,\,\,,\,\,\tilde{m}_{\ell i}=\frac{\langle\tilde{f}_{\ell},k_{i}\rangle}{\langle k_{i},k_{i}\rangle},\,\,\mathfrak{K}_{\ell}=\frac{k_{\ell}}{\vert \vert k_{\ell}\vert \vert}~,
    \end{split}
\end{align}
where this procedure stops upon reaching a vanishing $k_{l}$ for some $l\geq 1$. In this case, the set of functions $\lbrace \mathfrak{K}_{n}\rbrace$ forms an orthonormal basis $\langle \mathfrak{K}_{m},\mathfrak{K}_{n}\rangle = \delta_{mn}$ of a subset of all measurable observables $\Sigma_{f}\subset\hat{\Sigma}_{\mathcal{S}}$ starting from $f$ (with respect to the measure $\mu$). These orthonormal functions are the classical phase space analogue of the quantum Krylov basis, with $\Sigma_{f}$ being the classical counterpart to the quantum Krylov subspace $\mathcal{K}$.\\
$\phantom{word}$ The Gramm--Schmidt procedure~\eqref{eq:GrammSchmidtFunctions} can typically be simplified into a classical analogue of the Lanczos algorithm. Starting from
\begin{align}
\label{eq:ClassLanczInitial}
\begin{split}
&\mathfrak{f}_{0}=f_{0}~, \quad \mathfrak{b}_{0}=:\sqrt{\left\langle \mathfrak{f}_{0},\mathfrak{f}_{0}\right\rangle}\equiv \vert \vert \mathfrak{f}_{0}\vert \vert~, \quad \mathfrak{r}_{0}=\mathfrak{f}_{0}/\mathfrak{b}_{0}~,
\end{split}
\end{align}
for $n\geq 1$, we define
\begin{align}
\label{eq:ClassLanczRecursion}
\begin{split}
    &\mathfrak{f}_{n}=-\{H,\mathfrak{r}_{n-1}\}+\mathfrak{b}_{n-1}\mathfrak{r}_{n-2}~,\\
    &\mathfrak{b}_{n}=\sqrt{\left\langle \mathfrak{f}_{n},\mathfrak{f}_{n}\right\rangle}\equiv \vert \vert \mathfrak{f}_{n}\vert \vert~,\\
    & \mathfrak{r}_{n}=\mathfrak{f}_{n} / \mathfrak{b}_{n}~,
\end{split}
\end{align}
where $\mathfrak{r}_{-1}\equiv 0$ and where we assume $\mathfrak{f}_{n} \cancel{\propto} \mathfrak{f}_{j} $ with $0\leq j\leq n-1$. Analogously to the general Gramm--Schmidt procedure described above, this process stops once we reach an $l\geq 1$ such that $\mathfrak{b}_{l}=0$. By direct comparison, the $\lbrace \mathfrak{r}_{n}\rbrace$ with $ 0 \leq n \leq l-1$ form an orthonormal basis  $\langle \mathfrak{r}_{n},\mathfrak{r}_{m} \rangle = \delta_{nm}$ of the classical Krylov subspace $D_{\Sigma_{f}}$ that is isomorphic to $\lbrace \mathfrak{K}_{n}\rbrace$ $ 0 \leq n \leq l-1$. In this approach, the $\lbrace \mathfrak{b}_{n} \rbrace$ are the classical Lanczos coefficients and similarly to the quantum case, carry information about the time evolution of the observable $f_{0}$.
Their properties have been amply discussed in~\cite{RecursionBook}. The only difference between~\eqref{eq:GrammSchmidtFunctions} and~\eqref{eq:ClassLanczRecursion} is that the orthogonality between different Krylov basis elements is directly implemented in the former, while in the latter it is assumed that the Poisson brackets generate a new direction in the subalgebra $\hat{\Sigma}_{f}$, which is not always guaranteed. Thus, in~\eqref{eq:ClassLanczRecursion} one needs to verify (often manually) that each new basis element $\mathfrak{f}_{n}$ is linearly independent from all previous ones. \\

$\phantom{word}$A general class of classical observables are polynomials of the phase space coordinates 
\begin{align}
\label{eq:ClassicalObservable}
\begin{split}
f=&a^{(0)}+\sum_{i}a^{(1)}_{i}x^{i}+\sum_{j,k}a^{(2)}_{jk}x^{j}x^{k}+\ldots+\\
&\sum_{i_{1},\ldots,i_{2N}}a^{(2N)}_{i_{1}\ldots i_{2N}}x^{i_{1}}\cdots x^{i_{2N}}~,
\end{split}
\end{align}
with $i_{1}+\ldots +i_{2N}=2N$ and where for bosonic phase space, only the completely symmetric part of $a^{(n)}_{i_{1}\ldots i_{n}}$ ($1\leq n \leq 2N$ and $i_{1}+\ldots + i_{n}\leq 2N$) is non-vanishing. For a given polynomial of degree $0 \leq d \leq 2N$ that depends on $k\leq \mathrm{dim}(\mathcal{S})=2N$ phase space coordinates, the Krylov basis will generically consist of at most $(k+d)!/(k!d!)$ elements, and thus for this type of classical observables, the dimension of the classical Krylov subspace satisfies the inequalities 
\begin{equation}
   1\leq D_{\Sigma_{f}}:=\mathrm{dim}(\Sigma_{f}) \leq \frac{(k+d)!}{k!d!}\leq \frac{(4N)!}{((2N)!)^2}~.
   \label{eq:DimKrylovClassical}
\end{equation}
Thus, the problem of finding a classic Krylov basis for a given classical observable directly translates to the problem of finding a specific family of multivariate orthogonal polynomials with respect to a given phase space measure~\cite{Lee:1982ort,FlorencioLee:1985har}. \\
C. \textit{Classical Krylov and logK-Complexity}\label{subsec:ClassicalKRylov} -- Since the functions~\eqref{eq:GrammSchmidtFunctions} (or equivalently~\eqref{eq:ClassLanczRecursion}) form a complete basis in $\Sigma_{f}$, this means that we can write the time-evolved $f_{t}$ as a linear combination of them, namely
\begin{align}
    \label{eq:ClassicEvolf}
    f_{t}=\sum_{n\geq0}^{}c_{n}(t)\mathfrak{K}_{n}\,\,,\,\,c_{n}(t):=\langle f_{t},\mathfrak{K}_{n} \rangle~.
\end{align}
We can also define the classical analogue of the spreading operator $\mathcal{N}:\hat{\Sigma}_{\mathcal{S}}\rightarrow \Sigma_{f}$ by
\begin{align}
    \label{eq:ClassSpreadingOp}
    \mathcal{N}(\cdot):=\sum_{n\geq 0}^{}n \,\mathfrak{K}_{n}\langle \,\mathfrak{K}_{n},\,\cdot\, \rangle~.
\end{align}
With it, and in analogy to the definition~\eqref{eq:KrylovComp}, we define the classical Krylov complexity of the classical observable $f(\mathbf{x})$ as the ``expectation value'' of the spreading operator $\mathcal{N}$ with respect to $f_{t}$, in other words, by
\begin{align}
    \label{eq:ClassicalKrylovComplexity}
    \begin{split}
    K_{f}(t):&=\frac{\langle f_{t},\mathcal{N}(f_{t})\rangle}{\langle f_{t},f_{t}\rangle}\equiv\frac{\sum_{n\geq 0}n\vert\langle f_{t},\mathfrak{K}_{n} \rangle\vert^{2}}{\vert \vert f_{t}\vert \vert^{2}}\\
    &=\frac{\sum_{n\geq 0} n\vert c_{n}(t) \vert^{2}}{\sum_{n\geq 0}\vert c_{n}(t) \vert^{2}}~.
    \end{split}
\end{align}
In this classical approach, we can still think that the orthonormal basis $\lbrace \mathfrak{K}_{n}\rbrace$ defines an auxiliary one-dimensional and discrete space where each index $n$ labels the ``position'' along this space; a classical version of the Krylov chain. Thus, we can still think of the operational interpretation of Krylov complexity $K_{f}(t)$ as measuring the mean position of the observable $f_{t}$ in the Krylov chain generated by $\lbrace \mathfrak{K}_{n}\rbrace$. Substituting~\eqref{eq:ClassicEvolf} into ~\eqref{eq:TimeEvolClassic} and using the recursion relation~\eqref{eq:ClassLanczRecursion}, where we identify $\mathfrak{r}_{n}\equiv \mathfrak{K}_{n}$, we can find a recursion relation for the coefficients $c_{n}(t)$ 
\begin{align}
    \label{eq:RecursionClassicCoefs}
    \frac{\mathrm{d} c_{n}(t)}{\mathrm{d}t}=\mathfrak{b}_{n}c_{n-1}(t)-\mathfrak{b}_{n+1}c_{n+1}(t)~,
\end{align}
for $n\geq 0$ with $c_{-1}(t)=0$ and $c_{n}(t=0)\propto\delta_{n0}$. \\
$\phantom{word}$A replica generalization of the classical spreading operator~\eqref{eq:ClassSpreadingOp} is the $m$-th order classical spreading operator~\eqref{eq:ClassSpreadingOp}, which we define as
\begin{align}
    \label{eq:ClassGenSpreadingOp}
    \begin{split}
    \mathcal{N}^{m}(f_{t})&:=\underbrace{\mathcal{N}(\cdots\mathcal{N}(f_{t})\cdots)}=\sum_{n\geq 0}n^{m} \,c_{n}(t)\,\mathfrak{K}_{n}~.\\
    &\quad \quad \quad \,\, m \,\,\mathrm{ times}
    \end{split}
\end{align}
With it, we define the classical higher-order Krylov and logK complexity following Eqs.~\eqref{eq:HigherOrdKcomp} and~\eqref{eq-Log-K replica}, namely
\begin{subequations}
    \label{eq:ClassicalmthKrylovandLogK}
    \begin{equation}
        \label{eq:ClassicalmthKrylov}
        \begin{split}
        K_{f}^{(m)}(t):&=\frac{\langle f_{t},\mathcal{N}^{m}(f_{t})\rangle}{\langle f_{t},f_{t}\rangle}=\frac{\sum_{n\geq 0} n^{m}\vert c_{n}(t) \vert^{2}}{\sum_{n\geq 0}\vert c_{n}(t) \vert^{2}}~,
        \end{split}
    \end{equation}
    \begin{equation}
        \label{eq:ClassicalLogK}
        \mathbf{L}_{f}(t):= \frac{\partial K^{(m)}_{f}(t)}{\partial m}\Big\vert_{m\rightarrow 0}\equiv\frac{\sum_{n\geq 1} \log(n)\vert c_{n}(t) \vert^{2}}{\sum_{n\geq 0}\vert c_{n}(t) \vert^{2}}~,
    \end{equation}
\end{subequations}
where we used the fact that the classical Krylov subspace of an observable of the form~\eqref{eq:ClassicalObservable} is always finite-dimensional. We remark that in the above construction, we take $\mathfrak{K}_{n}$ to be functions of the initial phase space coordinates $\mathbf{x}(0)=\mathbf{x}$. Thus, when computing the coefficients $c_{n}(t)$ in Eq.~\eqref{eq:ClassicEvolf}, we evolve the observable $f$ in time via $f_{t}(\mathbf{x}):=f(\mathbf{x}(t))$ while keeping the basis functions $\lbrace \mathfrak{K}_{n}\rbrace$ fixed at the initial time. This leads to the question of whether we can consider the time dependence in the basis functions $\mathfrak{K}_{n}$ while keeping the initial observable $f$ fixed. Due to the invariance of the measure $\mu(\mathbf{x})$ with respect to the Hamiltonian phase flow, we can show that $\langle f\cdot g_{t}\rangle_{\mathcal{S}}\equiv\langle f_{-t}\cdot g\rangle_{\mathcal{S}}$, for any $f,g\in \hat{\Sigma}_{\mathcal{S}}$ and as a consequence,$\langle f,g_{t}\rangle=\langle f_{-t},g \rangle$. Thus, for any $\mathfrak{K}_{n}$
\begin{align}
    \label{eq:ClassicalKMS}
    \begin{split}
    c_{n}(t)=\langle f_{t},\mathfrak{K}_{n} \rangle \equiv \langle f,\mathfrak{K}_{n}(-t) \rangle :=\tilde{c}_{n}(-t)~.
    \end{split}
\end{align}
In other words, these two kinds of coefficients are, in general, related to each other by time reflection. As a consequence, if instead we evolve the Krylov basis in time, the time-evolved spreading operator $\mathcal{N}_{t}=\sum_{n\geq 0}\mathfrak{K}_{n}(t)\langle \mathfrak{K}_{n}(t),\cdot \rangle$ gives rise to the Krylov complexity
\begin{align}
    \label{eq:ClassicKrylovTimeEvolBasis}
    \tilde{K}_{f}(t):=\frac{\langle f,\mathcal{N}_{t}(f)\rangle}{\langle f,f\rangle}=\frac{\sum_{n\geq 0}n \vert \tilde{c}_{n}(t)\vert^2}{\sum_{n \geq 0}\vert c_{n}(0)\vert^2}~,
\end{align}
which is in general related to~\eqref{eq:ClassicalKrylovComplexity} by $K_{f}(t)=\tilde{K}_{f}(-t)$, since $\vert \vert f_{t} \vert \vert^{2}\equiv \vert \vert f_{} \vert \vert^{2}$, given that $\langle f_{t}(\mathbf{x}),f_{t}(\mathbf{x}) \rangle$ = $ \langle f(h^{-1}_{t}(h_{t}(\mathbf{x}))),f(\mathbf{x})\rangle$ = $\langle f(h_{-t}(h_{t}(\mathbf{x}))),f(\mathbf{x})\rangle$ = $\langle f(\mathbf{x}),f(\mathbf{x})\rangle$. We discuss these subtleties in detail and provide examples of the Krylov formalism in classical phase space in the supplemental material~\hyperref[app:ClassPhaseSpace]{D.}\\
D. \textit{Quantum to Classical Transition}\label{subsec:QuantToClassicalTr} -- A different question that is independent of the classical phase space formalism described above is the quantum-to-classical transition. This transition can be achieved in different ways, two of which are the Husimi $Q$ representation~\cite{Husimi1940:kod} and the Wigner--Weyl quasi-probability method~\cite{Wigner1932:Prob}. Generally speaking, the Husimi $Q$ representation maps a quantum Hamiltonian $\hat{H}_{\mathrm{Q}}$ through a coherent basis $\{ \ket{s,m}\}$ to a classical version of said Hamiltonian $H_{\mathrm{C}}$. This approach works when we are given a quantum Hamiltonian $\hat{H}_{\mathrm{Q}}$ that is constructed from SU$(2)$ (pseudo-)spin operators $\lbrace \hat{S}_{i}\rbrace$ ($i\in \lbrace x,y,z \rbrace$) satisfying the $\mathfrak{su}(2)$ algebra. The key idea being that taking the limit of the spin representation $s$ to infinity $s\rightarrow \infty$ amounts to taking a semiclassical limit $\hbar_{\mathrm{eff}}\sim1/s\rightarrow 0$, which, when applied to the expectation value of the Hamiltonian with respect to the coherent states, gives the semiclassical limit of the quantum Hamiltonian. Consider spin coherent states defined by 
\begin{equation}
    \ket{\a}=e^{\a^* \hat{S}_+}\ket{s,-s}~,
\end{equation}
where $\{ \ket{s,m}\}$ are eigenstates of $\{\hat{S}^2,\hat{S}_z\}$. Then, the quantum Hamiltonian $\hat{H}_{\mathrm{Q}}$ can be mapped to its classical version $H_{\mathrm{C}}$ by taking the following limit
\begin{equation}
    \hat{H}_{\mathrm{Q}}\mapsto H_{\mathrm{C}}(\a,\a^*):=\lim_{s \ra \infty} \frac{\bra{\a} \hat{H}_{\mathrm{Q}}\ket{\a}}{\braket{\a}{\a}}~.
\end{equation}
Observables in the operator algebra can be mapped to classical phase space observables in the same way. In this case, the dynamics of phase space coordinates $\lbrace \mathrm{Re}(\alpha),\mathrm{Im}(\alpha)\rbrace $ are governed by the Hamilton equations obtained from $H_{\mathrm{C}}(\alpha,\alpha^{\ast})$. See e.g.~\cite{Ribeiro:2008ltc} for details of this approach in the context of the LMG model.\\
$\phantom{word}$A different, but related approach to map quantum operators to classical observables is through the Wigner--Weyl transform \footnote{See \cite{Pal:2026otf} for a recent work on studying Krylov complexity from Wigner's method. It will be interesting to explore connections between their work and our formalism.}, which schematically takes the following form
\begin{equation}
    A(\mathbf{q},\mathbf{p}):=\frac{1}{\pi \hbar}\int \textrm{d}\mathbf{y}\,\left\langle  \mathbf{q}+\frac{\mathbf{y}}{2}\right\vert \hat{A}\left\vert  \mathbf{q}-\frac{\mathbf{y}}{2}\right\rangle e^{-i \mathbf{p}\cdot \mathbf{y}/\hbar}~,
    \label{eq:WignerWeylTrafo}
\end{equation}
where $\lbrace\mathbf{q},\mathbf{p}\rbrace$ are classical phase space coordinates and where we temporarily reinstated $\hbar$. Eq.~\eqref{eq:WignerWeylTrafo} becomes the Wigner function whenever $\hat{A}$ is a density operator $\hat{\rho}$. The Wigner representation naturally induces a Moyal bracket (see e.g.~\cite{Cotler:2017myn}):
\begin{equation}
    \{\!\{A,B\}\!\}=-\{A,B\}_{\mathrm{PB}}+O(\hbar^2)~.
\end{equation}
The equations of motion derived from this representation naturally include quantum corrections on top of the classical function:
\begin{equation}
    A_{t}(\mathbf{q},\mathbf{p})=A^{\mathrm{C}}_{t}(\mathbf{q},\mathbf{p})+\sum_{i=1}^{\infty} \hbar^{2i}A^{(i)}_{t}(\mathbf{q},\mathbf{p})~,
\end{equation}
which is a solution of the Moyal equations of motion:
\begin{equation}
    \dot{A}_{t}=\{\!\{{H,A_{t}\}}\!\}=-\{H,A_{t} \}_{\mathrm{PB}}+O(\hbar^2)~,
    \label{eq:MoyalEoM}
\end{equation}
where $H$ is a classical Hamiltonian. Eq.~\eqref{eq:MoyalEoM} reduces to~\eqref{eq:TimeEvolClassic} in the limit $\hbar\rightarrow 0$. One strength of the Wigner function is that it allows us to find the classical analog of quantum states. Intuitively, given a density operator $\hat{\rho}$, its classical analogue is a (quasi-)probability measure $\mu_{\hat{\rho}}$. For more details of the Wigner approach to the study of operator growth in phase space, see~\cite{Pal:2026otf,Shrestha:2026wbb} and~\cite{Basu:2024tgg}.\\
E. \textit{Classical Measures arising from Quantum States}\label{subsec:ClassState} --
$\phantom{word}$In this direction, we can also ask what is the classical analogue of the Wightman inner product~\eqref{eq:WightmanInProd}. Given $\hat{\rho}_{\beta}=e^{-\beta \hat{H}}$ and $\mathcal{Z}_{\beta}:=\mathrm{tr}(\hat{\rho}_{\beta})$, it is straightforward to verify using the Wigner function formalism that the thermal density operator $\hat{\rho}_{\beta}$ induces the statistical Boltzmann--Gibbs measure in classical phase space given by
\begin{align}
    \label{eq:BoltGibbsMeasure}
    \mathrm{d}\mu_{\beta}(\mathbf{q},\mathbf{p}):=e^{-\beta H(\mathbf{q},\mathbf{p})}\mathrm{d}\mathbf{q}\mathrm{d}\mathbf{p}~,
\end{align}
where $H$ is the corresponding classical Hamiltonian to the quantum Hamiltonian $\hat{H}$. Thus, the thermal expectation value $\langle \cdot \rangle_{\beta}=\mathrm{tr}(\hat{\rho}_{\beta}\,\cdot)/\mathcal{Z}_{\beta}$ in quantum mechanics induces the statistical (canonical) Boltzmann--Gibbs expectation value in classical phase space
\begin{align}
    \label{eq:BoltGibbsExpVal}
    \langle f\rangle^{\beta}_{\mathcal{S}}:=\frac{1}{\mu_{\beta}(\mathcal{S})}\int_{\mathcal{S}}\,\mathrm{d}\mathbf{q}\mathrm{d}\mathbf{p}\, e^{-\beta H(\mathbf{q},\mathbf{p})}\,f(\mathbf{q},\mathbf{p})~,
\end{align}
with $\mu_{\beta}(\mathcal{S})\equiv \int_{\mathcal{S}}\,\mathrm{d}\mathbf{q}\mathrm{d}\mathbf{p}\, e^{-\beta H(\mathbf{q},\mathbf{p})} $ being the classical analogue of the thermal partition function $\mathcal{Z}_{\beta}$. As a consequence, the Wightman inner product, as well as other thermal inner products such as the Kubo inner product, become
\begin{align}
    \label{eq:BoltGibbsInnProd}
    \begin{split}
   & \langle f,g\rangle_{\beta}:= \langle f\cdot g\rangle_{\mathcal{S}}^{\beta}\\
   &\equiv\frac{1}{\mu_{\beta}(\mathcal{S})}\int_{\mathcal{S}}\,\mathrm{d}\mu_{\beta}(\mathbf{q},\mathbf{p})f(\mathbf{q},\mathbf{p})g(\mathbf{q},\mathbf{p})~,
    \end{split}
\end{align}
where here $f,g \in\hat{\Sigma}_{\mathcal{S}}$ are \emph{centered} observables (vanishing expectation value $\langle f\rangle_{\mathcal{S}}=0=\langle g\rangle_{\mathcal{S}}$). Therefore, the Wightman inner product~\eqref{eq:WightmanInProd} becomes the Boltzmann--Gibbs inner product~\eqref{eq:BoltGibbsInnProd} in the classical limit: $(f\vert g)^{W}\mapsto \langle f,g\rangle_{\beta}$, where $\vert f) ,\vert g)$ are the GNS states corresponding to the operators $\hat{f},\hat{g}$ and where $f,g$ are the corresponding classical observables, which we can also obtain using the Wigner--Weyl transform. In contrast, the density operator corresponding to a single energy eigenstate $\hat{\rho}_{E}=\vert E\rangle\langle E\vert$ gives rise to the microcanonical measure $\mu_{E}(\mathbf{q},\mathbf{p})$ given by
\begin{align}
    \label{eq:MicrocanMeasure}
    \mathrm{d}\mu_{E}(\mathbf{q},\mathbf{p}):=\delta(H(\mathbf{q},\mathbf{p})-E)\mathrm{d}\mathbf{q}\mathrm{d}\mathbf{p}~,
\end{align}
which intuitively corresponds to the constraint that the orbit in phase space $\lbrace \mathbf{q}(t),\mathbf{p}(t)\rbrace$ remains confined to a constant-energy hypersurface $E=H(\mathbf{q},\mathbf{p})$. It is worth noting that the phase space trajectories in integrable systems do not go through the whole energy surface; while in chaotic systems, it could be ergodic in the long-time limit. In this case, $\mu_{E}(S)=\int \textrm{d}\mu_{E}(\mathbf{q},\mathbf{p})\equiv \mathcal{Z}(E)$. Finally, the symplectic form $\omega=\mathrm{d}\mathbf{p}\wedge \mathrm{d}\mathbf{q}$ induces a natural symplectic volume over the phase space, given by the Liouville measure $\mathrm{d}\mu_{L}(\mathbf{q},\mathbf{p}):=\omega^{\wedge ^{n}}/n!=\mathrm{d}q^{1}\wedge\mathrm{d}p_{1}\wedge\ldots\wedge \mathrm{d}q^{N}\wedge\mathrm{d}p_{N}=\mathrm{d}\mathbf{q}\mathrm{d}\mathbf{p}$. It represents a uniform distribution over the whole phase space $\mathcal{S}$. In this case, $\mu_{L}(\mathcal{S}):=\int \mathrm{d}\mu_{L}(\mathbf{q},\mathbf{p})\equiv \mathrm{Vol}(\mathcal{S})$ represents the volume of $\mathcal{S}$. Finally, a key observation is that the measures $\lbrace \mu_{\beta}$~\eqref{eq:BoltGibbsMeasure}$,\mu_{E}$~\eqref{eq:MicrocanMeasure}$,\mu_{L}\rbrace$ are invariant under Hamiltonian phase flow. This is true also for $\mu_{L}$ even though it does not depend explicitly on the Hamiltonian. The reason is that the symplectic form $\omega$ is indeed preserved under Hamiltonian phase flow~\cite{BookArnold}. On the other hand, $\mu_{\beta}$ and $\mu_{E}$ are invariant under phase flow if the Hamiltonian present in their expression is the same one that gives rise to the phase flow. 

We emphasize that our description of the classical phase space approach is independent of choices of (semi-)classical mapping from the quantum level. Therefore, in principle, our classical algorithm applies to any well-defined and bounded functions in classical phase space. See for example~\cite{Haneder:2025obr} for an application in the context of OTOCs.
%

\begin{figure}
    \centering
    \includegraphics[width=0.5\linewidth]{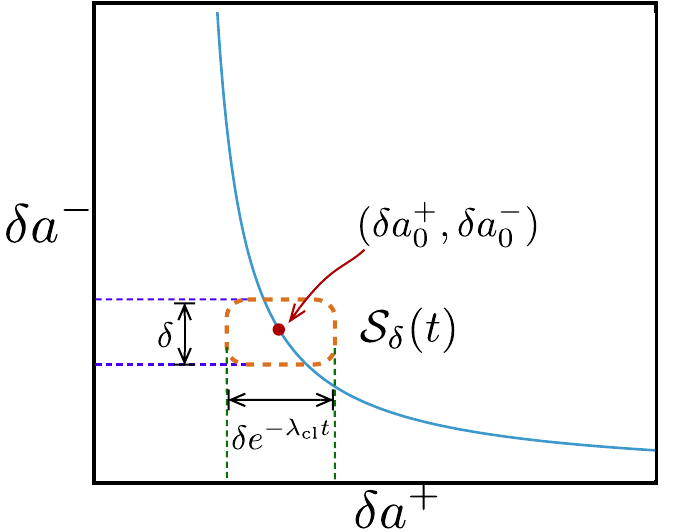}
\caption{Schematic figure for the phase space trajectory of the linearized dynamics around the unstable saddle point. The blue curve represents the solution~\eqref{eq:UnstableSaddleSol}. The orange and dashed region is the strip ~\eqref{eq:PhaseSpSaddleRegion} within which the orbit of the initial phase space point $(\delta a^{+}_{0},\delta a^{-}_{0})$ evolves exponentially.}
    \label{fig:DiagSaddlePhaseSpace}
\end{figure}

\phantomsection
VIII. \textit{Applications to Classical Saddle Dominated Scrambling}\label{sec:ClassKrylovExampleSaddle} -- Given this general formalism, in this section, we discuss its application to identify scrambling coming from a single isolated saddle. For simplicity, consider a dynamical system with a two-dimensional phase space $\mathcal{S}$, which has an isolated unstable saddle at $(a^{+}_{\mathrm{s}},a^{-}_{\mathrm{s}})$. By this we mean that the solutions to the linearized equations of motion around this point $(a^{+}_{\mathrm{s}},a^{-}_{\mathrm{s}})+(\d a^{+}, \d a^{-})$, given by
\begin{align}
    \label{eq:UnstableSaddleEoM}
    \frac{\mathrm{d} \delta a^{\pm}(t)}{\mathrm{d}t}\approx\pm \lambda_{\mathrm{cl}}\delta a^{\pm}(t)~,
\end{align}
have the following exponential behavior:
\begin{equation}
    \d a^{+}(t)= \delta a^{+}_0\,e^{+\lambda_{\mathrm{cl}}t}, ~~~\d a^{-}(t)= \delta a^{-}_0\,e^{-\lambda_{\mathrm{cl}}t}~,
    \label{eq:UnstableSaddleSol}
\end{equation}
with $\delta a^{\pm}_{0}:=\delta a^{\pm}(t=0)$, but behave regularly away from $(a^{+}_{\mathrm{s}},a^{-}_{\mathrm{s}})$. Such would be the case in a classically integrable system that has a single isolated unstable saddle, such as the LMG model. In this context, $( \delta a^{+},\delta a^{-} )$ are called \emph{normal} coordinates. The linearized Hamiltonian giving rise to~\eqref{eq:UnstableSaddleEoM} is of the form $H_{\mathrm{L}}\approx\lambda_{\mathrm{cl}}\delta a^{+}\delta a^{-}$, which is related to the Hamiltonian of an \emph{inverted} harmonic oscillator, $H_{\mathrm{IHO}}=\lambda_{\mathrm{cl}}(p^{2}-q^{2})/2$, by a $\pi/4$-rotation: $\delta a^{+}=-(q+p)/\sqrt{2}$, $\delta a^{-}=(q-p)/\sqrt{2}$. One can verify that in these coordinates the Poisson bracket has the same form as~\eqref{eq:PoissonBracket}, where $\delta a^{+}$ acts as a generalized coordinate and $\delta a^{-}$ as its conjugate momentum. 
\\
$\phantom{word}$ The exponential behavior~\eqref{eq:UnstableSaddleSol} is typically constrained to occur near the saddle point $(a_{s}^{+},a_{s}^{-})$ within a narrow strip $S_{\delta}(t)\subset \mathcal{S}$ of the full phase space~\cite{Xu:2019lhc, Trunin:2023xmw, Trunin:2023rwm}
\begin{equation}
     S_{\delta}(t)=\lbrace ( \delta a^{+},\delta a^{-})\in \mathcal{S}\,\big\vert\,\,\vert \delta a^{+} \vert<\delta e^{-\lambda_{\mathrm{cl}} t}/2,~\vert \delta a^{-} \vert<\delta/2\rbrace~,
    \label{eq:PhaseSpSaddleRegion}
\end{equation}
with a volume $\mathrm{Vol}(S_{\delta}(t))= \d^2 e^{-\lambda_{\mathrm{cl}} t}$, for early times $t<t_{\ast}\sim \log(S)$ and sufficiently close to the saddle point $\delta \ll 1$. A schematic plot of this growth is shown in Figure~\ref{fig:DiagSaddlePhaseSpace}.
Note the exponential decay of the volume $\mathrm{Vol}(S_{\delta}(t))$ around the saddle point. This is to counteract the exponential growth of the positive normal coordinate $\delta a^{+}(t) \propto e^{\lambda_{\mathrm{cl}}t}$ and is an implementation of the constraint that $\delta a^{+}(t)$ grows exponentially only sufficiently near the saddle point $(a^{+}_{s},a^{-}_{s})$. The authors in~\cite{Xu:2019lhc} argue that the fact that the volume of the strip decays exponentially along the solution~\eqref{eq:UnstableSaddleSol} is the reason why the classical \emph{unnormalized} ``OTOC", computed by the \emph{unnormalized} phase space average of the following Poisson bracket squared,
\begin{align}
    \label{eq:OTOCSaddleScr}
    \begin{split}
    &C_{\pm}(t)=\mathrm{Vol}(S_{\delta}(t))\times\left\langle \left\vert\left\lbrace \frac{\delta a^{+}(t)}{\sqrt{2}},\frac{\delta a^{-}}{\sqrt{2}}\right\rbrace_{\mathrm{PB}}\right\vert^{2}\right\rangle_{S_{\delta}(t)}\\
    &=\frac{\mathrm{Vol}(S_{\delta}(t))}{\mathrm{Vol}(S_{\delta}(t))}\int_{S_{\delta}(t)}\mathrm{d}\mu(\delta a^{+},\delta a^{-})\left\vert \frac{1}{2}\frac{\delta a^{+}(t)}{\delta a^{+}_{0}}\right\vert^{2}\\
    &=\mathrm{Vol}(S_{\delta}(t))\times \left(\frac{e^{+2\lambda_{\mathrm{cl}}t}}{4}\right)=\frac{\delta^{2}}{4}\,e^{+\lambda_{\mathrm{cl}}t}~,
    \end{split}
\end{align}
grows as $\propto e^{+\lambda_{\mathrm{cl}}t}$ in the LMG model and not as $\propto e^{+2\lambda_{\mathrm{cl}}t}$, as it would in a truly chaotic system~\cite{Xu:2019lhc}. In~\eqref{eq:OTOCSaddleScr}, we used the Liouville measure $\mathrm{d}\mu_{L}=\mathrm{d}(\delta a^{+})\,\mathrm{d}(\delta a^{-})$ in the phase space average~\eqref{eq:PhaseSpaceAve} constrained to the strip $S_{\delta}(t)$ to compute the expectation value of the Poisson bracket. In their case, the phase space average is normalized with respect to the surface area of the unit $2$-sphere $\mathbb{S}^{2}$, which highlights an important difference between our approaches. To be precise, it is clear that if we normalize the OTOC~\eqref{eq:OTOCSaddleScr} by the volume of the strip, this would contribute a positive exponential factor: $\overline{C}_{\pm}(t)=\left\langle \left\vert\left\lbrace \delta a^{+}(t),\delta a^{-}\right\rbrace_{PB}\right\vert^{2}\right\rangle_{S_{\delta}(t)}=\vert \delta a^{+}(t)/\delta a^{+}_{0}\vert^2=e^{2\lambda_{\mathrm{cl}}t}$, leading to a different semiclassical bound on the quantum Lyapunov exponent $\lambda_{L}\geq 2\lambda_{\mathrm{cl}}$, contradicting their claim. One way to make sense of this apparent contradiction is that the authors in~\cite{Xu:2019lhc} use coordinates defined in $\mathbb{S}^{2}$ to compute an equivalent expectation value in the strip, where normalizing with respect to the surface area of $\mathbb{S}^{2}$ is the natural choice in their setup. \\ 
$\phantom{word}$ We now ask whether we can compute the Krylov and logK-complexity of classical observables in this setup, and how does it compare with the behavior of the classical unnormalized OTOC: $C_{\pm}(t) = \delta^{2} e^{\lambda_{\mathrm{cl}}t}/4$. An important remark is that, regardless of the normalization factor in the phase space average, integrating functions $f(\delta a^{+},\delta a^{-})$ with respect to the Liouville measure over the strip will lead to additional time-dependent factors in their associated Krylov bases, while at the same time spoiling the equivalence $\vert \vert f_{t} \vert \vert=\vert \vert f \vert \vert$. In other words, even though the measure $\mathrm{d}\mu_{L}$ is invariant under phase flow, integrating observables or Krylov basis elements over the strip $S_{\delta}(t)$ produces additional and exponentially suppressing factors. Thus, omitting the normalization factor in the phase space average may not be enough to remove the additional exponentially-suppressing factors arising from integrating over the strip~\eqref{eq:PhaseSpSaddleRegion}, as we discuss below. \\
$\phantom{word}$Ideally, we would like to compute these complexities for an observable that is linear in normal coordinates $f\sim \delta a^{+},\delta a^{-}$. Choosing $f_{a^{+}}=\delta a^{+}$ or $f_{a^{-}}=\delta a^{-}$ leads to a one-dimensional classical Krylov subspace $\mathrm{dim}(\Sigma_{f_{a^{-}}})=1=\mathrm{dim}(\Sigma_{f_{a^{+}}})$ (as can be seen in~\eqref{eq:DimKrylovClassical}) with vanishing Krylov and logK complexities. This is simply because the sequence of nested Poisson brackets~\eqref{eq:SequencefPB} satisfies $\tilde{f}_{n}\propto \lambda_{\mathrm{cl}}^{n}\tilde{f}_{0}$ $\forall n\geq 1$ and the resulting Krylov basis elements vanish identically $\mathfrak{K}_{n}\equiv 0 $ $\forall n \geq 0$, with $\mathfrak{K}_{0}:=\tilde{f}_{0}/(\vert \vert \tilde{f}_{0}\vert \vert)$ being the only non-vanishing element. This makes complete sense, since a non-zero Krylov/logK complexity requires at least a two/three dimensional classical Krylov subspace $\Sigma_{f}$ and we need a single properly normalized polynomial to describe the full time evolution of such type of observables. The same is true for observables of the form $f^{\ell}_{a^{+}}=(\delta a^{+})^{\ell}$ or $f_{a^{-}}^{\ell}=(\delta a^{-})^{\ell}$ with $\ell \geq 1$, as in this case one has $D_{\Sigma_{f}}=\mathrm{dim}(\Sigma_{f})\leq (1+(\ell-1))!/\ell !\equiv 1$ (where we removed the contribution from the $0$-th order monomial, i.e. the constant term in~\eqref{eq:ClassicalObservable}). \\
$\phantom{word}$Thus, the simplest observable with a non trivial classical Krylov complexity is $f(\delta a^{+},\delta a^{-})=A_{+} \delta a^{+}+A_{-}\delta a^{-}$. To make a connection with the canonical coordinate of the inverted harmonic oscillator $q=(\delta a^{-}-\delta a^{+})/\sqrt{2}$, we choose $A_{-}=1/\sqrt{2}=-A_{+}$, that is, $f(\delta a^{+},\delta a^{-}):=q(\delta a^{+},\delta a^{-})$. In this case, a straightforward computation of the Krylov basis~\eqref{eq:GrammSchmidtFunctions}, which as mentioned above carries an inherent time dependence due to the phase space average being computed in the strip~\eqref{eq:PhaseSpSaddleRegion}, leads to the following classical Krylov complexity~\eqref{eq:ClassicalKrylovComplexity}
\begin{align}
    \label{eq:ClassKrylovSaddleDomExq}
    K_{q}(t)=\frac{\vert c_{1}(t)\vert^{2}}{\vert c_{0}(t)\vert^{2}+\vert c_{1}(t)\vert^{2}}=\tanh(\lambda_{\mathrm{cl}} t)^{2}~,
\end{align}
which is independent of $\delta$ and, moreover, is also independent of the choice of normalization factor $1/\mu(S_{\delta}(t))=1/\mathrm{Vol}(S_{\delta}(t))$ in the phase space average~\eqref{eq:PhaseSpaceAve}. This Krylov complexity has an initial time behavior $K_{q}(t)\approx \lambda_{\mathrm{cl}}^{2}t^{2}+O(t^{4})$, consistent with expectations, but does not grow according to $\sinh(\lambda_{\mathrm{cl}} t)^{2}$ beyond this early time behavior. However, an interesting feature of this quantity is that it saturates to a value $\overline{K_{q}}=D_{\Sigma_{f}}/2=1$, which signals the finiteness of the Krylov subspace of $f_{q}$. \\ 
$\phantom{word}$If we do not normalize the Krylov complexity~\eqref{eq:ClassicalKrylovComplexity} by the factor $1/(\vert \vert f_{t}\vert \vert^{2})$ and if we keep a generic normalization of the phase space average with $1/\mathrm{Vol}(\mathcal{S})$, we instead find a behavior of the form
\begin{align}
    \label{eq:ClassKrylovSaddleDomExqGeneral}
    \widetilde{K}_{q}(t)=\vert c_{1}(t)\vert^{2}=\frac{\delta^{4} e^{-2\lambda_{\mathrm{cl}} t}}{12 \,\mathrm{Vol}(\mathcal{S})}\frac{\sinh(\lambda_{\mathrm{cl}} t)^{2}}{\cosh(\lambda_{\mathrm{cl}} t)}~.
\end{align}
Choosing $\mathrm{Vol}(\mathcal{S})\mapsto\mathrm{Vol}(\mathbb{S}^{2})=1$ as in~\cite{Xu:2019lhc} leads to slower early-time growth $ \widetilde{K}_{q}(t)\approx \delta^{4}((\lambda_{\mathrm{cl}}t)^2/12-(\lambda_{\mathrm{cl}}t)^3/6+ O(t^{4}))$ and late-time vanishing $\widetilde{K}_{q}(t)\xrightarrow{t\rightarrow \infty} 0$. Instead, choosing $\mathrm{Vol}(\mathcal{S})\mapsto\mathrm{Vol}(S_{\delta}(t))=\delta^{2}e^{-\lambda_{\mathrm{cl}} t}$ in~\eqref{eq:ClassKrylovSaddleDomExqGeneral} leads to the following expression
\begin{align}
    \label{eq:ClassKrylovSaddleDomExqVolStrip}
    \widetilde{K}_{q}(t)\Big\vert_{S_{\delta}(t)}=\frac{\delta^{2}}{12}\sinh(\lambda_{\mathrm{cl}} t)^{2}(1-\tanh(\lambda_{\mathrm{cl}} t))~.
\end{align}
This expression has an initial growth $\widetilde{K}_{q}(t)\vert_{S_{\delta}(t)}\approx (\delta^{2}/12)((\lambda_{\mathrm{cl}} t)^2-(\lambda_{\mathrm{cl}} t)^3+O(t^{4}/3))$ and a late time saturation $\overline{\widetilde{K}_{q}}\vert_{S_{\delta}(t)}=\delta^{2}/24$. Both~\eqref{eq:ClassKrylovSaddleDomExqGeneral} and~\eqref{eq:ClassKrylovSaddleDomExqVolStrip} grow slower than~\eqref{eq:ClassKrylovSaddleDomExq}. Thus, none of these quantities has an exponentially growing behavior similar to $C_{\pm}(t)$. But is it a sensible comparison? The unnormalized OTOC~\eqref{eq:OTOCSaddleScr} is computed for the Poisson brackets of the normal coordinates $\lbrace \delta a^{+}(t),\delta a^{-}(0)\rbrace_{PB}$. However, the Krylov complexity~\eqref{eq:ClassKrylovSaddleDomExq} is computed for the observable $q(\delta a^{+},\delta a^{-})=(\delta a^{-}-\delta a^{+})/\sqrt{2}$. Thus, a more sensible comparison would be between~\eqref{eq:ClassKrylovSaddleDomExq} and the following unnormalized OTOC
\begin{align}
    \label{eq:OTOCSaddleScrqp}
    \begin{split}
    &C_{qp}(t)=\mathrm{Vol}(S_{\delta}(t))\times\left\langle \left\vert\left\lbrace q(t),p(0) \right\rbrace_{\mathrm{PB}}\right\vert^{2}\right\rangle_{S_{\delta}(t)}\\
    &=\frac{\mathrm{Vol}(S_{\delta}(t))}{\mathrm{Vol}(S_{\delta}(t))}\int_{S_{\delta}(t)}\mathrm{d}\mu(\delta a^{+},\delta a^{-})\left\vert \cosh(\lambda_{\mathrm{cl}}t)\right\vert^{2}\\
    &=\mathrm{Vol}(S_{\delta}(t))\times \left(\cosh(\lambda_{\mathrm{cl}}t)^{2}\right)=\delta^{2}\,e^{-\lambda_{\mathrm{cl}}t}\cosh(\lambda_{\mathrm{cl}}t)^{2}\\
    &= \frac{\delta^{2}}{4}\left(e^{-3\lambda_{\mathrm{cl}}t}+e^{-\lambda_{\mathrm{cl}}t}+e^{\lambda_{\mathrm{cl}}t}\right)~,
    \end{split}
\end{align}
where we still find a late time exponential growth $\propto e^{+\lambda_{\mathrm{cl}}t}$ beyond the initial exponentially decaying-behavior governed by $\propto e^{-3\lambda_{\mathrm{cl}}t}$. So, even though at small times the classical Krylov complexity~\eqref{eq:ClassKrylovSaddleDomExq} grows faster than the OTOC~\eqref{eq:OTOCSaddleScrqp}, the latter will continue to grow exponentially at a timescale where the former has already saturated. These observations suggest that the classical version of Krylov complexity~\eqref{eq:ClassicalKrylovComplexity} is insensitive to the exponentially-growing contribution from the unstable saddle.

\begin{figure}[htbp]
\begin{center}
        {\includegraphics[width=4.1cm]{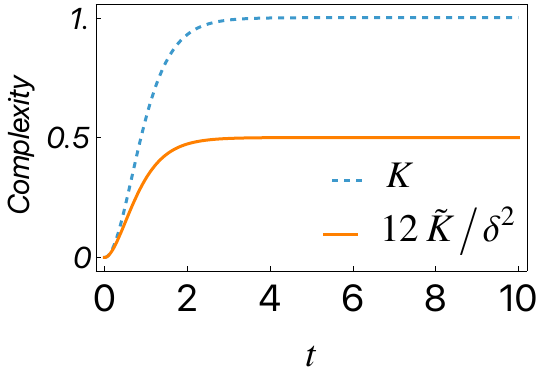}\hspace{3mm}}
        {\includegraphics[width=3.9cm]{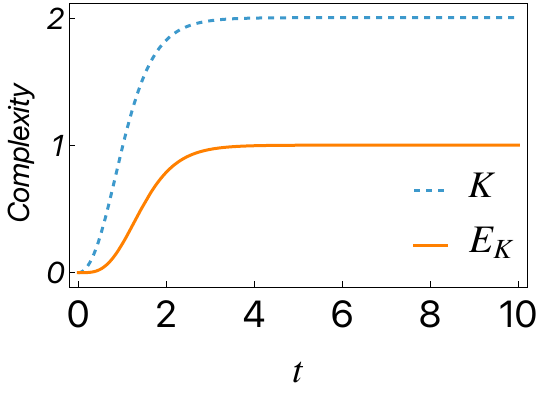}}
\end{center}
\caption{\textbf{Left panel:} Normalized Krylov complexity $K\left(t\right)$ (\ref{eq:ClassKrylovSaddleDomExq}) (blue) and un-normalized Krylov complexity (\ref{eq:ClassKrylovSaddleDomExqVolStrip}) with strip constraint $12\tilde{K}\left(t\right)/\d^2$ (orange) for initial function $q$. \textbf{Right panel:} Normalized Krylov complexity $K\left(t\right)$ (\ref{eq: NormKq2}) (blue) and elogK complexity (orange) $\mathbf{E}_K(t)$ from (\ref{eq: NormLKq2}) (orange) for initial function $q^2$. We chose $\lambda_{\text{cl}}=1$ for both plots.}
\label{fig:ClassKandlogKSaddleDomScr}
\end{figure}

For completeness, it is also illustrative to compute the logarithmic complexity. As discussed above, the first non-trivial example where we can compute logK involves quadratic polynomials of phase-space coordinates. For simplicity, we choose the initial function as $f\left(p,q\right)=q^{2}$. Following the classical Krylov formalism, we find the associated normalized Krylov and logK complexities:
\begin{align}
K_{q^{2}}\left(t\right)&=\frac{4\sinh^{2}{\left(t\lambda_{\text{cl}}\right)}}{2+5\cosh{\left(2t\lambda_{\text{cl}}\right)}}\nonumber\label{eq: NormKq2}\\
&\times\frac{\left(9+79\cosh{\left(2t\lambda_{\text{cl}}\right)}+10\cosh{\left(4t\lambda_{\text{cl}}\right)}\right)}{\left(5+2\cosh{\left(2t\lambda_{\text{cl}}\right)}\right)^{2}}~,\\
\mathbf{L}_{q^{2}}\left(t\right)&=\frac{40\log{\left(2\right)}\sinh^{4}{\left(t\lambda_{\text{cl}}\right)}}{29\cosh{\left(2t\lambda_{\text{cl}}\right)}+5\left(3+\cosh{\left(4t\lambda_{\text{cl}}\right)}\right)}~.\label{eq: NormLKq2}
\end{align}
Their behavior is shown in Fig.~\ref{fig:ClassKandlogKSaddleDomScr} for $\lambda_{\text{cl}}=1$.
From this Figure we see that, similarly to the numerical results in Sec.~\hyperref[sec:NumericalAnalysis]{V.}, the elogK is suppressed at early times compared to Krylov complexity. Thus, both $K_{q^{2}}(t)$ and $\mathbf{L}_{q^{2}}(t)$ behave as they would in an otherwise integrable classical system; see the Supplemental Material~\hyperref[app:ClassPhaseSpace]{D.} for a comparison. In the next section, we discuss a possible resolution of the insensitivity of logK to suppress the exponential growth in the infinite-dimensional quantum cases of the integrable $q=2$ SYK and the inverted harmonic oscillator. 

%

\phantomsection
IX. \textit{Revisiting the Replica Approach and a New Operator Growth Measure}\label{sec:ReplicaGeneralization} -- 

In this section, we discuss a possible resolution to the observations in Sections.~\hyperref[sec:ExamplelogKCFT]{IV.},~\hyperref[sec:QuantKrylovIHO]{VI.} regarding the behavior of logK complexity in the conformal limit of the SYK and in the inverted harmonic oscillator, respectively. We outline a direction that goes beyond the universal information accessible to Krylov complexity by taking into account details of the underlying theory and operator. At the end of this section, we also connect it to the higher-order Krylov complexities used to define the logarithmic Krylov complexity via the replica trick.

A. \textit{Basic Idea and Intuition}\label{subsec:RenOperator} -- One of the key differences between the SYK$_{q}$ and the inverted harmonic oscillator and those described in Sec.~\hyperref[sec:NumericalAnalysis]{V.} is the dimensionality of the Krylov space. This suggests that the implementation of the replica trick~\eqref{eq-Log-K replica} in infinite-dimensional Krylov spaces is still unable to accurately capture the integrable properties of the underlying theory from the perspective of operator growth. At the same time, it is also unable to offer a precise distinction between scrambling coming from an unstable saddle, as in the case of the inverted harmonic oscillator, and scrambling associated with a genuine chaotic theory in the spectral sense, as in the $q>2$ SYK$_{q}$ case. 

In the case of the SYK$_{q}$ at low energies and temperatures, the description of the Krylov complexity and its higher-order generalizations relied entirely on the $\mathrm{SL}(2,\mathbb{R})$ symmetries of the chiral CFT$_{2}$ which gave an exact form of the wavefunctions $\varphi_{n}(t)$. In such a derivation, no details of the CFT are present, namely whether it is an integrable or a large-$c$ holographic CFT$_{2}$. Similarly, in the case of the inverted harmonic oscillator, the computation of the Krylov complexity for Gaussian operators relied on the fact that we could find an exact form of the wavefunctions which in this case relied on the special choice of initial operator. It would seem that we need to incorporate more fine-grained information about the integrability details of the theory or about the particular choice of operator if we want to solve this issue.

A way to do this is to consider a function of the spreading superoperator $\breve{n}(t)=e^{-i\breve{\mathcal{L}
}t}\breve{n}e^{i\breve{\mathcal{L}}t}$ that is sensitive to theory and operator-dependent information $\delta_{\hat{\mathcal{O}}}(\breve{n},\mathrm{info}(\breve{\mathcal{L}}))$. In general, we propose that such a function $\delta_{\hat{\mathcal{O}}}(\breve{n},\mathrm{info}(\breve{\mathcal{L}}))$ can be written as a subtraction of two terms. The first one consisting of a function of the spreading superoperator that weighs directly the locality of the Hamiltonian through an exponent that takes into account the particular interplay between a given operator and the Liouvillian, and the second one a regularized version version of the first one, that takes into account universal (operator-independent) information about the underlying theory. Schematically, this can be expressed by the following formula
\begin{align}
    \label{eq:DeltaqSensOpGen}
    \delta_{\hat{\mathcal{O}}}\left(\breve{n},\mathrm{info}(\breve{\mathcal{L}})\right)\sim \mathrm{Bare}(\breve{n}^{F(\hat{\mathcal{O}},\breve{\mathcal{L}})})-\mathrm{Reg}(\breve{n}^{F(\hat{\mathcal{O}},\breve{\mathcal{L}})})~,
\end{align}
where $F(\hat{\mathcal{O}},\breve{\mathcal{L}})$ is a function that depends on the particular class of operators and details of the theory captured by the Liouvillian. For example, in $k$-local theories, it would be desirable for such a function to weight the change in operator size $\Delta s$ increasingly more as the operators in the theory become more non-local in the whole operator Hilbert space and not just in its own Krylov subspace. In this way, one could gain better insight into how to engineer concrete probes of operator growth that accurately capture the dynamics.

B. \textit{General Idea and Realization in the SYK$_{q}$ Model}\label{subsec:QSensCompSYK} -- We now outline concretely what we mean by this in the case of the SYK$_{q}$. Recall the $q$-body SYK Hamiltonian for $N_{f}$ Majorana fermions given by~\eqref{eq:SYKqHam}. Let us focus on the case where $N_{f}$ and $q$ are even. We propose a $q$-sensitive measure of operator growth that accurately distinguishes between the integrable $q=2$ and the chaotic $q\geq4$ cases
\begin{align}
    \label{eq:qSensComplexity}
    Q^{(q)}_{\hat{\mathcal{O}}}(t)= (\mathcal{O}(t)\vert \delta_{\hat{\mathcal{O}}}(\breve{n},q) \vert \mathcal{O}(t))~,
\end{align}
where $\vert \mathcal{O} )$ is a GNS state corresponding to an operator $\hat{\mathcal{O}}$ in the algebra of bounded linear operators $\mathcal{B}(\mathcal{H})$ in the SYK Hilbert space $\mathcal{H}$, and where $\delta_{\mathcal{O}}(\breve{n},q)$ is a function of the spreading superoperator $\breve{n}$ that crucially takes into account the details of the precise theory, encoded in the $q$-body interaction, as well as the specific details about the operator, such as its relative \emph{size} compared to the basis elements of  $\mathcal{B}(\mathcal{H})$. To be concrete, in the case of operators of the form $\hat{\mathcal{O}}\propto \hat{\psi}_{i}$, we propose a precise form of $\delta_{\hat{\mathcal{O}}}(\breve{n},q)$ to be given by
\begin{align}
    \label{eq:DeltaqSensOp}
    \delta_{\hat{\mathcal{O}}}(\breve{n},q):=\upsilon_{\hat{\mathcal{O}}}(\breve{n},q)-\gamma_{\hat{\mathcal{O}}}(\breve{n},q)~,
\end{align}
where $\upsilon_{\hat{\mathcal{O}}}(\breve{n},q)$ and $\gamma_{\hat{\mathcal{O}}}(\breve{n},q)$ are $q$-weighted functions of the spreading operator given by
\begin{align}
    \label{eq:UpsilonqSensOp}
    \upsilon_{\hat{\mathcal{O}}}(\breve{n},q):=\Lambda \breve{n}^{\frac{q-2}{2}}~,
\end{align}
\begin{align}
    \label{eq:GammaqSensOp}
    \begin{split}
    \gamma_{\hat{\mathcal{O}}}(\breve{n},q):=&\Lambda\Bigg( \breve{n}^{\frac{q-2}{2}}-\left(1-I(q)\right)\frac{q}{2}\breve{n}\\
    &-I(q)\left(1-\frac{(0)_{n}}{n!}\right)\breve{\mathbf{1}}\Bigg)~,
    \end{split}
\end{align}
respectively, where $n:=\langle \breve{n}\rangle_{\mathcal{K}}$ is the expectation value of the spreading superoperator in the Krylov basis, $(a)_{k}:=\Gamma(a+k)/\Gamma(a)$ is the Pochhammer symbol, $\Lambda$ is related to the effective dimension of $\mathcal{B}(\mathcal{H})$ probed by $\hat{\mathcal{O}}$, $\breve{\mathbf{1}}$ is the identity operator in the GNS Hilbert space and where $I(q)$ is a selector defined by
\begin{align}
    \label{eq:qSensSelector}
    I(q):=\frac{\sin(\pi(q-2))}{\pi(q-2)}~,
\end{align}
such that $I(2):=\lim_{q\rightarrow 2}I(q)=1$ and $I(q)=0$ for $q\geq 4$ and hence $I(q)=\delta_{q2}$ for integer $q\geq 2$. The $q$-weighted functions~\eqref{eq:UpsilonqSensOp} and~\eqref{eq:GammaqSensOp} similarly define $q$-sensitive notions of operator growth
\begin{align}
    \label{eq:UpsilonqSensComplexity}
    \Upsilon^{(q)}_{\mathcal{O}}(t):=(\mathcal{O}(t)\vert  \upsilon_{\hat{\mathcal{O}}}(\breve{n},q) \vert\mathcal{O}(t))~,
\end{align}
\begin{align}
    \label{eq:GammaqSensComplexity}
    \Gamma^{(q)}_{\mathcal{O}}(t):=(\mathcal{O}(t)\vert  \gamma_{\hat{\mathcal{O}}}(\breve{n},q) \vert\mathcal{O}(t))~.
\end{align}
To see what these $q$-weighted functions $\upsilon$~\eqref{eq:UpsilonqSensOp} and $\gamma$~\eqref{eq:GammaqSensOp} physically mean, let us study separately the $q=2$ and $q\geq 4$ cases. For $q=2$, their expectation values with respect to the time evolved operator $\hat{\mathcal{O}}(t)$ using the Wightman inner product yield
\begin{align}
    \label{eq:UpsilonGammaqSensComplexityq2}
    \begin{split}
    &\Upsilon^{(2)}_{\mathcal{O}}(t)=(\mathcal{O}(t) \vert \Lambda \breve{n}^{0}\vert \mathcal{O}(t))=\Lambda~,\\
    &\Gamma^{(2)}_{\mathcal{O}}(t)=(\mathcal{O}(t) \vert \Lambda (\breve{n}^{0}-(1-(0)_{n}/n!)\breve{\mathbf{1}})\vert \mathcal{O}(t))\\
   & =\Lambda\left(1-(1-\vert \varphi_{0}(t)\vert^{2})\right)=\Lambda\cosh^{-2}(\pi t/\beta)~,
    \end{split}
\end{align}
where we used the wavefunctions~\eqref{eq:PhinCFT} with $\eta=2/q$ and $\alpha = \pi/\beta$, corresponding to the wavefunctions of an initial operator $\hat{\mathcal{O}}_{0}\propto \hat{\psi}_{1}$ with respect to the Wightman inner product at low temperatures, namely
\begin{align}
    \label{eq:PhinqSYK1Maj}
    \varphi_{n}(t)=\sqrt{\frac{\Gamma(n+2/q)}{n!\Gamma(2/q)}}\frac{\tanh^{n}(\pi t / \beta)}{\cosh^{2/q}(\pi t /\beta)}~.
\end{align}
Thus, in this case, Eq.~\eqref{eq:qSensComplexity} yields
\begin{align}
    \label{eq:q2SensComplexity}
    Q^{(2)}_{\hat{\psi}_{1}}(t)=\Lambda \tanh^{2}(\pi t /\beta)\quad (q=2)~,
\end{align}
where $\Lambda$ sets the late-time saturation value of $Q^{(2)}_{\hat{\psi}_{1}}(t)$. Similarly, for $q\geq 4$, we have 
\begin{align}
    \label{eq:UpsilonGammaqSensComplexityq4}
    \begin{split}
    &\Upsilon^{(q)}_{\mathcal{O}}(t)=(\mathcal{O}(t) \vert \Lambda \breve{n}^{\frac{q-2}{2}}\vert \mathcal{O}(t))=\Lambda K^{\left(\frac{q-2}{2}\right)}_{\mathcal{O}}(t)~,\\
    &\Gamma^{(q)}_{\mathcal{O}}(t)=\left(\mathcal{O}(t) \left\vert \Lambda \left(\breve{n}^{\frac{q-2}{2}}-\frac{q}{2}\breve{n}\right)\right\vert \mathcal{O}(t)\right)\\
    &=\Lambda\left(K^{\left(\frac{q-2}{2}\right)}_{\mathcal{O}}(t)-\frac{q}{2} K_{\mathcal{O}}(t)\right)~,
    \end{split}
\end{align}
and thus 
\begin{align}
    \label{eq:q4SensComplexity}
    Q^{(q)}_{\hat{\psi}_{1}}(t)=\Lambda \sinh^{2}(\pi t /\beta)\quad (q\geq 4)~,
\end{align}
for $q\geq 4$. Here, $K^{(q-2)/2}(t)$ are the higher-order complexities with integer $m=(q-2)/2$. In the expressions above, we used again the fact that for operators of the form $\hat{\mathcal{O}}_{0}\propto \hat{\psi}_{1}$, the wavefunctions are unchanged and given by~\eqref{eq:PhinqSYK1Maj}. Now let us analyse these results.

$\Upsilon^{(q)}_{\mathcal{O}}(t)$ is a \emph{bare} $q$-sensitive complexity that weights directly the $q$-locality of the SYK Hamiltonian. The exponent $(q-2)/2$ is no accident: it represents the increase in \emph{size} that a single Majorana, our initial operator $\hat{\mathcal{O}}\propto \hat{\psi}_{i}$, undergoes as it commutes with the SYK Hamiltonian~\eqref{eq:SYKqHam} initially. Before discussing this in more detail below, let us point out that however, this quantity by itself has two issues: 1) it remains constant and equal to the scale $\Lambda$ for $q=2$, and 2) at late times, it grows according to $\propto e^{\pi(q-2)t/\beta}$, by analogy with the higher-order complexities. Thus, for $q>4$, the growth rate of this bare complexity exceeds the (conjectured) generalized chaos bound $\lambda_{K}\leq 2\pi/\beta$, where $\lambda_{K}$ is the Krylov exponent. To account for this exceeding growth, we have to subtract a \emph{regularized} $q$-sensitive complexity $\Gamma^{(q)}_{\mathcal{O}}(t)$, which has a leading growth similar to $\Upsilon^{(q)}_{\mathcal{O}}(t)$, but from which we subtract the universal aspect of the operator growth: 1) for $q=2$ the probability amplitude of the autocorrelation function $\vert \varphi_{0}(t)\vert^{2}$, and 2) for $q>4$ the Krylov complexity $K_{\mathcal{O}}(t)$. Subtracting $\Upsilon^{(q)}_{\mathcal{O}}(t)$ and $\Gamma^{(q)}_{\mathcal{O}}(t)$ thus provides a $q$-sensitive quantity~\eqref{eq:qSensComplexity} that retains the universal aspect of the usual Krylov complexity, while also accurately capturing the integrable properties of the theory. Computationally, it would have made sense to just consider the difference~\eqref{eq:DeltaqSensOp}, but conceptually it is important to note that it arises from the difference of two $q$-sensitive notions of operator growth that nevertheless, by themselves, suffer from similar issues to the usual higher-order Krylov complexities.

An important aspect to note is that in this case, where the initial operator is a simple Majorana fermion $\hat{\mathcal{O}}\sim \hat{\psi}_{i}$, the $q$-sensitive complexity~\eqref{eq:qSensComplexity} reduces to a subtraction of higher-order complexities regulated by universal information: the probability amplitude of the autocorrelation function for $q=2$ and the Krylov complexity for $q \geq 4$. If we instead wanted to engineer the function $\delta_{\mathcal{O}}(\breve{n},q)$ for different initial operators, say, for \emph{strings} of Majorana fermions, we would need to account for this through the exponent in $\upsilon_{\mathcal{O}}(\breve{n},q)$ and through the subtraction in $\gamma_{\mathcal{O}}(\breve{n},q)$, while at the same time considering that in this case the wavefunctions would certainly be different from~\eqref{eq:PhinqSYK1Maj}. For example, as we will discuss below, one would change $\breve{n}^{(q-2)/2}\mapsto \breve{n}^{\kappa(\Delta s_{\mathrm{typ}}/2)}$, where $\Delta s_{\mathrm{typ}}$ reflects the ``typical'' increase in operator size induced by the action of the Liouvillian and $\kappa$ is an effective exponent. At the same time, the selector $I(q)$ and universal subtractions would need to be adjusted.

C. \textit{Details of Operator Size in the SYK$_{q}$ Model}\label{subsec:OperatorSizeSYK} -- Let us now be precise on where the factor $(q-2)$ comes from in~\eqref{eq:UpsilonqSensOp}, ~\eqref{eq:GammaqSensOp} and the selector~\eqref{eq:qSensSelector}. To do this, we have to discuss the increase in size of a strings of Majorana fermions through their commutation with the $q$-body Hamiltonian~\eqref{eq:SYKqHam}. We follow the general arguments outlined in~\cite{Roberts:2018mnp,Qi:2018bje} (see also~\cite{Li:2026pim} for a recent application of these ideas in Brownian Spin SYK models). Consider the $q$-Majorana monomials
\begin{align}
    \label{eq:MajoranaMonomial}
    \hat{\psi}_{I}:=\hat{\psi}_{i_{1}}\cdots\hat{\psi}_{i_{q}}\,\,\,,\,\,\, I=\lbrace i_{1}<\cdots < i_{q} \rbrace\,~,
\end{align}
which contain ordered products of $q$-Majorana fermions. We will use them to define a basis for the operator algebra in terms of \emph{Majorana strings}. For each subset $A\subset S_{N_{f}}:=\lbrace 1,\ldots,N_{f}\rbrace\,$ define a Majorana string $\hat{\Psi}_{A}$ associated with the subset $A$ by
\begin{align}
    \label{eq:MajoranaString}
    \hat{\Psi}_{A}:=i^{\vert A\vert(\vert A \vert -1)/2}\prod_{a\in A}\hat{\psi}_{a}~,
\end{align}
where we define $\hat{\Psi}_{\emptyset}:=\hat{1}$. The \emph{size} of the Majorana string $\hat{\Psi}_{A}$ is defined as the cardinality of the subset
\begin{align}
    \label{eq:OpSizeSYK}
    \mathrm{s}(\hat{\Psi}_{A}):=\vert A \vert~.
\end{align}
The Majorana strings are Hermitian operators $\hat{\Psi}_{A}^{\dagger}=\hat{\Psi}_{A}$ that are orthonormal with respect to the Hilbert--Schmidt inner product $(\hat{\Psi}_{A} \vert \hat{\Psi}_{B} )^{\mathrm{HS}}:=2^{N_{f}/2}\mathrm{tr}(\hat{\Psi}_{A}\hat{\Psi}_{B})=\delta_{AB}$. Moreover, they form a complete basis of the algebra of linear bounded operators in the SYK Hilbert space $\mathcal{H}$ with respect to the HS inner product, $\mathcal{B}(\mathcal{H})$. Since there is a single basis element for a given subset $A\subset S_{N_{f}}=\lbrace 1,\ldots , N_{f}\rbrace$, this means that there are $2^{N_{f}}$ basis elements in $\mathcal{B}(\mathcal{H})$. Because of this, the operator size $s(\hat{\Psi}_{A})=\vert A\vert$ induces a direct-sum decomposition of $\mathcal{B}(\mathcal{H})$
\begin{align}
    \label{eq:DirectSumDecSYKOperGNS}
    \mathcal{B}(\mathcal{H})=\bigoplus_{s=0}^{N_{f}}V_{s}\,\,\,\,,\,\,\,V_{s}=\mathrm{span}\lbrace \hat{\Psi}_{A}\,:\, \vert A \vert = s \rbrace~,
\end{align}
where $\mathrm{dim}(V_{s})=N_{f}!/(s!(N_{f}-s)!)$.
Intuitively, this implies that ``most'' operators live near $s\sim N_{f}/2$, while simple operators are supported at small $s$.\\
At the same time, any operator $\hat{\mathcal{O}}\in \mathcal{B}(\mathcal{H})$ can be uniquely expanded in the Majorana string basis
\begin{align}
    \label{eq:OpSYKExpBasisString}
    \hat{\mathcal{O}}=\sum_{A \subset S_{N_{f}}}(\hat{\Psi}_{A} \vert \hat{O})^{\mathrm{HS}}\hat{\Psi}_{A}~,
\end{align}
and can be projected to a fixed size $s$ sector by the projection operator
\begin{align}
    \label{eq:ProjSYKsizes}
    \hat{\Pi}_{s}\hat{\mathcal{O}}:=\sum_{\vert A\vert=s}(\hat{\Psi}_{A} \vert \hat{O})^{\mathrm{HS}}\hat{\Psi}_{A}~.
\end{align}
Now, given two subsets $A,B\subset S_{N_{f}}$, the Clifford algebra for the Majorana fermions $\lbrace \hat{\psi}_{i}\rbrace$ implies that the products of Majorana strings satisfy
\begin{align}
    \label{eq:ProdMajoranaString}
    \hat{\Psi}_{A}\hat{\Psi}_{B}=\sigma(A,B)\hat{\Psi}_{A\triangle B}~,
\end{align}
where $\sigma(A,B)\in \lbrace \pm 1\rbrace$ is just a sign that can be directly computed, and where $A\triangle B:=(A\cup B)- (A\cap B)$ is the symmetric difference~\cite{Xu:2024owa}. As a consequence, the size of $\hat{\Psi}_{A}\hat{\Psi}_{B}$ is determined by
\begin{align}
    \label{eq:SizeProMajStrings}
    \begin{split}
    s(\hat{\Psi}_{A}\hat{\Psi}_{B})&=\vert A\triangle B\vert=\vert A\vert + \vert B \vert - 2\vert A\cap B \vert \\
    &= s(\hat{\Psi}_{A})+s(\hat{\Psi}_{B})-2r_{AB}~,
    \end{split}
\end{align}
where we defined $r_{AB}:=\vert A\cap B \vert$. Since the SYK$_{q}$ Hamiltonian can be seen as a sum of size-$q$ Majorana strings, it can be schematically written as
\begin{align}
    \label{eq:SYKqHamMajString}
    \hat{H}^{(q)}_{\mathrm{SYK}}=i^{q/2}\sum_{\vert I \vert = q}J_{I}\hat{\Psi}_{I}~.
\end{align}
Now, taking a single Majorana string $\hat{\Psi}_{I}$ with $\vert I \vert = q$ and commuting it with a basis element $\hat{\Psi}_{A}$ we get
\begin{align}
    \label{eq:CommSykMajStrings}
    [\hat{\Psi}_{I},\hat{\Psi}_{A}] = \left\{ \begin{array}{ccc} 2\sigma(I,A)\hat{\Psi}_{I\triangle A} & \mbox{for} & \mbox{odd }r_{IA}~, \\ 0 & \mbox{for} & \mbox{even }r_{IA} ~.
    \end{array}\right.
\end{align}
As a consequence, if the initial size of $\hat{\Psi}_{A}$ is $s$, and if its size after acting on it with the Liouvillian
\begin{align}
    \label{eq:SYKqMajStringLiouvillian}
    \breve{\mathcal{L}}^{(q)}_{\mathrm{SYK}}(\hat{\Psi}_{A}):=[ \hat{H}^{(q)}_{\mathrm{SYK}},\hat{\Psi}_{A}]\equiv i^{q/2}\sum_{\vert I \vert = q}J_{I}[\hat{\Psi}_{I},\hat{\Psi}_{A}]~,
\end{align}
is given by $s'=\vert I\triangle A\vert $, then the \emph{increase in the size} of the operator $\hat{\Psi}_{A}$ through the action of the Liouvillian is given by
\begin{align}
    \label{eq:SizeIncreaseMajString}
    \begin{split}
    \Delta s&:=s'-s=\vert I\triangle A \vert - \vert A \vert\\
    &=\vert I \vert-2\vert I\cap A\vert = q-2r_{IA}~.
    \end{split}
\end{align}
Moreover, since $\hat{\Pi}_{s}\breve{\mathcal{L}}^{(q)}_{\mathrm{SYK}}\hat{\Pi}_{s'}=0$ unless $\Delta s = s'-s=q-2r_{IA}$ for odd $r_{IA} \in [1,\min(s,q)]$, this means that the Liouvillian only connects size sectors separated exactly by
\begin{align}
    \label{eq:RuleIncreaseSize}
    \Delta s=q-2r_{IA}\in \lbrace q-2,q-6,q-10,\ldots \rbrace~,
\end{align}
and only through odd-overlap channels. In particular, this means that for operators $\hat{\Psi}_{A}$ of size $s=\vert A \vert$ with $s\ll N_{f}$, overlaps with $r_{IA}\geq 3$ are combinatorially suppressed. This implies that the dominant early time growth channel  in this case, the $r=1$ channel, gives $\Delta s =q-2$. A way to state this fact schematically is 
\begin{align}
    \label{eq:SchematicActionLiouvillian}
    \breve{\mathcal{L}}^{(q)}_{\mathrm{SYK}}:V_{s}\rightarrow \bigoplus_{1\leq r\leq \min(s,q)}^{}V_{s+q-2r}~,
\end{align}
where $r$ is odd. Thus, for $s\ll N_{f}$ the dominant block resulting from the action of the Liouvillian is $V_{s}\mapsto V_{s+q-2}$. This fact becomes a sharp statement if we consider a single Majorana fermion $\hat{O}=\hat{\psi}_{i}$, i.e. $s=1$. A term in the Hamiltonian $\hat{\Psi}_{I}$ contributes in the Liouvillian only if $i\in I$, in which case we have $r=1$. Thus, acting once with the Liouvillian on $\hat{\psi}_{i}$ sends a size $1$ string to a $q-1$ string: $\Delta s = (q-1)-1=q-2$. Moreover, in the large-$N_{f}$ limit, finite-size operators will increase their size to a good approximation by $\Delta s \approx q-2$. However, for finite $N_{f}$ and for initial operators of size $s\sim O(N_{f})$, the overlaps with $r=3,5,\ldots$ are no longer combinatorially rare and the Liouvillian can increase the operator size by~\eqref{eq:RuleIncreaseSize} including negative values, i.e. via shrinking channels. As the initial operator continues to commute with the Hamiltonian, the possible outcomes in operator size begin to resemble a biased random walk~\footnote{See~\cite{Patramanis:2026jtp} for a discussion of the emergence of Krylov complexity from quantum random walks on graphs.}. This is shown schematically in Fig. $2$ of~\cite{Roberts:2018mnp}. In such a case, it is no longer true that the increase in size is fixed to be $\Delta s =q-2$, and thus a correct $q$-sensitive notion of operator growth, such as~\eqref{eq:UpsilonqSensComplexity} and~\eqref{eq:GammaqSensComplexity} will have to be modified in some way to account for this biased random walk. Therefore, the factors of $q-2$ appearing in~\eqref{eq:UpsilonqSensOp}, ~\eqref{eq:GammaqSensOp} and the selector~\eqref{eq:qSensSelector} represent the dominant small-size/early-time growth channel generated by the SYK Liouvillian. \\

D. \textit{The Inverted Harmonic Oscillator}\label{subsec:OperatorSizeQuantIHO} -- A similar logic can be applied to the case of the inverted harmonic oscillator~\eqref{eq:QuantumIHOHam} for simple operators constructed from $\lbrace \hat{x},\hat{p}\rbrace$. Since the Liouvillian $\breve{\mathcal{L}}_{\mathrm{IHO}}=[\hat{H}_{\mathrm{IHO}},\cdot]$ does not generate new operator structures beyond the linear span of $\lbrace \hat{x},\hat{p}\rbrace$ for any given initial operator written as a linear combination of these two operators, repeated action by the Liouvillian will never leave the size $s=1$ subspace, where here by size we mean the minimal number of factors of $\lbrace \hat{x},\hat{p}\rbrace$ in a normally-ordered monomial basis. This means that the increase in size will always be $\Delta s = 0$. In this sense, a correct probe of operator growth should not grow in complexity under the Liouvillian, thus the only sensible choice for~\eqref{eq:qSensComplexity}  would be
\begin{align}
    \label{eq:qSensComplexityIHO}
    \begin{split}
    &Q^{\mathrm{IHO}}_{\mathcal{O}}(t)= \left(\hat{\mathcal{O}}(t)\vert \delta_{\mathcal{O}}(\breve{n},\Delta s) \vert\hat{\mathcal{O}}(t)\right)^{\mathrm{HS}}\\
   & =\left(\hat{\mathcal{O}}(t)\vert \Lambda (\breve{n}^{\Delta s}-\breve{n}^{\Delta s}+(1-(0)_{n}/n!)) \vert\hat{\mathcal{O}}(t)\right)^{\mathrm{HS}}\\
    &=\Lambda (1-\vert \varphi_{0}(t)\vert^{2})~,
    \end{split}
\end{align}
which in the case where $\hat{O}=\hat{x}$ yields
\begin{align}
    \label{eq:qSensComplexityIHOx}
    Q^{\mathrm{IHO}}_{\hat{x}}(t)=\Lambda (1-\sech(\lambda t))~.
\end{align}
This behavior is qualitatively similar to the results from the classical phase space analysis described in the previous section for classical integrable systems with unstable saddles~\eqref{eq:ClassKrylovSaddleDomExq},
\begin{align}
    \label{eq:RelqSensIHOClassK}
    (1-Q^{\mathrm{IHO}}_{\hat{x}}(t)/\Lambda)^2\leftrightarrow1-K_{q}(t)~,
\end{align}
where we set $\lambda=\lambda_{\mathrm{cl}}$ for the comparison. This is perhaps the cleanest connection between a classical notion of Krylov complexity $K_{q}(t)$ and its regularized quantum counterpart $Q_{\hat{x}}$, which agree on the integrable properties of the inverted harmonic oscillator.

E. \textit{Connection to the Replica Trick}\label{subsec:qSensReplicaRelation} --
We end this section with a brief discussion of how this approach could mathematically arise from the replica trick. In Sections.~\hyperref[sec:ExamplelogKCFT]{IV.} and~\hyperref[sec:QuantKrylovIHO]{VI.}, we discussed the analytic continuation of the higher-order spreading superoperator via the standard analytic continuation $\breve{n}^{m}\mapsto e^{m\log(\breve{n})}$, where $\mathbb{Z}^{+}\ni m \mapsto m \in \mathbb{R}$. However, it is also possible to consider different analytic continuations, such as
\begin{align}
    \label{eq:GenAnnContReplica}
    \breve{n}^{m}\mapsto e^{m\log(\breve{n})}+\frac{\sin(m\pi)}{\pi}F(\breve{n}):=\breve{\mathfrak{N}}^{(m)}(F,\breve{n})~,
\end{align}
which, for integer $m\in \mathbb{Z}^{+}$ reduces to the usual higher-order spreading superoperator, but which for real $m\in \mathbb{R}$ picks up an additional phase proportional to a function $F(\breve{n})$. In particular, taking the derivative of $\breve{\mathfrak{N}}^{(m)}(F,\breve{n})$ with respect to $m$ and subsequently the limit $m\rightarrow 0$ yields
\begin{align}
    \label{eq:GenAnnContReplicaDerLimit}
    \lim_{m\rightarrow 0}\frac{\partial}{\partial m}\breve{\mathfrak{N}}^{(m)}(F,\breve{n})=\log(\breve{n})-F(\breve{n})~.
\end{align}
In other words, the specific analytic continuation of $\breve{n}^{m}$ to real $m\in \mathbb{R}$ depends on the continuation scheme. This is similar to computations of entanglement entropies, where the integer values of the replica index are fixed by path integrals but the analytic continuation to real numbers is not fixed or unique without additional physical input, and where subtractions correspond to defining a particular renormalized entropy, e.g. vacuum-subtracted, or area-law subtracted. In this case, the additional physical input is represented by the function $F(\breve{n})$. One advantage of this generalized replica trick is that we can directly ``remove'' the logarithmic divergence at $\langle \breve{n}\rangle_{\mathcal{K}}=n\rightarrow 0$, which we previously removed by subtracting it from the finite $n\geq 1$ contribution. Shifting $F(\breve{n})$ to $\tilde{F}(\breve{n}):=F(\breve{n})-\log(\breve{n})$ not only removes this divergence, but also completely changes the behavior of the logarithmic Krylov complexity subject to $F$. Therefore, a way to connect our discussion of the theory- and operator-sensitive complexities and the replica approach is to take
\begin{align}
    \label{eq:ConnectqSensGenReplica}
    \tilde{F}(\breve{n})\equiv \delta_{\hat{\mathcal{O}}}(\breve{n},\mathrm{info}(\breve{\mathcal{L}}))~,
\end{align}
with $\delta_{\mathcal{O}}$ given by~\eqref{eq:DeltaqSensOp}  and~\eqref{eq:qSensComplexityIHO} in the specific cases of the SYK$_{q}$ and the inverted harmonic oscillator. In this way, the replica trick applied to the spreading operator can be turned into a probe of operator growth that is sensitive to physical information about the system and class of operators beyond the universal behavior of Krylov complexity, namely
\begin{align}
    \label{eq:logKqSensComplexityRelationReplica}
    \mathbf{L}_{\mathcal{O}}^{(\delta)}(t)\equiv \left.\frac{\partial}{\partial m}(\mathcal{O}(t)|\breve{\mathfrak{N}}^{m}(\delta_{\mathcal{O}},\breve{n})|\mathcal{O}(t))\right\vert_{m\rightarrow 0}\equiv Q_{\mathcal{O}}^{\delta}(t)~.
\end{align}

We end this section with a remark. Our motivation to define~\eqref{eq:DeltaqSensOpGen} came from the particular cases of the SYK$_{q}$ and the inverted harmonic oscillator studied in the present manuscript. It is unclear to us at the moment whether this approach yields consistent results in other cases not describable by a $k$-local Hamiltonian. A particular reason why this approach has succeeded for these two models is that we were able to reduce the information about integrability/chaoticity to a single parameter, which we used to construct the change in operator size $\Delta s$. This is generally not possible. For example, in CFT$_{2}$, integrability/rationality relies on more details than just the ratio of the central charge $c$ to the operator scaling dimension $\Delta$, for example. As a consequence, in such a case, it is not possible to directly use our proposed approach. Nevertheless, we would like to understand if similar approaches can be used in such systems where the breaking of integrability and emergence of chaotic behavior depend on the interplay of more parameters. We leave this direction for future work.


\phantomsection
X.\textit{ Discussion and Future Directions}\label{sec:Discussion} --

\phantomsection
1.\textit{The Definition of LogK Complexity}\label{subsec:DiscussionReplica} --
Let us recapitulate the regularization procedure of logK-complexity described in Sec.~\hyperref[sec:LogKandElogK]{III}. The spreading superoperator $\breve{n}$ is a positive and Hermitian operator on $\mathcal{K}$, with a spectrum contained in $\lbrace 0,1,2,\ldots\rbrace$. Applying our working definition of logK-complexity through the replica trick~\eqref{eq-Log-K replica} required a regularization of the logarithmic divergence for the ``zeroth''-mode in the spectrum through~\eqref{eq:ShiftedSpreadingOp}. Another way to perform the regularization of $\log(\breve{n})$ would be to consider a finite small shift in the spreading superoperator $\breve{n}+\breve{\epsilon}=\breve{n}_{\epsilon}$ with $\Vert \breve{\epsilon} \Vert  \ll \Vert \breve{n}\Vert$ with respect to the canonical norm induced by the inner product. Then, taking $\lim_{\breve{n}\rightarrow 0}\log(\breve{n}+\breve{\epsilon})=\log(\breve{\epsilon})$ absorbs the logarithmic divergence into the regulator $\breve{\epsilon}$. To be precise, for any $\mathbb{R}\ni \epsilon >0$, let us define the shifted superoperator
\begin{align}
    \label{eq:ShiftSpreadOp}
    \breve{n}\mapsto \breve{n}_{\epsilon}=\breve{n}+\epsilon\breve{\mathbf{1}}~,
\end{align}
which is positive and bounded from below. In this case, $\log(\breve{n}_{\epsilon})$ is a well-defined Hermitian superoperator for all $n$, which in the Krylov basis takes the form
\begin{align}
    \label{eq:lognepsilon}
    \log(\breve{n}_{\epsilon})=\sum_{n\geq 0}\log(n+\epsilon)\vert \mathcal{K}_{n})(\mathcal{K}_{n}\vert~,
\end{align}
and whose expectation value is given by
\begin{align}
    \label{eq:logKnepsilon}
    \begin{split}
    \mathbf{L}^{\mathrm{bare}}_{K}(t,\epsilon)&=(\mathcal{O}(t)\vert \log(\breve{n}_{\epsilon})\vert\mathcal{O}(t))\\
    &=\sum_{n\geq 0}\log(n+\epsilon)\vert \varphi_{n}(t)\vert^{2}\\
    &=\log(\epsilon)\vert \varphi_{0}(t)\vert^{2}+\sum_{n\geq 1}\log(n+\epsilon)\vert \varphi_{n}(t)\vert^{2}~.
    \end{split}
\end{align}
Thus, in the limit, $\epsilon\rightarrow 0$ the only divergence is given precisely by $\log(\epsilon)\vert \varphi_{0}(t)\vert^{2}$. Our regularized logK-complexity~\eqref{eq-NaiveLogK} is equivalent to the renormalized expression
\begin{align}
    \label{eq:RenormlogKnepsilon}
    \begin{split}
    \mathbf{L}^{\mathrm{ren}}_{K}(t):&=\lim_{\epsilon\rightarrow 0}\left(\mathbf{L}^{\mathrm{bare}}_{K}(t,\epsilon)-\log(\epsilon)\vert \varphi_{0}(t)\vert^{2}\right)\\
    &=\sum_{n\geq 1}\log(n)\vert \varphi_{n}(t)\vert^{2}~.
    \end{split}
\end{align}


\phantomsection
2.\textit{Path Integral LogK-Complexity}\label{subsec:DiscussionPathIntlogK} --
Our discussions on Krylov complexity in classical phase space rely on the Hamiltonian formalism. Therefore, it would be interesting to consider the construction of the Krylov algorithm, Krylov complexity, and the logK complexity in the Lagrangian formalism by using the path integral in classical phase space~\cite{Gozzi:1994path}. In the quantum realm, the path-integral description of Krylov complexity was discussed in~\cite{Beetar:2025erl, Aguilar-Gutierrez:2025pqp, Aguilar-Gutierrez:2025mxf, Aguilar-Gutierrez:2025hty, Aguilar-Gutierrez:2025sqh}. More recently, the Schwinger--Keldyish formulation of Krylov complexity using a path-integral approach was discussed in~\cite{Murugan:2026yyu}. It would be interesting to compare our approaches from classical and quantum aspects, with possible applications to understand our observations on (e)logK complexity. Further directions would be to seek insights into path-integral complexity, which could potentially build interconnections between different notions of quantum complexities. We leave this for future work.


\phantomsection
3.\textit{Pollicott--Ruelle Resonances}\label{subsec:DiscussionPollRuelle} --
In Section.~\hyperref[sec:ClassicalPSKrylov]{VII.}, we provide a classical description of the Lanczos algorithm and construct classical notions of Krylov and logK complexities. Such a construction seems to be general and applicable to study the classical Krylov complexity of any quantum system with a well-defined classical limit. Although in our context, we applied it mainly to saddle-dominated systems, we see no obstruction in using it to study quantum systems with a classical limit where the spectrum gap indicates the Pollicott--Ruelle resonances~\cite{Pollicott1985, ruelle1986locating} that are known to govern the late-time behavior of OTOCs after the Lyapunovian regime \cite{Garcia-Mata:2018slr}. Since the classical Krylov complexity and the elogK complexity saturate in finite-size systems, we expect that it should be possible, at least in principle, to study how the Policott--Ruelle resonances are encoded in the classical Krylov (and logK) complexities. This is because the wavefunctions used to construct both Krylov and logK complexity depend on the two-point (autocorrelation) function, which also governs Pollicott--Ruelle resonances. 


\phantomsection
4.\textit{The Thermodynamic Limit of the LMG and the Mixed-Field Ising}\label{subsec:Thermolimit} --
In Section.~\hyperref[sec:NumericalAnalysis]{V.}, we discussed the numerical computation of the logarithmic K-complexity in finite-dimensional many-body systems. In particular, we focus on the LMG and the mixed-field Ising model at the chaotic point for a finite number of degrees of freedom $S$. This allowed us to accurately study their early-time behavior and observe a clear distinction in how close logK-complexity tracks the usual Krylov complexity, despite both Lanczos sequences displaying a similar linear behavior for $n\lesssim \log(S)$. However, if we take the thermodynamic limit $S\rightarrow \infty, V\rightarrow \infty, S/V$ finite, then both Lanczos sequences $b_{n}$ are expected to become indistinguishable. Then it is natural to ask whether logK complexity would behave differently in these two systems, since after all, from the perspective of the Lanczos sequences they would be indistinguishable. Assuming the thermodynamic limit entails a well-defined limit $\lim_{D_{\mathcal{K}}\rightarrow \infty}$ of the higher-order Krylov complexities, it would be interesting to understand what happens to the regularized logarithmic K-complexity in such a limit. We leave this question for future work.


5.\textit{Logarithmic Krylov Complexity and OTOCs}\label{subsec:logKandOTOCs} --
In his works on logarithmic OTOCs~\cite{Trunin:2023rwm,Trunin:2023xmw}, D. Trunin claims that the refined Lyapunov exponent $\overline{\lambda}_{L}$ obtained from the logarithmic OTOC is bounded by $2 \pi T$, similarly to how the usual Lyapunov exponent $\lambda_{L}$ satisfies the Maldacena--Shenker--Stanford (MSS) chaos bound~\cite{Maldacena:2015waa}. It has been conjectured~\cite{Parker:2018yvk,Avdoshkin:2019trj} that the Krylov exponent $\lambda_{K}$ (or twice the growth rate of the Lanczos coefficients $2\alpha$) provides a tighter bound to the Lyapunov exponent than $2 \pi T$. It is then natural to ask whether the \emph{refined} Krylov exponent, defined through the logarithmic Krylov complexity
\begin{align}
\label{eq:RefinedKrylovExponent}
\lambda_{\mathbf{L}_{K}}:=\lim_{t\rightarrow \infty}\frac{\mathrm{d}(\mathbf{L}_{K}(t))}{\mathrm{d}t}~,
\end{align}
provides a tighter bound on the refined Lyapunov exponent:
\begin{align}
\label{eq:RefinedChaosBound}
\overline{\lambda}_{L}\leq\lambda_{\mathbf{L}_{K}}\leq 2\pi T~.
\end{align}
Since generally OTOCs cannot be determined by the information of two-point functions (which are the basis for computing the Krylov and logK-complexities), we currently do not possess strong arguments that support this conjecture, and we leave its study for future work. 


6.\textit{Logarithmic Krylov Complexity and $C=\mathrm{Anything}$}\label{subsec:logKandCequalsEnything} --
Krylov complexity has been recently explored within the AdS/CFT correspondence, where it has been shown that in the context of two-dimensional gravity models (such as Jackiw--Teitelboim gravity) in anti de-Sitter space, it can be represented as the length of the two-sided wormhole ~\cite{Rabinovici:2023yex,Balasubramanian:2024lqk,Aguilar-Gutierrez:2025pqp,Aguilar-Gutierrez:2025mxf}. This connection between Krylov and a particular notion of holographic complexity begs the question whether logarithmic K-complexity is a type of holographic ``\emph{Complexity=Anything}'' measure~\cite{Myers:2024vve}. We would like to understand the connection better in future work.

\phantomsection
XI. \textit{Conclusion}\label{sec:Conclusion} -- 

In this manuscript, we proposed and tested a new notion of complexity, the logarithmic Krylov complexity, along with its exponentiated form. Our motivation to propose this notion was to offer a plausible resolution to the ``fake" chaos signatures that arise in saddle-dominated scrambling systems. In practice, we compute (a regularized version of) the logarithmic Krylov complexity through a replica trick applied to the higher-order generalizations of Krylov complexity.

For a better understanding of this new quantity, we provide an analytical analysis of the universal initial-time growth of (e)logK complexity and the late-time saturation in thermalizing many-body systems. Moreover, we examine (e)logK complexity in systems where calculations can be carried out analytically, such as the SYK model and the inverted harmonic oscillator. In the conformal limit of the SYK$_{q}$ model, we find early-time exponential growth in both Krylov and elogK complexities, regardless of the value of $q$. In the inverted harmonic oscillator, which is dominated by unstable saddles, the Krylov complexity resembles the ones from the SYK case with $\eta=1/2$, and higher-order Krylov and elogK complexities match with the SYK case for $\eta=1$. Such a result seems to put in question the usefulness of elogK complexity in resolving the instability arising from unstable saddles. We argue that this is in part due to the infinite-dimensional nature of the GNS Hilbert space in these cases, which leads to issues with the replica trick in practically defining the logK complexity. To solve this issue without imposing an artificial cut-off, we propose a new definition of the Krylov spreading operator in Section. \hyperlink{sec:ReplicaGeneralization}{IX}. By using a new definition (\ref{eq:DeltaqSensOp}), which retains universal information about the operator growth, but is also sensitive to details of the systems and operators, we are able to find early-time sub-exponential growth of Krylov complexity and late-time saturation for SYK$_2$, and exponential growth for $q\geq4$. Such a definition of the spreading operator has its origins in previous considerations about operator size in the SYK and similar systems.

Beyond the cases with infinite-dimensional GNS Hilbert spaces, for finite-dimensional systems, our definition of (e)logK complexity seems to resolve the problem of early-time scrambling from unstable saddles. This is supported by our numerical analyses of both conventional Krylov complexity and elogK complexity in the LMG model, which is integrable but with an unstable saddle point. Our results show a significant distinction between elogK complexity from the conventional Krylov complexity. To further test the validity of our proposal, we examined the mixed-field Ising model at the chaotic point. In this case, both measures of complexity exhibit early-time exponential growth with negligible deviation between them. Taken together, our numerical results support the idea that (e)logK complexity successfully captures the correct dynamical behavior in both saddle-dominated scrambling and truly chaotic finite-dimensional systems. While we cannot guarantee that logK-complexity is free from ``false-positives'', our results suggest at least a partial resolution to the issue of early-time scrambling in finite-dimensional saddle-dominated systems. We thus believe that (e)logK complexity offers a good starting point for a refined definition of quantum complexity that can be used as a reliable indicator of scrambling.

%

\textbf{Note added}. In recent work~\cite{Pal:2026otf} the authors analyze the problem of state/operator spreading under time evolution described in terms of a Krylov basis in the Wigner--Weyl phase space formulation of quantum mechanics. It would be interesting to compare our approaches and determine how theirs connects to ours. We thank the authors of~\cite{Pal:2026otf} for related discussions. 
\\
\textbf{Second Note added}. As we were completing our manuscript, we became aware of~\cite{DeRo:2026mlc}, which also studies classical notions of Krylov complexity. We thank the authors for the discussions related to logK-complexity. 
\\

%

\phantomsection
\textit{Acknowledgments}\label{sec:Acknowledgements}--
We are grateful to Sergio E. Aguilar--Gutierrez, Viktor Jahnke,  Mitsuhiro Nishida, Kuntal Pal, Adri\'{a}n S\'{a}nchez--Garrido and to Dmitrii A. Trunin  for valuable discussions, comments on the draft, and correspondence.
This work was supported by the Basic Science Research Program through the National Research Foundation of Korea (NRF) funded by the Ministry of Science, ICT \& Future Planning (NRF-2021R1A2C1006791), the Korea government(MSIT)(RS-2025-02311201), (RS-2024-00445164) and the framework of international cooperation program managed by the NRF of Korea (RS-2025-02307394), the Creation of the Quantum Information Science R\& D Ecosystem (Grant No. RS-2023-NR068116) through the National Research Foundation of Korea (NRF) funded by the Korean government (Ministry of Science and ICT). 
This research was also supported by GIST research fund (Future leading Specialized Resarch Project, 2026, and the Regional Innovation System \& Education(RISE) program through the (Gwangju RISE Center), funded by the Ministry of Education(MOE) and the (Gwangju Metropolitan City), Republic of Korea.(2025-RISE-05-001)
H.~A. Camargo was supported by the Basic Science Research Program through the National Research Foundation of Korea (NRF) funded by the Ministry of Education (NRF-2022R1I1A1A01070589) and by the National Science and Technology Council, the Ministry of Education (Higher Education Sprout Project NTU-114L104022-1), and the National Center for Theoretical Sciences of Taiwan. Hugo A. Camargo, Yichao Fu and Yeong Han Park are considered co–first authors and contributed equally to this work.
%

\bibliographystyle{apsrev4-1}
\bibliography{ref}

\appendix

\section{Supplemental Material}

\subsection{A. Initial growth of logK Complexity}
\label{app:InitiallogK}--
In this Appendix, we examine the behavior of (e)logK complexity around $t=0$ using the generic properties of the wavefunctions $\varphi_{n}(t):=i^{-n}(\mathcal{K}_{n}\vert \mathcal{O}(t))$. This approach is universal and does not depend on the dynamics $\hat{\mathcal{L}}$ nor on the choice of initial operator $\hat{\mathcal{O}}_{0}$. In the Lanczos algorithm, an orthonormal basis in Krylov space $\mathcal{K}$ called the Krylov basis $\{ |\mathcal{K}_n)\}$ is explicitly constructed according to the Gramm--Schmidt procedure
\begin{eqnarray}
    |A_{n+1})&:=&(\breve{\mathcal{L}}-a_n)|\mathcal{K}_{n})-b_n|\mathcal{K}_{n-1})~,\nonumber\\
    |\mathcal{K}_n)&:=&b_n^{-1}|A_n)~,
    \label{App-eq-KAlg}
\end{eqnarray}
where two sequences $\lbrace a_n, b_n\rbrace $
\begin{equation}
    a_n=(\mathcal{K}_n|\hat{\mathcal{L}}|\mathcal{K}_n)~, ~~~ b_n=\sqrt{(A_n|A_n)}~,
\end{equation}
are the Lanczos coefficients, with the convention $a_{-1}=b_0=0$. This algorithm stops whenever we find an $N\geq 1$ such that $b_N=0$. From (\ref{App-eq-KAlg}),~\eqref{eq:TimeEvolOpKrylovBasis} and the Heisenberg equation, the wavefunctions $\varphi_{n}(t)$ can be shown to satisfy the Schr\"{o}dinger-like equation:
\begin{equation}
   \partial_t \varphi_n(t)=ia_n\varphi_n(t)+b_n\varphi_{n-1}(t)-b_{n+1}\varphi_{n+1}(t)~.
   \label{App-eq-SchrodingerEq}
\end{equation}
Given the conditions $\varphi_{-1}(t)=0=b_{0}$ \footnote{The definition here is a bit different from the main text (\ref{eq:ClassLanczInitial}), where we set $b_{-1}=0$.} and $\varphi_n(0)=\delta_{n0}$, the equation~\eqref{App-eq-SchrodingerEq} implies the following conditions at $t=0$ for $\varphi_{n}$ and their derivatives
\begin{eqnarray}
   && \varphi_{n}(0)=\delta_{n0}~,~~~\dot{\varphi}_{n}(0)=ia_0\delta_{n0}+b_1\delta_{n1}~,\nonumber\\
    &&\ddot{\varphi}_{n}(0)= -(a^2_0+b_1^2) \delta_{n0}+i(a_0+a_1)b_1 \delta_{n1}+b_2b_1\delta_{n2}~,\nonumber\\
    &&\dots
\end{eqnarray}
Using these, we can solve the Schr\"{o}dinger equation~\eqref{App-eq-SchrodingerEq} order by order in $t$ to approximate the Krylov complexity around $t=0$:
\begin{eqnarray}
   &&\varphi_{0}(t)\approx 1+ia_{0}t-\frac{1}{2}(a_{0}^{2}+b_{1}^{2}) t^{2}\nonumber\\
   &&-\frac{1}{6}i \left((a_{0}^{3}+(2a_{0}+a_{1})b_{1}^{2}\right)t^{3}\nonumber\\
 &&+\frac{1}{24}\left(a_{0}^{4}+3a_{0}^{2}b_{1}^{2}+2a_{0}a_{1}b_{1}^{2}+b_{1}^{2}(a_{1}^{2}+b_{1}^{2}+b_{2}^{2})\right)t^{4}~,\nonumber\\
 \\
  &&\varphi_{1}(t)\approx b_{1}t+\frac{1}{2}i(a_{0}+a_{1})b_{1} t^{2}\nonumber\\
  &&+\frac{1}{6} \left(a_{0}^{2}+a_{0}a_{1}+a_{1}^{2}+b_{1}^{2}+b_{2}^{2}\right)b_{1}t^{3}~,\\
   &&\varphi_{2}(t)\approx \frac{1}{2}b_{1}b_{2}t^{2}~.
\end{eqnarray}
After re-summing the factors of $a_0$ and up to the leading terms in $\lbrace b_{n}\rbrace$, the wavefunctions can be shown to behave near $t=0$ according to:
\begin{eqnarray}
    \varphi_0(t)&=&e^{ia_0 t}+\cdots,~~~\varphi_1(t)=e^{ia_0 t}b_1t +\cdots,\nonumber\\
    \varphi_2(t)&=&\frac{1}{2}e^{ia_0t}b_1b_2t^2+\cdots,~~~ \dots.
\end{eqnarray}
Note that for Hermitian initial operators, the coefficients $a_n$ vanish, which survive in open systems. However, here, we keep our analysis general, and properties of $a_n$ will not be used in our discussions.
These $0\lesssim t$ solutions have been used to show that conventional Krylov complexity generically exhibits a power-law behavior around $t=0$: $K(t\rightarrow0)=b^2_1t^2+\dots$~\cite{Fan:2022xaa}. However, for logK complexity, this is no longer true. As discussed in the main text, using the replica trick (\ref{eq-Log-K replica}), the logK complexity can be written as
\begin{equation}
   \mathbf{L}_K(t)=\left.\frac{\partial}{\partial m}\left( \sum_n n^m |\varphi_n(t)|^2\right)\right\vert_{m\rightarrow0}~,
\end{equation}
where we used the canonical analytic continuation $n^m\mapsto e^{m\log(n)}$. Note that this is in general not equivalent to $\sum_{n\geq 0}\log(n)|\varphi_n(t)|^2$ due to the divergence for $n=0$. A straightforward computation shows that the first wavefunction contributing non-trivially to logK is $\varphi_2(t)$, unlike in $K(t)$ where $\varphi_1(t)$ dominates around $t=0$. Therefore, the initial growth of logK complexity, as well as the elogK complexity, is given by
\begin{equation}
    \mathbf{E}_{K}(t\rightarrow0)\equiv \mathbf{L}_K(t\rightarrow0)=\frac{\log(2)}{4}b_1^2b_2^2t^4+O(t^{6})~.
\end{equation}
This initial behavior has been verified in our numerical computations for $t\lesssim 1$. This quartic initial growth $O(t^{4})$ moreover indicates their inadequacy as distance measures between operators, following the arguments of the authors in~\cite{Aguilar-Gutierrez:2023nyk}.
\subsection{B. Long-time average of logK-Complexity}
\label{app:LongTimelogK}--
Here we provide details on the long-time average of logK-complexity, following the arguments detailed in~\cite{Rabinovici:2021qqt}. In this analysis, we restrict ourselves to the time evolution in the Krylov subspace $\mathcal{K}$. Consider the eigenvalues and eigenstates of the Liouvillian $\hat{\mathcal{L}}\vert \omega_{i} )=\omega_{i}\vert \omega_{i} )$ with $i=0,\ldots,D_{\mathcal{K}}-1$, after resolving degeneracies in the spectrum arising from equal energy differences. Expanding the wavefunctions in the Liouvillian eigenbasis $\vert \omega_{i} )$
\begin{align}
    \label{eq:ProbAmpLiouvPhin}
    \begin{split}
    \varphi_{n}(t)&=i^{-n}(\mathcal{K}_{n}\vert \mathcal{O}(t))\\
    &=i^{-n}\sum_{j=0}^{D_{\mathcal{K}}-1}e^{i\omega_{j}t}(\omega_{j}\vert \mathcal{O}_{0})(\mathcal{K}_{n}\vert \omega_{j})~.
    \end{split}
\end{align}
Thus, the probabilities $\vert \varphi_{n}(t)\vert^{2}$ can now be interpreted as transition amplitudes from $\vert \mathcal{O}_{0})$ to $\vert \mathcal{K}_{n})$ at time $t$ and are given by
\begin{align}
    \label{eq:AmpLiouvPhin}
    \begin{split}
    \vert \varphi_{n}(t)\vert^{2}=&\sum_{i,j=0}^{D_{\mathcal{K}}-1}e^{i(\omega_{j}-\omega_{i})t}\\
    &\times(\omega_{j}\vert\mathcal{O}_{0})(\mathcal{K}_{n}\vert\omega_{j})(\omega_{i}\vert \mathcal{K}_{n})(\mathcal{O}_{0}\vert\omega_{i})~.
    \end{split}
\end{align}
The long-time average of the probabilities is then given by
\begin{align}
    \label{eq:LongTAverPhiSq}
    \begin{split}
    \overline{\vert \varphi_{n} \vert^{2}}&:=\lim_{T\rightarrow \infty}\frac{1}{T}\int_{0}^{T}\textrm{d}t\,\vert \varphi_{n}(t) \vert^{2}\\
    &=\sum_{i=0}^{D_{\mathcal{K}}-1}\vert(\mathcal{O}_{0}\vert \omega_{i})\vert^{2}\vert(\mathcal{K}_{n}\vert \omega_{i})\vert^{2}~,
    \end{split}
\end{align}
since the phase differences $i\neq j$ cancel out and only diagonal terms $i=j$ contribute to the integral. This is because, by construction, the spectrum of the Liouvillian restricted to the Krylov subspace has no degeneracies. If the Liouvillian eigenstates are fully delocalized on the Krylov basis, this implies $\vert (\mathcal{K}_{n}\vert \omega_{i})\vert^{2}\propto 1/D_{\mathcal{K}}$ for all $i,n$, and thus $\overline{\vert \varphi_{n} \vert^{2}}\sim(1/D_{K}^{2})*(D_{\mathcal{K}})=1/D_{\mathcal{K}}$ for all $n$. In this case, the long-time average of the K and logK complexities is given by
\begin{eqnarray}
\label{eq-K-longtime}
    \overline{K}=\sum^{D_\mathcal{K}-1}_{n=0} n \overline{\vert \varphi_{n} \vert^{2}}=\frac{(D_\mathcal{K}+1)}{2}\approx \frac{D_{\mathcal{K}}}{2}~,
\end{eqnarray}
and
\begin{align}
\label{eq-logK-longtime}
\begin{split}
    \overline{\mathbf{L}}_{K}&=\lim_{m\rightarrow 0}\frac{\partial}{\partial m}\left. \left(\sum^{D_\mathcal{K}-1}_{n=0} \frac{n^m}{D_\mathcal{K}} \right)\right.\\
    &=\lim_{m\rightarrow 0}\frac{\partial}{\partial m}\left. \left( \frac{H^{(-m)}_{D_\mathcal{K}-1}}{D_\mathcal{K}} \right)\right.=\frac{1}{D_\mathcal{K}} \sum^{D_\mathcal{K}-1}_{j=2}\log(j)\\
    &=\frac{1}{D_\mathcal{K}} \log((2)_{D_\mathcal{K}-2})~,
    \end{split}
\end{align}
where $H^{(r)}_{l}\equiv\sum^l_{j=1}1/j^r, \quad l\in \mathbb{N}$ is the (generalized) harmonic number of order $r$ and where we used the fact that
\begin{equation}
\label{eq-DerivHarmonicNumber}
    \frac{\partial H^{(r)}_{l}}{\partial r}=-\sum^l_{j=2}\frac{\log(j)}{j^r}~,
\end{equation}
and
\begin{equation}
    \sum^a_{j=2} \log(j) \equiv \log((2)_{a-1})~,
\end{equation}
and where $(x)_n\equiv\Gamma(x+n)/\Gamma(x)$ is the Pochhammer symbol.
In order to have a sensible comparison of their growth rates, consider the large-time behavior of the elogK-complexity:
\begin{align}
\label{eq-EK-longtime}
\overline{\mathbf{E}}_{K}=e^{\overline{\mathbf{L}}_{K}}-1\approx \Gamma(D_{\mathcal{K}})^{1/D_{\mathcal{K}}}~.
\end{align}
The ratio of the long-time averages of these two quantities in the large $D_{\mathcal{K}}$ limit is given exactly by
\begin{equation}
    \left.\frac{\overline{K}}{\overline{\mathbf{E}}_{K}}\right.=\frac{\mathfrak{e}}{2}>1~,
\end{equation}
where $\mathfrak{e}$ is Euler's number. It is also important to note that the large time averages~\eqref{eq-K-longtime} and ~\eqref{eq-logK-longtime} should be considered as upper bounds, which will in principle only be saturated in maximally thermalizing (chaotic) quantum systems where the Liouvillian eigenbasis completely delocalizes in the Krylov basis, such as in the SYK model \cite{Rabinovici:2020ryf}. However, some integrable systems with a right-biased Krylov chain may over-saturate these bounds \cite{Aguilar-Gutierrez:2025hbf}.

\subsection{C. Details in the Inverted Harmonic Oscillator}
\label{app:MathDetailsIHO}

In this appendix, we provide mathematical details on the main computations in Sec.~\hyperref[sec:QuantKrylovIHO]{VI.}. First, we derive the expression for the wavefunctions~\eqref{eq:ProbAmpQuantIHO}. Consider the equation of motion of the $\hat{x}$ operator:
\begin{align}
    \frac{d\hat{x}}{dt}=\frac{i}{\hbar}[\hat{H}_{\mathrm{IHO}},\hat{x}]=\lambda\hat{x}~,
\end{align}
whose general solution is given by
\begin{align}
    \hat{x}(t)=e^{\lambda t}\hat{x}(0)~.
\end{align}
This implies that the Gaussian operator $\hat{\mathcal{O}}_0=(2/\pi\alpha)^{1/4}e^{-\hat{x}^{2}/\alpha}$ has the following Heisenberg evolution
\begin{align}
    \hat{\mathcal{O}}_0 \mapsto \hat{\mathcal{O}}_0(t)=\left(\frac{2}{\pi\alpha}\right)^{1/4} e^{\lambda t/2}e^{- e^{2\lambda t}\hat{x}^2/\alpha}~,
\end{align}
where $e^{\lambda t/2}$ is the normalization factor that comes from the condition $\vert \vert\hat{\mathcal{O}}_{0}(t)\vert \vert ^{2}=(\hat{\mathcal{O}}_{0}(t)\vert \hat{\mathcal{O}}_{0}(t))^{\mathrm{HS}}=1$. Next, we compute the wavefunctions $\varphi_{n}(t)$ given by
\begin{align}
\begin{split}
    \varphi_n(t)&:=(\hat{\mathcal{K}}_{n}\vert \hat{\mathcal{O}}_{0}(t))^{\mathrm{HS}}=\int^{\infty}_{-\infty}\mathrm{d}x\, \langle x\vert\hat{\mathcal{K}}_{n}^{\dagger}\,\hat{\mathcal{O}}_0(t) \vert x \rangle \\
    &=\frac{1}{\sqrt{(2n)!}2^n}\left(\frac{2}{\pi\alpha}\right)^{1/2}e^{\lambda t/2} \\
    &\times \int^{\infty}_{-\infty}\mathrm{d}x\,H_{2n}\left(\sqrt{\frac{2}{\alpha}}x\right)e^{-(1+e^{2\lambda t})x^2/\alpha}~,\\
    \end{split}
\end{align}
where we omit the $i^{-n}$ prefactor and where the Krylov basis is given by~\eqref{eq:KrylovBasisHermiteIHO}. The integral above can be represented schematically in the following way
\begin{align}
    I_{H}(q,p)=\int^{\infty}_{-\infty}\mathrm{d}x\, H_{2n}(qx)e^{-p x^2}~,
\end{align}
where $p=(e^{2\lambda t}+1)/\alpha$ and $q=\sqrt{2/\alpha}$. To evaluate this integral, consider the generating function of the Hermite polynomials:
\begin{align}
    \sum_{m=0}^{\infty}\frac{H_m(y)}{m!}z^m=e^{2zy-z^2}~.
\end{align}
From this, we choose $y=qx$ and multiply both sides by $e^{-px^2}$. We have
\begin{align}
    \sum_{m=0}^{\infty}\frac{H_m(qx)}{m!}z^m e^{-px^2}=e^{2qxz-z^2-px^2}~.
\end{align}
We can now perform the integral over $x$
\begin{align}
   \int_{-\infty}^{+\infty}\mathrm{d}x\, \sum_{m=0}^{\infty}\frac{H_m(qx)}{m!}z^m e^{-px^2} = \sqrt{\frac{\pi}{p}}e^{\left(\frac{q^{2}}{p}-1\right)z^2}~.
\end{align}
Performing a Taylor series expansion on the right-hand side, we have
\begin{align}
\begin{split}
  & \sum_{n=0}^{\infty}\left(\int_{-\infty}^{+\infty}\mathrm{d}x\, \frac{H_{2n}(qx)}{(2n)!} e^{-px^2}\right)z^{2n} \\
   &=\sum_{n=0}^{\infty} \sqrt{\frac{\pi}{p}} \frac{(q^2/p-1)^n}{n!}z^{2n}~.
   \end{split}
\end{align}
From this expression, we can identify the series coefficients of $z^{2n}$ on both sides of the equality
\begin{align}
   \int_{-\infty}^{+\infty}\mathrm{d}x\,\frac{H_{2n}(qx)}{(2n)!} e^{-px^2}= \sqrt{\frac{\pi}{p}} \frac{(q^2/p-1)^n}{n!}~.
\end{align}
We thus find the result of the integral we are interested in
\begin{align}
    I_{H}(q,p)= \sqrt{\frac{\pi}{p}} \frac{(2n)!}{n!}\left(\frac{q^2}{p}-1\right)^n~.
\end{align}
Returning to the wavefunction, after substituting in $p,q$ and the integral independent factor, we find:
\begin{align}
\varphi_n(t)&=\frac{\sqrt{2}\sqrt{(2n)!}}{n!2^n}\sqrt{\frac{e^{\lambda t}}{1+e^{2\lambda t}}}\left(\frac{1-e^{2\lambda t}}{1+e^{2\lambda t}}\right)^n\\
&=(-1)^{n}\frac{\sqrt{(2n)!}}{n!2^n}\frac{(\tanh(\lambda t))^{n}}{(\cosh(\lambda t))^{1/2}}~.
\end{align}
We can also verify that the probability amplitudes are conserved
\begin{align}
    \sum_{n=0}^{\infty}|\varphi_n(t)|^2=1~.
\end{align}

\subsection{D. Examples of the Krylov Formalism in Classical Phase Space}
\label{app:ClassPhaseSpace}

\begin{figure}[htbp]
    \centering
        {\includegraphics[width=4.0cm]{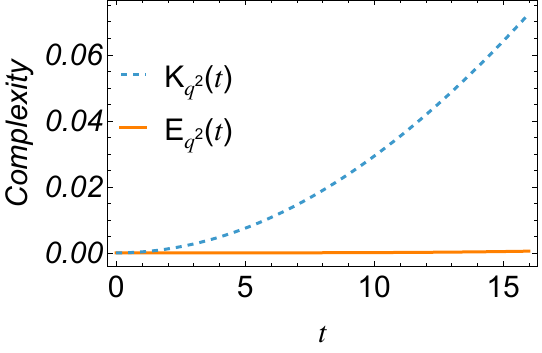}\hspace{3mm}}
        {\includegraphics[width=4.0cm]{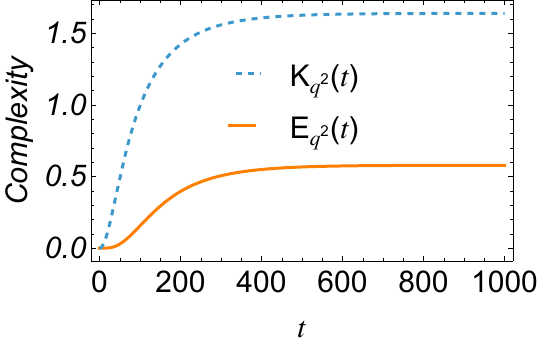}}
    \caption{\textbf{Left panel:} $K_{q^{2}}(t)$ (blue, dashed) and \(\mathbf{E}_{q^{2}}(t)=e^{\mathbf{L}_{q^{2}}\left(t\right)}-1\) (orange, solid) at early times for parameters $\epsilon=1/100$, $U_{0}=1/10$, $m=1/5$, $\omega=1/10$, and $\beta=1$. \textbf{Right panel:} $K_{f}(t)$ (blue, dashed) and \(e^{\mathbf{L}_{f}\left(t\right)}-1\) (orange, solid)  at long times for the same choice of parameters.}
    \label{fig-RepPotKvsLogK}
\end{figure}

In this appendix, we give further examples of the Krylov formalism discussed in the context of bosonic classical phase space.\\ \textit{I. Integrable System.}-- Consider the classical Hamiltonian of a one-dimensional simple harmonic oscillator written in terms of canonical coordinates $p,q$ as
\begin{align}
    \label{eq:SHOHam}
    H=\frac{1}{2m}p^{2}+\frac{1}{2}m\omega^{2}q^{2}~.
\end{align}
For arbitrary initial position $q(0)=q_{0}$ and conjugate momentum $p(0)=p_{0}$, the solution to Hamilton's equations is given by
\begin{align}
    \label{eq:SolHamiltonSHO}
    \begin{split}
   & q(t)=q_{0}\cos(\omega t)+\frac{p_{0}\sin(\omega t)}{m\omega}\,\,\,,\\
    &p(t)=p_{0}\cos(\omega t)-m \omega q_{0}\sin(\omega t)~.
    \end{split}
\end{align}
Consider the Boltzmann--Gibbs measure $\textrm{d}\mu_{\beta}(p,q):=e^{-\beta H(q,p)}\textrm{d}p\,\textrm{d}q$ corresponding to the thermal probability distribution in the canonical ensemble~\eqref{eq:BoltGibbsMeasure} as discussed in the main text. In this case, the phase-space average of functions $f\in \Sigma_{\mathcal{S}}$~\eqref{eq:PhaseSpaceAve} becomes the thermal Boltzmann--Gibbs ensemble average~\eqref{eq:BoltGibbsExpVal}, which is given by
\begin{align}
    \label{eq:PhaseSpaceAveSHOTHermal}
    \langle f\rangle_{\beta}=\frac{\beta \,\omega}{2\pi}\int_{\mathcal{S}}\,\textrm{d}p\,\textrm{d}q \,e^{-\beta H(q,p)}\,f(q,p)\,~.
\end{align}
Consider, for example, the quadratic observable $f(q,p):=q^{2}$. Using the inner product $\langle f,g\rangle=\langle f\cdot g\rangle_{\beta}-\langle f\rangle_{\beta}\langle g\rangle_{\beta}$~\eqref{eq:BoltGibbsInnProd}, we can find the orthonormal basis of functions $\lbrace \mathfrak{K}_{n}\rbrace$ starting from $f_{0}$ using~\eqref{eq:GrammSchmidtFunctions} and where the integral over $p,q$ becomes an integral over initial conditions $q_{0},p_{0}$. We find that the only non-vanishing elements of the basis are given by
\begin{align}
    \label{eq:GrammSchmidtFuncExampleSHO}
    \begin{split}
    &\mathfrak{K}_{0}=\frac{m\,\beta\, \omega^{2}}{\sqrt{2}}q_{0}^{2}~,\\
    &\mathfrak{K}_{1}=\,\beta\,\omega\,q_{0}\,p_{0}~,\\
    &\mathfrak{K}_{2}=\frac{\beta}{\sqrt{2}m}p_{0}^{2}~,
    \end{split}
\end{align}
where $f(q_{0},p_{0})\equiv f_{0}=\sqrt{2}\,\mathfrak{K}_{0}/(m\beta \omega^{2})$. Thus, the time evolution of $f(p,q)$, $f_{t}(p,q):=f(q(t),p(t))$ along curves of constant energy in phase space, given by $f_{t}(p,q)=q(t)^{2}$ with $q(t)$ given by~\eqref{eq:SolHamiltonSHO}, satisfies:
\begin{align}
    \label{eq:TimeEvolfSHO}
    \begin{split}
    &f_{t}(q_{0},p_{0})=c_{0}(t)\mathfrak{K}_{0}+c_{1}(t)\mathfrak{K}_{1}+c_{2}(t)\mathfrak{K}_{2}\\
    &=\frac{\sqrt{2}\cos(\omega t)^{2}}{m\beta \omega^{2}}\mathfrak{K}_{0}+\frac{\sin(2\omega t)}{m\beta \omega^{2}}\mathfrak{K}_{1}+\frac{\sqrt{2}\sin(\omega t)^{2}}{m\beta \omega^{2}}\mathfrak{K}_{2}~.
    \end{split}
\end{align}
Then, the classical Krylov complexity of the function $f_{t}(q,p):=q(t)^{2}$~\eqref{eq:ClassicalKrylovComplexity} is given by
\begin{align}
    \label{eq:ClassKrylovExampleSHO}
    K_{q^{2}}(t)=2\sin(\omega t)^{2}~.
\end{align}
Note that the dependence on the thermal scale $m\beta$ disappears due to the presence of the normalization of the observable $\vert \vert  f_{t}\vert \vert ^{2}\equiv \vert \vert  f_{0}\vert \vert ^{2} =2/(m\beta \omega^{2})^{2}$ in the denominator of the Krylov complexity. Also note that~\eqref{eq:ClassKrylovExampleSHO} is proportional to the \emph{quantum} Krylov complexity of the position operator $\hat{x}$ of the quantum simple harmonic oscillator given by $K_{\hat{x}}(t)=\sin^{2}(\omega t)$, but not of the squared position operator $\hat{x}^{2}$, which instead has a non-trivial temperature dependence (see App. A of ~\cite{Camargo:2023eev}). In contrast, the classical logK-complexity~\eqref{eq-Log-K replica} is given by
\begin{align}
    \label{eq:ClassLogKExampleSHO}
    \mathbf{L}_{q^{2}}(t)=\log(2)\sin(\omega t)^{4}~.
\end{align}
\textit{II. Integrable System with an Unstable Saddle.}-- As an example of a bosonic classical system with an unstable saddle, consider the repulsive potential
\begin{align}
    \label{eq:RepPot}
    U(\tilde{q}):=\frac{a^{2}U_{0}}{a^{2}+\tilde{q}^{2}}~,
\end{align}
with $a^{2}=1/(m\,\omega^{2})$ and $U_{0}:=U(\tilde{q}=0)>0$. For $\tilde{q}\sim \sqrt{\e} \,q$ and small $\epsilon$ ($1\gg \epsilon>0$), we can write~\eqref{eq:RepPot} as a power series in $\epsilon$
\begin{align}
    \label{eq:RepPotExpansion}
    U^{\epsilon}(q)\approx U_{0}\left(1-\frac{q^{2}}{a^{2}}\epsilon+ \frac{q^{4}}{a^{4}}\epsilon^{2}-O(\epsilon^{3})\right)~.
\end{align}
Now, take the Hamiltonian of a particle moving in the approximate potential~\eqref{eq:RepPotExpansion} up to quadratic order in $\epsilon$
\begin{align}
    \label{eq:RepPotHam}
    H^{\epsilon}=\frac{1}{2m}p^{2}+U_{0}\left(1-\frac{q^{2}}{a^{2}}\epsilon^{}+ \frac{q^{4}}{a^{4}}\epsilon^{2}\right)~.
\end{align}
For an arbitrary initial position $q(0)=q_{0}$ and momentum $p(0)=p_{0}$, Hamilton's equations can be solved up to linear order in $\epsilon$ exactly, and the solutions are given by 
\begin{align}
    \label{eq:SolHamiltonRepPotApproxEps}
    \begin{split}
   & q^{\epsilon}(t)=q_{0}\cosh( \tilde{\Omega}^{\epsilon}_{0}\,t)+\frac{p_{0}\sinh( \tilde{\Omega}^{\epsilon}_{0}\,t)}{m\tilde{\Omega}^{\epsilon}_{0}\,}\,\,\,,\\
    &p^{\epsilon}(t)=p_{0}\cosh( \tilde{\Omega}^{\epsilon}_{0}\, t)+m\tilde{\Omega}^{\epsilon}_{0}\, \,q_{0} \sinh( \tilde{\Omega}^{\epsilon}_{0}\,t)~,
    \end{split}
\end{align}
where we defined $\tilde{\Omega}^{\epsilon}_{0}:=\omega\,\sqrt{2\,U_{0}\epsilon}$. In other words,~\eqref{eq:SolHamiltonRepPotApproxEps} provide an approximate solution to Hamilton's equations for the repulsive potential~\eqref{eq:RepPot} near $\tilde{q}=0$. Because of this, the Hamiltonian~\eqref{eq:RepPotHam} evaluated for~\eqref{eq:SolHamiltonRepPotApproxEps} is only approximately conserved for small $\epsilon$
\begin{align}
    \label{eq:RepPotHamSolApprox}
    H_{0}^{\epsilon}\approx\frac{p_{0}^{2}}{2m}+U_{0}\left(1-m\omega^{2}q_{0}^{2}\epsilon^{}+\ldots\right)~,
\end{align}
where the ellipsis denotes terms of order $O(\epsilon^{2})$ and higher, and which are time-dependent. The unstable saddle point is located at $\lbrace q_{S},p_{S} \rbrace=\lbrace 0,0 \rbrace$, around which the solution has a classic Lyapunov exponent $\lambda_{cl}\equiv \tilde{\Omega}_{0}^{\epsilon}=\omega\,\sqrt{2\,U_{0}\epsilon}$. Here, $\epsilon$ is used to keep track of the perturbation order around the saddle point $\lbrace q_{S},p_{S} \rbrace$, but we can absorb it into the coordinates $\lbrace q^{\epsilon},p^{\epsilon}\rbrace$ by defining $\delta q=\sqrt{\epsilon} q^{\epsilon}, \delta p=\sqrt{\epsilon} p^{\epsilon}$ and solving Hamilton's equations up to linear order in $\delta q$. \\
$\phantom{word}$Keeping the quartic term $q^4$ in~\eqref{eq:RepPotHam} allows us to consider the same thermal measure in the phase space average as in~\eqref{eq:PhaseSpaceAveSHOTHermal}. In this case,
\begin{align}
    \label{eq:ThermalMeasureRepPot}
    \begin{split}
   & \mu(\mathcal{S})\equiv Z_{\beta}^{\epsilon}=\int\textrm{d}q\,\textrm{d}p \,e^{-\beta H^{\epsilon}(p,q)}\\
   &=\frac{\pi^{3/2}e^{-\frac{7\beta\, U_{0}}{8}}}{2\omega\sqrt{\beta\epsilon}\,}\left(I_{\frac{1}{4}}\left(\frac{\beta\, U_{0}}{8}\right)+I_{-\frac{1}{4}}\left(\frac{\beta \,U_{0}}{8}\right)\right)~,
    \end{split}
\end{align}
where $I_{n}(z)$ is the modified Bessel function of the first kind. Here we once again consider the Boltzmann--Gibbs ensemble average 
\begin{align}
    \label{eq:PhaseSpaceAveSHOTHermalRepPot}
    \langle f\rangle_{\beta}^{\epsilon}=\frac{1}{Z^{\epsilon}_{\beta}}\int\,\textrm{d}p\,\textrm{d}q \,e^{-\beta H^{\epsilon}(q,p)}\,f(q,p)\,~.
\end{align}
Similarly to the integrable case, we consider the quadratic observable $f(q,p)=q^{2}$ and construct the Krylov basis of functions using~\eqref{eq:GrammSchmidtFunctions}, but limiting ourselves to contributions at most linear in $\epsilon$ and quadratic in combinations of $q_{0},p_{0}$
\begin{align}
    \label{eq:GrammSchmidtFuncExampleUnstSaddle}
    \begin{split}
    &\mathfrak{K}_{0}=A_{0}\,q_{0}^{2}~,\\
    &\mathfrak{K}_{1}=A_{1}\,q_{0}\,p_{0}~,\\
    &\mathfrak{K}_{2}=\frac{\beta}{\sqrt{2}m}\,p_{0}^{2}~,
    \end{split}
\end{align}
where $A_{0}=A_{0}(\beta,U_{0},\omega,\epsilon)$ and $A_{1}=A_{1}(\beta,U_{0},\omega,\epsilon)$ involve sums of modified Bessel functions of the first kind. Thus, the time evolution of $f(q,p)$ evaluated along the approximate solution~\eqref{eq:SolHamiltonRepPotApproxEps}, $f(q^{\epsilon}(t),p^{\epsilon}(t))=f_{t}(q,p)$, is given by
\begin{align}
    \label{eq:TimeEvolfRepPotAPp}
    \begin{split}
    &f_{t}(q_{0},p_{0})=c_{0}(t)\mathfrak{K}_{0}+c_{1}(t)\mathfrak{K}_{1}+c_{2}(t)\mathfrak{K}_{2}\\
    &=\tilde{A}_{0}\cosh(\tilde{\Omega}^{\epsilon}_{0} t)^{2}\mathfrak{K}_{0}+\tilde{A}_{1}\sinh(2\tilde{\Omega}^{\epsilon}_{0} t)\mathfrak{K}_{1}\\
   & +\frac{\sqrt{2}\sinh(\tilde{\Omega}^{\epsilon}_{0} t)^{2}}{m\beta (\tilde{\Omega}^{\epsilon}_{0})^{2}}\mathfrak{K}_{2}~,
    \end{split}
\end{align}
where $\tilde{A}_{0}=\tilde{A}_{0}(\beta,U_{0},\omega,\epsilon)$$:=\langle f_{t}, \mathfrak{K}_{0}\rangle/(\cosh(\tilde{\Omega}^{\epsilon}_{0} t)^{2})$ and $\tilde{A}_{1}=\tilde{A}_{1}(\beta,U_{0},\omega,\epsilon)=\langle f_{t}, \mathfrak{K}_{1}\rangle/(\sinh(2\tilde{\Omega}^{\epsilon}_{0} t))$. From ~\eqref{eq:TimeEvolfRepPotAPp} we can compute the classical K and logK complexity of $f=q^{2}$ using Eqs.~\eqref{eq:ClassicalKrylovComplexity} and \eqref{eq:ClassicalLogK}. This can be done analytically, although their functional form is not particularly illuminating. Instead, in Figure~\ref{fig-RepPotKvsLogK} we display the behavior of $K_{f}(t)$ and $\mathbf{E}_{f}(t)=e^{\mathbf{L}_{f}(t)}-1$ computed for a choice of parameters.
From this Figure we see that, as expected for a system with a single unstable saddle (see Table~\ref{tab:KrylovTab}), at early times the elogK-complexity grows slower than K-complexity and at long times it saturates to a smaller value. This behavior can also be seen for other choices of parameters, suggesting that logK complexity successfully avoids the exponential growth coming from the unstable saddle point.



\subsection{E. Classical K and logK Complexity in the LMG Model}
\label{app:ExampleClassicLMG}
In this section, we provide an example of Krylov complexity in the LMG model using the classical phase space formalism and provide definitions for higher-order Krylov complexities. To begin with, we briefly review the phase space dynamical analysis. The classical Hamiltonian of the LMG model is given as  
\begin{equation}
    H=x+2z^2~,
\end{equation}
where $x$, $y$, $z$ are the classical correspondences of generators of $SU(2)$ spin that live on a unit sphere, satisfying:
\begin{equation}
    x^2+y^2+z^2=1~.
    \label{App-eq-SU2sphere}
\end{equation}
This is a non-canonical formulation in the spin classical phase space. Therefore, instead of the Poisson bracket, they satisfy the Lie-Poisson bracket:
\begin{equation}
    \{x_i,x_j\}=\epsilon_{ijk}x_k~,
\end{equation}
in our case of $SU(2)$. These coordinates live on a compact $S^2$ sphere in contrast with the canonical coordinates. As a consequence, the symplectic manifold is not flat, but controlled by the Kirillov–Kostant–Souriau form. The integral measure can be written as
\begin{equation}
    d\mu\propto \delta(x^2+y^2+z^2-1)dxdydz~,
\end{equation}
subject to the constraint (\ref{App-eq-SU2sphere}). One main inference from this is that in contrast with canonical phase space, the phase space volume in this formalism is finite due to its compactness, which is set to be 1 in the main text. 
The dynamics of these coordinates are controlled by Hamilton's equations:
\begin{eqnarray}
    \dot{x}&=&\{H,x\}=4yz~,\\
    \dot{y}&=&\{H,y\}=z-4xz~,\\
    \dot{z}&=&\{H,z\}=-y~.
\end{eqnarray}
By evaluating the Jacobian, we could find the fixed point with positive eigenvalues:
\begin{equation}
    (x,y,z)=(1,0,0)~.
\end{equation}
The dynamics of variation of coordinates near this unstable saddle-point is:
\begin{equation}
    \dot{\d x}=0~,~~~\dot{\d y}=-3\d z~,~~~\dot{\d z}=-\d y~.
\end{equation}
The solution of such equations of motion can be found as:
\begin{equation}
    \d x(t)=\d x_0~,~~~\d Y(t)=\d Y_0 e^{-\l_{\mathrm{cl}}t}~,~~~\d Z(t)=\d Z_0 e^{\l_{\mathrm{cl}}t}~,
\end{equation}
where $Y=\l_{\mathrm{cl}}z+y$,  $Z=\l_{\mathrm{cl}}z-y$, and $\l_{\mathrm{cl}}=\sqrt{3}$. 
Re-write the solutions in $(x,y,z)$ coordinate:
\begin{eqnarray}
    \d z(t)&=&\d z_0 \cosh(\l_{\mathrm{cl}} t)-\frac{\d y_0}{\l_{\mathrm{cl}}} \sinh(\l_{\mathrm{cl}} t)~,\\
    \d y(t)&=&\d y_0 \cosh (\l_{\mathrm{cl}} t)-\l_{\mathrm{cl}} \d z_0 \sinh(\l_{\mathrm{cl}} t)~.
\end{eqnarray}
Following our numerical study in Section.~\hyperref[subsec:NumericalAnalysisLMG]{V.A.}, we choose $\hat{z}$ as the initial operator. Classically, we start with the initial function in phase space as 
\begin{equation}
    \d z(t)(\d y_0,\d z_0)~,
\end{equation}
which is not necessarily a function of $\d x_0$ due to the measure on $S^2$: $\d(\d x_0^2+\d y_0^2+\d z_0^2-1)$. 
By performing a change of phase space coordinates, one can find:
\begin{equation}
    \d x(t)=\d x_0~,~~~\d Y(t)=\d Y_0 e^{-\l_{\mathrm{cl}}t}~,~~~\d Z(t)=\d Z_0 e^{\l_{\mathrm{cl}}t}~,
\end{equation}
where $Y=\l_{\mathrm{cl}}z+y$,  $Z=\l_{\mathrm{cl}}z-y$, and $\l_{\mathrm{cl}}=\sqrt{3}$. The exponential growth is restricted to the strip: 
\begin{eqnarray}
    -\d/2&<\d Y_0<&\d/2~,\\
    -\d e^{-\l_{\mathrm{cl}} t}/2&<\d Z_0<&\d e^{-\l_{\mathrm{cl}} t}/2~.
\end{eqnarray} 
We can then find the strip in $(x,y,z)$ coordinates by performing a reverse coordinate change: $z=\frac{Y+Z}{2\l_{\mathrm{cl}}}$ and $y=\frac{Y-Z}{2}$. The determinant of the Jacobian is 
\begin{equation}
    J=\left\vert \begin{bmatrix}
\frac{\partial \d Z }{\partial z} & \frac{\partial \d Z }{\partial y} \\
\frac{\partial \d Y }{\partial z} & \frac{\partial \d Y }{\partial y} 
\end{bmatrix}
\right\vert=\left\vert \begin{bmatrix}
\l_{\mathrm{cl}} & -1 \\
\l_{\mathrm{cl}} & 1 
\end{bmatrix}
\right\vert=2\l_{\mathrm{cl}}~.
\end{equation}
The integration is thus changed as:
\begin{equation}
    dy_0dz_0=\frac{1}{2\l_{\mathrm{cl}}}dY_0dZ_0~,
\end{equation}
with the relation of the associated measure as
\begin{equation}
    \mu_{Y_0Z_0}=\d^2 e^{-\l_{\mathrm{cl}} t}=2\l_{\mathrm{cl}} \mu_{y_0 z_0}~.
\end{equation}
We can therefore perform the integral in terms of coordinates $\d Y_0, \d Z_0$ with the associated measure.

\begin{figure}[h]
    \centering
        {\includegraphics[width=4.2cm]{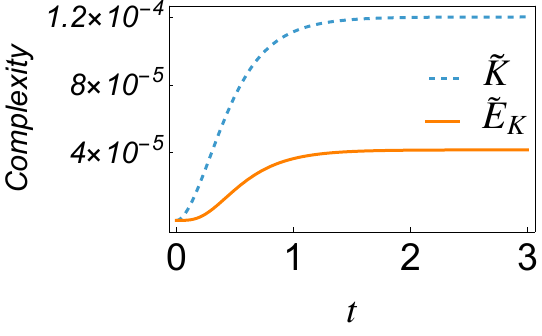}\hspace{3mm}}
        {\includegraphics[width=3.8cm]{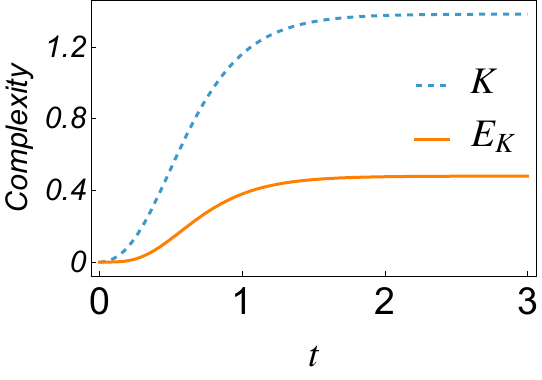}}
    \caption{ Un-normalized and normalized Krylov and elogK complexities in the LMG model for initial function $\d^2z(t)$ from classical phase space analysis.
    \textbf{Left panel:} un-normalized Krylov complexity (\ref{App-eq-UnnorK}) (blue, dashed) and Un-normalized elogK complexity using (\ref{App-eq-UnnorLK}) (orange), \textbf{Right panel:} normalized Krylov complexity (\ref{App-eq-NorK}) (blue, dashed) and normalized elogK complexity using (\ref{App-eq-NorLK}) (orange).}
    \label{fig:LMGz2}
\end{figure}

Following the classical Krylov algorithm (\ref{eq:GrammSchmidtFunctions}), we can find the following time-evolved Krylov basis:
\begin{eqnarray}
    \mathfrak{K_0}&=&\frac{2 \sqrt{3} e^{-\lambda_{\mathrm{cl}} t} \left(\d Y_0 e^{2 \lambda_{\mathrm{cl}}  t}+\d Z_0\right)}{ \sqrt{\delta ^2 e^{-4 \lambda_{\mathrm{cl}}  t} \left(e^{6 \lambda_{\mathrm{cl}}  t}+1\right)}}~,\\
    \mathfrak{K_1}&=&-\frac{2 \sqrt{3} e^{-\lambda_{\mathrm{cl}} t} \sqrt{\frac{\delta ^2 e^{2 \lambda_{\mathrm{cl}}  t}}{e^{6 \lambda_{\mathrm{cl}}  t}+1}} \left(\d Z_0 e^{4 \lambda_{\mathrm{cl}}  t}-\d Y_0\right)}{\delta ^2}~.\nonumber\\
\end{eqnarray}
After setting $\l_{\mathrm{cl}}=\sqrt{3}$ and $\d=1$, we can find the transition amplitudes
\begin{eqnarray}
    |c_0(t)|^2&=&\frac{1}{72} \left(1-\tanh \left(\sqrt{3} t\right)\right)~,\\
    |c_1(t)|^2&=&\frac{1}{72} \sinh ^2\left(\sqrt{3} t\right) \left(1-\tanh \left(\sqrt{3} t\right)\right)~.
\end{eqnarray}
Therefore, the un-normalized classical Krylov complexity is
\begin{eqnarray}
    \tilde{K}(t)&=&|c_1(t)|^2\\
    &=&\frac{1}{72} \sinh ^2\left(\sqrt{3} t\right) \left(1-\tanh \left(\sqrt{3} t\right)\right)~.
\end{eqnarray}
While the normalized one has a simpler form:
\begin{eqnarray}
    K(t)&=&\frac{|c_1(t)|^2}{|c_0(t)|^2+|c_1(t)|^2}\\
    &=& \tanh ^2\left(\sqrt{3} t\right)~.
\end{eqnarray}
Both complexities grow quadratically at the initial time as expected from their quantum correspondence:
\begin{equation}
    \tilde{K}(t)\vert_{t\ra 0}\approx\frac{t^2}{24}~,~~~K(t)\vert_{t\ra 0}\approx 3t^2~.
\end{equation}
Both $K(t)$ and $\tilde{K}(t)$ exhibit similar behavior with a scale difference. More importantly, they both show a sub-exponential growth and vanish at late times, which is the expected behavior from integrable systems. 
Because the phase space is two-dimensional, so based on the definition of logK complexity, it vanished identically, which produces no exponent. It is therefore fair to say that the logK definition indeed suppresses the exponent since no classical chaos exists in a 2-dimensional phase space due to the Poincaré–Bendixson theorem for isolated systems. Moreover, Krylov complexity itself in classical phase space already indicates no scrambling behavior, unlike OTOC. Therefore, classical Krylov complexity is a good enough indicator to eliminate fake scrambling, which thus tells us that quantum scrambling cannot imply classical scrambling or chaos. 
To consider non-trivial contributions from (e)logK, we study a quadratic function:
\begin{equation}
    \d z^2(t)~,
\end{equation}
for which the nested Lie-Poisson brackets give three independent functions:
\begin{equation}
    \{\d z^2(t),~~~-4\d y(t)\d z(t)),~~~2\d y^2(t)+6\d z^2(t)\}~.
\end{equation}
Similarly, after performing the coordinate change to $\{\d Y_0, \d Z_0\}$ coordinates, we follow the classical Krylov algorithm to obtain three orthonormal bases:
\begin{eqnarray}
\mathfrak{K}_0&=&\frac{4 \sqrt{15} e^{-2 \lambda_{\mathrm{cl}}  t} \left(\d Y_0+\d Z_0 e^{2 \lambda_{\mathrm{cl}}  t}\right)^2}{\delta ^2 \sqrt{3 e^{-4 \lambda_{\mathrm{cl}}  t}+10 e^{-2 \lambda_{\mathrm{cl}}  t}+3}}~,\\
    \mathfrak{K}_1&=&-6 \sqrt{10} e^{-\lambda_{\mathrm{cl}} t} \left(\d Y_0+\d Z_0 e^{2 \lambda_{\mathrm{cl}}  t}\right) \times\nonumber\\
    &&\frac{ \left((3 \d Y_0-5 \d Z_0) e^{2 \lambda_{\mathrm{cl}}  t}+5 \d Y_0-3 \d Z_0\right) }{\delta ^2 \sqrt{156 \cosh (2 \lambda_{\mathrm{cl}}  t)+45 \cosh (4 \lambda_{\mathrm{cl}}  t)+55}}~,\\
    \mathfrak{K}_2&=&\sqrt{\frac{6}{7}} \left[\frac{ \left(45 \d Y_0^2-56 \d Y_0 \d Z_0+45 \d Z_0^2\right) e^{2 \lambda_{\mathrm{cl}}  t}}{\delta ^2 \sqrt{2 e^{2 \lambda_{\mathrm{cl}}  t}+15 e^{4 \lambda_{\mathrm{cl}}  t}+15}}\right.\nonumber\\
    &&\left.-\frac{ 25 \d Y_0^2+25 \d Z_0^2 e^{4 \lambda_{\mathrm{cl}}  t}}{\delta ^2 \sqrt{2 e^{2 \lambda_{\mathrm{cl}}  t}+15 e^{4 \lambda_{\mathrm{cl}}  t}+15}}\right].
\end{eqnarray}
The transition amplitudes for each basis are given as
\begin{eqnarray}
    |c_0(t)|^2&=&\frac{e^{-2 \sqrt{3} t} \left(5 \cosh \left(2 \sqrt{3} t\right)+19\right)^2}{155520 \left(3 \cosh \left(2 \sqrt{3} t\right)+5\right)}~,\\
    |c_1(t)|^2&=&\frac{2 e^{-6 \sqrt{3} t} \left(e^{4 \sqrt{3} t}-1\right)^2}{405 \left(156 \cosh \left(2 \sqrt{3} t\right)+45 \cosh \left(4 \sqrt{3} t\right)+55\right)}\nonumber~,\\
    \\
    |c_2(t)|^2&=&\frac{7 e^{-6 \sqrt{3} t} \left(e^{2 \sqrt{3} t}-1\right)^4}{5184 \left(45 \cosh \left(2 \sqrt{3} t\right)+3\right)}~,
\end{eqnarray}
where we have specified $\d=1, \l_{\mathrm{cl}}=\sqrt{3}$. Therefore, we can explicitly give the results for both un-normalized and normalized classical Krylov complexities:
\begin{eqnarray}
\label{App-eq-UnnorK}
    \tilde{K}(t)&=&|c_1(t)|^2+2|c_2(t)|^2\nonumber\\
    &=&\frac{e^{-2 \sqrt{3} t} \sinh ^2\left(\sqrt{3} t\right)}{9720}\nonumber\\
   &\times& \frac{ \left(332 \cosh \left(2 \sqrt{3} t\right)+105 \cosh \left(4 \sqrt{3} t\right)-53\right)}{ \left(3 \cosh \left(2 \sqrt{3} t\right)+5\right) \left(15 \cosh \left(2 \sqrt{3} t\right)+1\right)}~,\nonumber\\
   \\
    K(t)&=&\frac{|c_1(t)|^2+2|c_2(t)|^2}{|c_0(t)|^2+|c_1(t)|^2+|c_2(t)|^2}\nonumber\\
    &=&\frac{16 \sinh ^2\left(\sqrt{3} t\right)}{9}\nonumber\\
    &\times&\frac{ \left(332 \cosh \left(2 \sqrt{3} t\right)+105 \cosh \left(4 \sqrt{3} t\right)-53\right)}{\left(3 \cosh \left(2 \sqrt{3} t\right)+5\right)^2 \left(15 \cosh \left(2 \sqrt{3} t\right)+1\right)}~.\nonumber\\
    \label{App-eq-NorK}
\end{eqnarray}
Their behaviors at the vicinity of $t=0$ are: $\tilde{K}(t)\vert_{t\ra0}\approx \frac{\delta ^4 t^2}{360 \lambda_{\mathrm{cl}} ^2}$, and $K(t)\vert_{t\ra0}\approx \frac{2 \lambda_{\mathrm{cl}} ^2 t^2}{3}$. In this case, we are able to compute logK complexity explicitly:
\begin{eqnarray}
\label{App-eq-UnnorLK}
    \tilde{\mathbf{L}}_K(t)&=&\log(2)|c_2(t)|^2\nonumber\\
    &=&\frac{7 e^{-6 \sqrt{3} t} \left(e^{2 \sqrt{3} t}-1\right)^4 \log (2)}{5184 \left(45 \cosh \left(2 \sqrt{3} t\right)+3\right)}~,\\
    \mathbf{L}_K(t)&=&\frac{\log(2)|c_2(t)|^2}{|c_0(t)|^2+|c_1(t)|^2+|c_2(t)|^2}\nonumber\\
    &=&\frac{2240 \log (2) \sinh ^4\left(\sqrt{3} t\right)}{9 \left(156 \cosh \left(2 \sqrt{3} t\right)+45 \cosh \left(4 \sqrt{3} t\right)+55\right)}~.\nonumber\\
    \label{App-eq-NorLK}
\end{eqnarray}
Their initial time behaviors are given as: $\tilde{\mathbf{L}}_K(t)\vert_{t\ra0}\approx \frac{7\log(2)}{1728}t^4$ and $\mathbf{L}_K(t)\vert_{t\ra0}\approx \frac{35\log(2)}{4}t^4$, which matches with our analytical analysis. Fig.~\ref{fig:LMGz2} compares the behavior of the un-normalized and normalized Krylov and log-Krylov complexities.

\end{document}